\documentclass[twocolumn]{aastex63}

\usepackage{graphicx}	% Including figure files
\usepackage{amsmath}	% Advanced maths commands
\usepackage{amssymb}	% Extra maths symbols
\usepackage{longtable}

\newcommand{ \Msun } {M$_{\odot}$}
\newcommand{ \MHtwo } {$M(H_2)$}
\newcommand{ \Herschel } {\textit{Herschel}}

\newcommand{ \coldgass} {xCOLD~GASS}

\newcommand{\ratio }{$r_{31}$}
\newcommand{\ratiotwo }{$r_{21}$}
\newcommand{\uclpdr}{{\tt UCL-PDR}}
\newcommand{\nH }{$n_\text{H}$}
\newcommand{\LCOone }{$L'_\text{CO(1-0)}$}

\newcommand{\LCOthree }{$L'_\text{CO(3-2)}$}
\newcommand{\kms }{km s$^{-1}$}
\newcommand{\Nbass}{36} 
\newcommand{\Nbasstot}{46}
\newcommand{\Nslugs}{38} 

\accepted{December 1, 2019}
\shorttitle{CO(3-2)/CO(1-0) line ratio }
\shortauthors{Lamperti et al.}
\begin{document}

\title{The CO(3-2)/CO(1-0) luminosity line ratio in nearby star-forming galaxies and AGN from \\xCOLD GASS, BASS and SLUGS}

\correspondingauthor{Isabella Lamperti}
\email{isabellalamperti@gmail.com}

%\author[0000-0002-0786-7307]{Isabella Lamperti}
%\affiliation{Department of Physics \& Astronomy, University College London, Gower Street, London WC1E 6BT, UK}

\author{Isabella Lamperti}
\affiliation{Department of Physics \& Astronomy, University College London, Gower Street, London WC1E 6BT, UK}

\author{Am{\'e}lie Saintonge}
\affiliation{Department of Physics \& Astronomy, University College London, Gower Street, London WC1E 6BT, UK}

\author{Michael Koss}
\affiliation{Eureka Scientific, 2452 Delmer Street, Suite 100, Oakland, CA 94602-3017, USA}

\author{Serena Viti}
\affiliation{Department of Physics \& Astronomy, University College London, Gower Street, London WC1E 6BT, UK}

\author{Christine D. Wilson}
\affiliation{Department of Physics and Astronomy, McMaster University, Hamilton, ON L8S 4M1, Canada}

\author{Hao He}
\affiliation{Department of Physics and Astronomy, McMaster University, Hamilton, ON L8S 4M1, Canada}

\author{T. Taro Shimizu}
\affiliation{Max-Planck-Institut f{\"u}r extraterrestrische Physik, Postfach 1312, 85741, Garching, Germany}

\author{Thomas R. Greve}
\affiliation{Department of Physics \& Astronomy, University College London, Gower Street, London WC1E 6BT, UK}
\affiliation{Cosmic Dawn Center}

\author{Richard Mushotzky}
\affiliation{Department of Astronomy, University of Maryland, College Park, MD 20742, USA}

\author{Ezequiel Treister}
\affiliation{Instituto de Astrof\'{\i}sica, Facultad de F\'{\i}sica, Pontificia Universidad Cat\'olica de Chile, Casilla 306, Santiago 22, Chile}

\author{Carsten Kramer}
\affiliation{Institut de RadioAstronomie Millim{\'e}trique (IRAM), Avenida Divina Pastora 7, 18012 Granada, Spain}

\author{David Sanders}
\affiliation{Institute for Astronomy, University of Hawaii, 2680 Woodlawn Drive, Honolulu, HI, 96822, USA}

\author{Kevin Schawinski}
\affiliation{Modulos AG, Technoparkstr. 1, CH-8005, Z{\"u}rich, Switzerland}

\author{Linda J. Tacconi}
\affiliation{Max-Planck-Institut f{\"u}r extraterrestrische Physik, Postfach 1312, 85741, Garching, Germany}

\nocollaboration{14}



\begin{abstract}
We study the $r_{31} =L'_\text{CO(3-2)}/L'_{\text{CO(1-0)}}$
 luminosity line ratio in a sample of nearby ($z<0.05$) galaxies: 25 star-forming galaxies (SFGs) from the xCOLD GASS survey, \Nbass\ hard X-ray selected AGN host galaxies from  BASS and 37 infrared luminous galaxies from SLUGS.
We find a trend for $r_{31}$ to increase with star-formation efficiency (SFE).
 We model $r_{31}$ using the \uclpdr\ code and find that the gas density is the main parameter responsible for variation of $r_{31}$, while the interstellar radiation field and cosmic ray ionization rate play only a minor role.
We interpret these results to indicate a relation between SFE and gas density.
 We do not find a difference in the $r_{31}$ value of SFGs and AGN host galaxies, when the galaxies are matched in SSFR ($<r_{31}> = 0.52\pm0.04$ for SFGs and $<r_{31}> = 0.53\pm0.06$ for AGN hosts).
According to the results of \uclpdr\ models, the X-rays can contribute to the enhancement of the CO line ratio, but only for strong X-ray fluxes and for high gas density ($n_\text{H}>10^4$ cm$^{-3}$).
 We find a mild tightening of the Kennicutt-Schmidt relation when we use the  molecular gas mass surface density traced by CO(3-2) (Pearson correlation coefficient $R=0.83$), instead of the molecular gas mass surface density traced by CO(1-0) ($R=0.78$), but the increase in correlation is not statistically significant ($p$-value = 0.06). This suggests that the CO(3-2) line can be reliably used to study the relation between SFR and molecular gas for normal SFGs  at high redshift, and to compare it with studies of low-redshift galaxies, as is common practice.
\end{abstract}

%\keywords{editorials, notices --- miscellaneous --- catalogs --- surveys}



\section{Introduction} 
\label{sec:intro}

Star formation in galaxies is closely related to their  gas content. This has been found in the correlation between the star-formation rate (SFR) surface density and  gas mass surface density \citep[Kennicutt-Schmidt (KS) relation, ][]{Kennicutt1998}. The relation between SFR  and molecular gas content is stronger than with the total gas content \citep{Bigiel2008, Leroy2008, Saintonge2017}.
However, there is some scatter in this relation: the SFR surface density can vary by an order of magnitude for the same molecular gas mass surface density, measured from the CO(1-0) luminosity \citep{Saintonge2012}. 
A possible explanation is that CO(1-0) is a good tracer of the total  molecular gas in massive galaxies, but it does not accurately trace the amount of gas located in the dense molecular cores where the formation of stars takes place \citep[e.g.,][]{Solomon1992, Kohno2002, Shibatsuka2003}.
Since stars form in dense molecular clouds, it is reasonable to expect the SFR to correlate better with  the amount of dense molecular gas than with the total (dense and diffuse) molecular gas.
Commonly used tracers of dense gas are HCN, HCO+ or CS \citep[e.g.,][]{Tan2018, Gao2004a, Gao2004b, Wu2010, Zhang2014}.

   Observations have shown that the HCN(1-0)/CO(1-0) ratio is enhanced in galaxies with high star-formation efficiency (SFE =  SFR/\MHtwo), like Luminous Infra-Red Galaxies \citep[LIRGs;][]{Gao2004a, Gracia-Carpio2008, Garcia-Burillo2012}. However, the HCN(1-0) line flux is usually fainter than CO by more than an order of magnitude, making surveys of large samples of normal star-forming galaxies very time consuming. Another option is to use higher CO transitions to trace the mass of dense molecular gas. 
 The ideal transition is CO(3-2): it does not trace low density gas \citep[critical density $n_{crit}=3.6\cdot 10^{4}$ cm$^{-3}$, calculated under the optically thin assumption,][]{Carilli2013} like the CO(1-0) and CO(2-1) transitions, and at the same time it does not require high temperatures to populate it \citep[the minimum gas temperature needed for significant excitation is $T_{min}=33$~K;][]{Mauersberger1999, Yao2003, Wilson2009}.
 If the gas density is the key quantity regulating the relation between molecular gas mass and SFR, then we expect to see a correlation between the SFE and the \ratio $=L'_\text{CO(3-2)}/L'_\text{CO(1-0)}$ luminosity line ratio, that can be interpreted as an indicator of the gas density.

The $r_{31}$ value has been measured in samples of luminous infrared  galaxies \citep{Leech2010, Papadopoulos2012}, 
 in the central regions of nearby galaxies \citep{Mauersberger1999, Mao2010}, in sub-millimeter galaxies \citep[SMGs,][]{Harris2010},  and in nearby galaxies \citep{Wilson2012}.
\cite{Yao2003} and \cite{Leech2010} found a trend for \ratio\ to increase with increasing star formation efficiency in samples of infrared luminous galaxies and LIRGs.
 This trend has also been found in spatially resolved observations of M 83, NGC 3627, and NGC 5055 \citep{Muraoka2007, Morokuma2017}.  \cite{Sharon2016} found a similar trend in a sample of sub-millimeter galaxies and AGN-hosts at redshift $z=2-3$.
Most studies of the \ratio\ line ratio focused on extreme objects, like LIRGs, or are limited to small samples. In this work, we collect CO observations for a homogeneous sample of  main-sequence galaxies to investigate the $r_{31}$ line ratio in more `normal' star-forming galaxies. 

We also analyse a sample of galaxies hosting active galactic nuclei (AGN), to investigate if the AGN has an effect on the \ratio\ line ratio of its host galaxy.
Several studies of the CO Spectral Line Energy Distribution (SLED) of AGN focused on the high-J rotational transition levels. For instance, \cite{Lu2017} studied the CO SLED in the GOALS sample \citep[The Great Observatories All-Sky LIRG Survey][]{Armus2009} and found that the presence of an AGN influences  only the very high J levels (J $>$ 10). \cite{Mashian2015} find that the CO SLED is not the same in all AGN and that the shape of the CO SLED of a galaxy is more related to the  content of warm and dense molecular gas than to the excitation mechanism.
\cite{Rosenberg2015}  analyse the CO ladder of 29 objects from the \textit{Herschel} Comprehensive ULIRG Emission Survey (HerCULES). They find that in objects with a large AGN contribution the CO ladder peaks at higher J levels, which means that in these objects the CO excitation is influenced by harder radiation sources (X-rays or cosmic rays).
 These studies focus mostly on the high J levels (J $>$ 4).
\cite{Rosario2018} studied the molecular gas properties, traced by CO(2-1), of a sample of 20 nearby (z $<$ 0.01)  hard X-ray selected AGN hosts from the LLAMA survey and compare it with a control sample of star-forming galaxies. They found similar molecular gas fraction and SFE in the central region of AGN and in the control galaxies. 
Also \cite{Sharon2016} compared the $r_{31}$ values of 15 SMGs and 13 AGN host galaxies at redshift $ z = 2-3$ and did not find a significant difference.

In this paper we study the the \ratio $=L'_\text{CO(3-2)}/L'_\text{CO(1-0)}$ luminosity line ratio line in a sample of nearby (z $<$ 0.05) star-forming galaxies and AGN.
In Sections \,\ref{sec:sample} and \ref{sec:CO_data} we describe the  sample and the CO observations. In Section\,\ref{sec:r31_results} we present the \ratio\ values and analyse the correlation with SFR, SSFR and SFE. We also compare the \ratio\ values for AGN and star-forming galaxies. In Section\,\ref{sec:uclpdr} we use modelling of the line ratio using a PDR (photo-dissociation region) code to test which parameters regulate the CO line ratios.
Finally in Section\,\ref{sec:KS_relation} we compare the Kennicutt-Schmidt relation with molecular gas masses derived using the CO(1-0) and CO(3-2) line emission.

Throughout this work, we assume a cosmological model with $\Omega_\lambda = 0.7$, $\Omega_\text{M}= 0.3$ , and $H_0 = 70$ km s$^{-1}$ Mpc$^{-1}$.

\newpage
\section{Sample}
\label{sec:sample}
\subsection{Star-forming galaxies: xCOLD GASS}
  The \coldgass\ survey \citep{Saintonge2011, Saintonge2017} was designed to observe the  CO(1-0) emission for $\sim$500 galaxies in order to establish the first unbiased scaling relations between the cold gas (atomic and molecular) contents of galaxies and their stellar, structural, and chemical properties. 
A sample of 25 galaxies from \coldgass\ also has JCMT observations of the CO(3-2) emission line. The sample was selected based on the following criteria:
\begin{itemize}
\item good detection of the CO(1-0) line (signal-to-noise of the line $>3$);
\item CO(3-2) luminosity high enough to require less than two hours of integration time with the JCMT in band 3 (opacity $\tau_{225 \text{GHz}} = 0.08-0.12$). Assuming \ratio = 0.5,  this requirement corresponds to CO(1-0) luminosities $L'_\text{CO(1-0)}> 10^8$ K \kms\ pc$^2$. 
\item  the targets were selected to span a broad range of specific star-formation rate ($\text{SSFR}=\text{SFR}/M_*$, $-10.5 < \log \text{SSFR/yr}^{-1}< -8.5$) and star-formation efficiency ($\text{SFE}=\text{SFR}$/\MHtwo, $ -9.5 < \log \text{SFE/yr}^{-1} < -8$).
\end{itemize}
  
 The galaxies in the sample are in the redshift interval $0.026 < z < 0.049$. They have stellar masses in the range $10 < \log M_{*}/M_{\odot} < 11$ and star-formation rates in the range $-0.05 < \log \text{SFR/[\Msun\ yr}^{-1}] < 1.54$.

All the galaxy properties are taken from the \coldgass\ catalogue \citep{Saintonge2017}. In particular, star-formation rates are calculated by combining the IR and UV based SFR components obtained from WISE and GALEX photometry, as described in Janowiecki et al. (2017). Stellar masses come from the SDSS DR7 MPA/JHU catalogue\footnote{http://home.strw.leidenuniv.nl/$\sim$jarle/SDSS/}.  
 The 25 galaxies with CO(3-2) observations are not classified as AGN by the optical emission line diagnostics BPT diagram \citep{Baldwin1981, Kewley2001, Kauffmann2003}. Four objects are classified as composite, one as LINER, and the remaining galaxies are classified as star-forming.
 The properties of the sample are summarized in Table\,\ref{tab:COLD_prop}.

\subsection{Active galactic nuclei: BASS} 
We include in our study a sample of AGN selected in the hard X-ray from the \textit{Swift}/BAT 70 Month survey \citep{Baumgartner2013}.
 We have CO(3-2) observations of \Nbasstot\ BAT AGN at redshift $<0.04$. In our analysis we focus on sources for which we also have observations of the CO(2-1) transition.
Additionally, we discard from our sample three AGN for which \Herschel\ FIR observations are not available and thus we cannot infer their SFR. Thus the final BASS sample that we use in our analysis consists of \Nbass\ objects.
 These sources are part of the BAT AGN Spectroscopic Survey (BASS\footnote{www.bass-survey.com}), for which ancillary information from optical and X-ray spectroscopic analysis is available \citep{Koss2017, Ricci2017}.
The AGN are in the redshift range $0.002 <\ z <\ 0.040$.

The SFR is inferred from the total (8-1000\micron) infrared (IR) luminosity due to star-formation given in \cite{Shimizu2017}, which was measured by decomposing the infrared SED in the AGN and host galaxy component.
 We use the following conversion from total infrared luminosity (3-1100\micron\ range) to SFR , calculated assuming a Kroupa IMF \citep{Hao2011, Murphy2011, Kennicutt2012}:
\begin{equation}
SFR  = 3.89\cdot 10^{-44}\cdot L_{\rm IR},
\label{eq:SFR_from_FIR}
\end{equation}
where the $SFR$ is in units of [M$_\odot$ yr$^{-1}$], and $L_{\rm IR}$ is the total infrared luminosity in [erg s$^{-1}$].

We use stellar masses measured for BAT AGN host galaxies from Secrest et al. (in prep.). They are derived by spectrally de-convolving the AGN emission from stellar emission via SED decomposition, combining near-IR data from 2MASS, which is more sensitive to stellar emission, with mid-IR data from the AllWISE catalog \citep{Wright2010}, which is more sensitive to AGN emission. % More details are in \cite{Secrest2018}.
The galaxies in the sample have stellar masses in the range $9.7 <\ \log M_*/M_\odot  <\ 11.1$ and SFR in the range $-0.83 <\ \log \text{SFR/[\Msun\ yr}^{-1}] < 1.75$.
 Table\,\ref{tab:BASS_prop} lists the properties of this sample.

\subsection{Infrared luminous galaxies: SLUGS}

We also include in our analysis a sample of infrared luminous galaxies ($L_{\rm FIR}>\ 10^{10}\ L_{\odot}$)  from  the SCUBA Local Universe Galaxy Survey \citep[SLUGS,][]{Dunne2000}.
 We include this sample in order to extend the parameter range to galaxies with higher SFR. We chose this sample over other samples available in the literature because it has beam matched observations and information about how to scale the total SFR to the SFR within the beam.
 
 We select the \Nslugs\  SLUGS galaxies with observations of both CO(3-2) and CO(1-0) available in  \citet{Yao2003}. 
These galaxies are in the redshift range $0.006 <\ z <\ 0.048$. 
Stellar masses from the SDSS DR7 MPA/JHU catalogue are available for only 22 galaxies of this  sample and are in the range $9.6 <\ \log M_*/M_\odot  <\ 11.4$. 

We use the optical emission line diagnostic \citep{Baldwin1981, Kewley2001, Kauffmann2003} from SDSS DR12 to distinguish between AGN and SFGs. 
Of the 22 galaxies with stellar masses from SDSS, two are classified as Seyferts (IRAS 10173+0828 and Arp 220), seven as Composite and 13 as star-forming galaxies. We include the galaxies classified as Composite in the star-forming galaxies sample.

The total SFR are derived from the total infrared luminosities $L_{\rm IR}$ using eq.~(\ref{eq:SFR_from_FIR}). We measure $L_{\rm IR}$ by integrating the SED, approximated by a modified black-body, in the range $8-1000\micron$. The parameters of the modified black-body (MBB) model are given in \citet{Dunne2000}. We calculate the uncertainties on $L_{\rm IR}$ by propagating the uncertainties on the MBB parameters given in \citet{Dunne2000}. The SFRs are in the range $0.18 < \log$ SFR/[\Msun\ yr$^{-1}] < 2.15$.
 \cite{Yao2003} also provide  the FIR luminosity and SFR corresponding to the 15" central part of the galaxy (equivalent to the size of the CO beam), obtained by applying a scale factor to the total FIR luminosity. This factor is derived from the original 850\micron\ SCUBA-2 images.
To calculate the SFE, we use the SFR in the 15" central part of the galaxy, since it matches the beam size of the CO observations.
 
We note that for this sample the molecular gas mass \MHtwo, and consequently also the star-formation efficiency (SFE),  represents only the value in the central 15" region of the galaxy, since no correction has been applied to extrapolate from the beam area to the total \MHtwo.

\subsection{Samples in the SFR-M$_*$ plane}
\begin{figure*}
\centering
\includegraphics[width=0.44\textwidth]{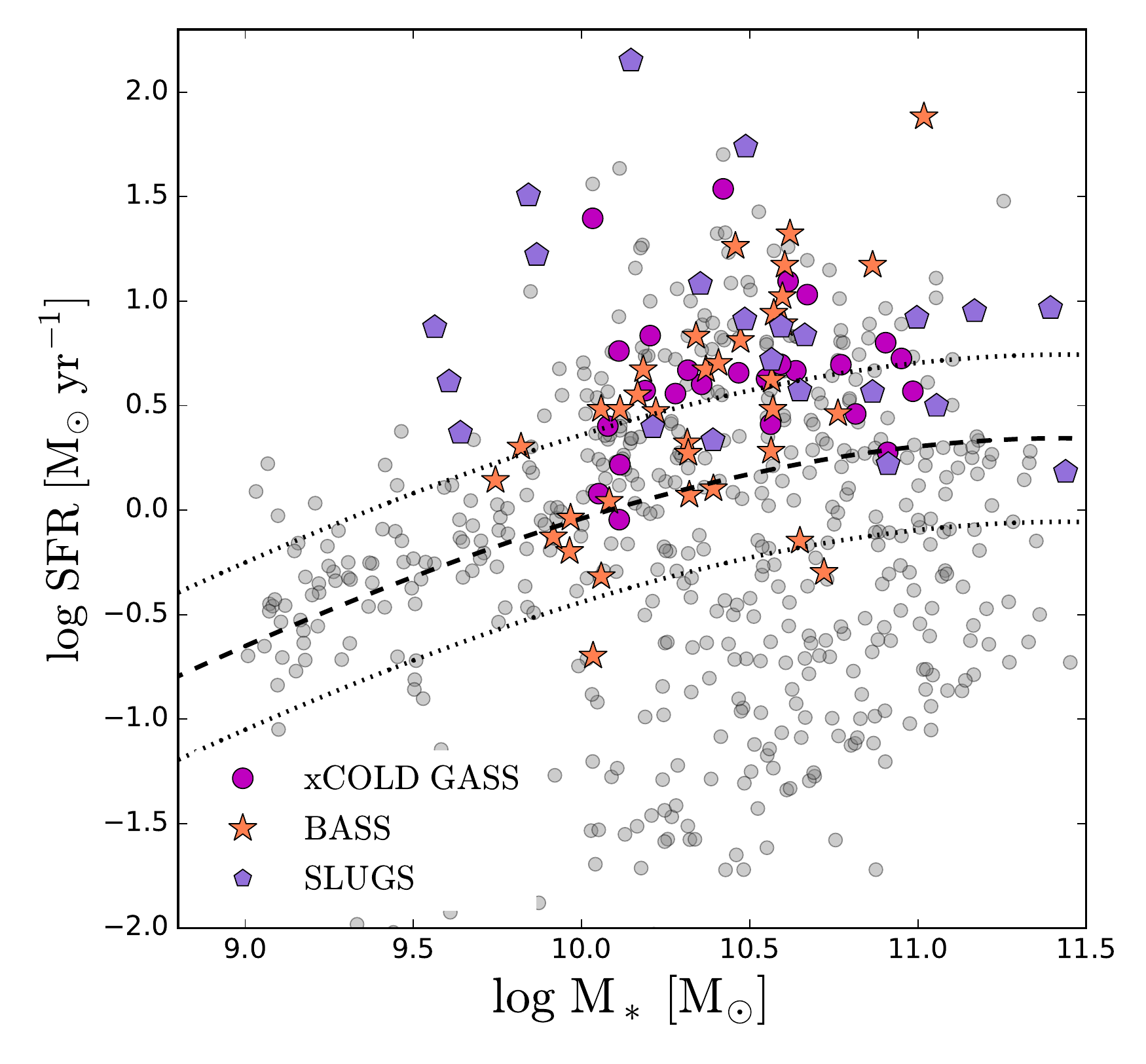}
\includegraphics[width=0.44\textwidth]{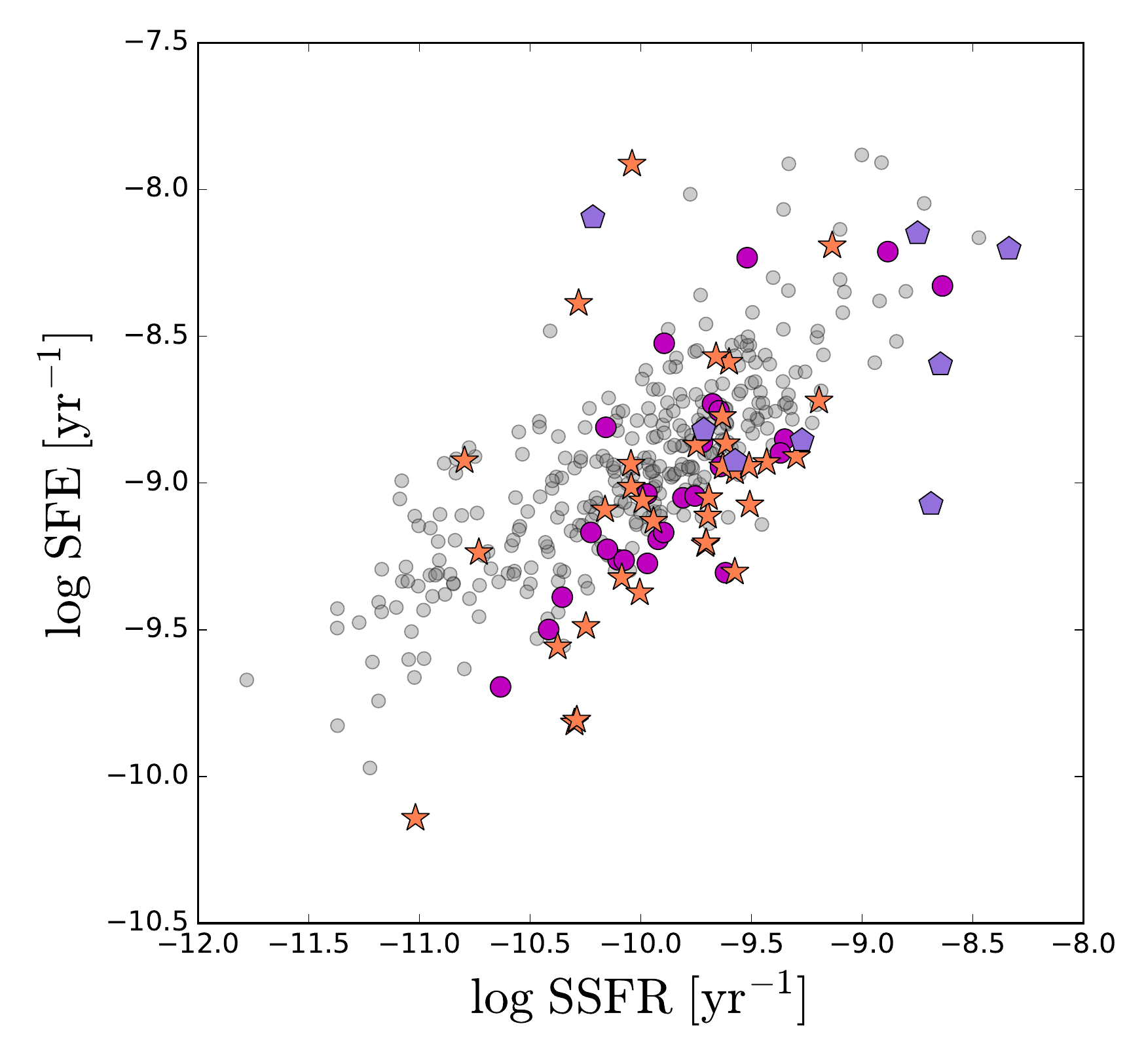}
\caption{\textit{Left:} Distribution of the \coldgass, BASS and SLUGS samples in the SFR-$M_*$ plane. The position of the star formation main sequence \citep{Saintonge2016} is shown by the dashed line, the 0.4 dex dispersion is shown by dotted lines. The full \coldgass\ sample is shown by the grey points, while the sub-sample with CO(3-2) observations is shown in magenta. \textit{Right:} SSFR versus star-formation efficiency (SFE = SFR/\MHtwo). Galaxies from the BASS sample have in general lower SFE than the \coldgass\ galaxies at the same SSFR. For the SLUGS sample, we plot only the galaxies with angular diameter $D<$ 100", since their SFE is measured within the beam, while the SSFR is the total value.} 
\label{fig:SFR_Mstar_plane}
\end{figure*}

Figure\,\ref{fig:SFR_Mstar_plane} shows the distribution of the \coldgass, BASS and SLUGS samples in the SFR-$M_*$ plane. The position of the star formation main sequence \citep{Saintonge2016} is shown by the dashed line, and the dotted lines show the 0.4 dex dispersion. The `full \coldgass' sample is shown by the grey points for reference. The three samples cover a similar range in stellar masses.  All galaxies from the \coldgass\ sample are on the main sequence or above, while the infrared luminous galaxies from the SLUGs sample are mostly above the main sequence. The BASS sample spans a broad range of SSFR, with $\sim$8 AGN below the main sequence and the rest of the sample overlapping in the parameter space with the \coldgass\ galaxies. The right panel of Fig.\,\ref{fig:SFR_Mstar_plane} shows the SSFR versus the star-formation efficiency SFE. 
 The three samples span a similar range of SSFR (-11 $<$ log SSFR/[yr$^{-1}$] $<$ -8.5).
The galaxies of the \coldgass\ sample have slightly higher SFE at the same SSFR than the BASS galaxies, but there is a good overlap with the BASS sample. The infrared luminous galaxies from SLUGS have in general high SSFR and high SFE.

\section{CO data, observations and data reduction}
\label{sec:CO_data}

\subsection{xCOLD GASS}
\subsubsection{xCOLD GASS: CO(1-0) data from the literature}

The CO(1-0) line luminosities \LCOone\ are taken from the \coldgass\ catalogue \citep{Saintonge2017}. The CO(1-0) line fluxes are observed with the IRAM 30m telescope (beam size: 22"). The 25 galaxies from \coldgass\ selected for the CO(3-2) observations all have S/N$>\ $3 in CO(1-0). We refer to \cite{Saintonge2017} for information about the observations and data reduction.

\subsubsection{xCOLD GASS: CO(3-2) observations}

The CO(3-2) observations are taken with the HARP instrument \citep[Heterodyne Array Receiver Program, beam size: 14",][]{Buckle2009} on the James Clerk Marxwell Telescope (JCMT, observing program  M14AU21, PI: A. Saintonge).
Theses observations took place between January and June 2014.

Each CO(3-2) spectrum was observed in a single HARP pointing  in `hybrid' mode,  which produces two spectra for every scan (in two spectral windows). 
The spectra were reduced using the {\tt Starlink} software  \citep{Currie2014}.
 First the two spectra within each scan were combined, after correcting for any baseline difference, and then all scans were combined together. A linear fit to the continuum was used to remove the baseline and then the spectrum was binned to a resolution of 40 \kms. 
 The HARP instrument has $4\times4$ receptors (pixels), each one with a half power beam width of 14". We extract the CO(3-2) spectrum only from the  pixel which is centred on the galaxy. The technique used to measure the flux from the reduced spectrum is the same used for the main \coldgass\ survey \citep{Saintonge2017}. 
 We convert the antenna temperature to flux units by applying the point source sensitivity factor 30 Jy/K recommended for HARP\footnote{\scriptsize{www.eaobservatory.org/jcmt/instrumentation/heterodyne/harp/,  www.eaobservatory.org/jcmt/instrumentation/heterodyne/calibration/}}.
 We measure the velocity-integrated line flux $S_{\text{CO}}$ in [Jy \kms]  by adding the signal within a spectral window. We initially set the width of the spectral window ($W_{\rm CO}$) equal to the FWHM of the CO(1-0) given in the \coldgass\ catalog. In case the CO(3-2) line is clearly wider, we extend $W_{\rm CO}$ to cover the total line emission. 
  We determine the center of the line based on the SDSS spectroscopic redshift. In two cases where the CO(3-2) is clearly shifted with respect to the position determined from the SDSS redshift, we use the redshift of the CO(1-0) line, which is shifted in the same direction of the CO(3-2) line, to center the CO(3-2) line.
 We measure the baseline rms noise of the line-free channels ($\sigma_{\rm CO}$) per 40 \kms\ channel in the spectral regions around the CO line.
 
 The beam-integrated CO(3-2) line luminosity in units of K km s$^{-1}$ pc$^2$ is defined following \cite{Solomon1997} as:
\begin{equation}
L'_{\text{CO}}= 3.25 \cdot 10^7  S_{\rm CO} \nu_{\rm obs}^{-2} D_L^2 (1+z)^{-3},
\end{equation}
where $S_{\rm CO}$ is the velocity-integrated CO(3-2) line flux within the HARP beam in units of Jy km s$^{-1}$, $\nu_{\rm obs}$ is the observed frequency of the CO(3-2) line in GHz, and $D_L$ is the luminosity distance in Mpc. The error on the line flux is defined as:
\begin{equation}
\epsilon_{\rm obs} = \frac{\sigma_{\rm CO} W_{\rm  CO}}{\sqrt{W_{\rm CO} \Delta w_{ch}^{-1}}},
\end{equation}
where $\sigma_{\rm  CO}$ is the rms noise achieved around the CO(3-2) line in spectral channels with width $\Delta w_{ch} = 40$ \kms, and  $W_{\rm  CO}$ the width (in \kms) of the spectral window where we integrate the CO(3-2) line flux.

  We use a detection threshold of signal-to-noise S/N$ >\ 3$, defined as S/N $ = S_{\rm CO}/\epsilon_{\rm obs}$, which is the same adopted for the main \coldgass\ catalogue. 
   In 7/25 galaxies the CO(3-2) line is not detected and  we use conservative upper limits equal to five times the error: $S_{\rm CO(3-2),limit} =5\cdot \epsilon_{\rm CO(3-2), obs}$. 
 The 5$\sigma$ upper limits correspond to a `false negative' fraction of 2\%, which is the probability that a source with `true' flux higher than this upper limit is not detected.
   To calculate $\epsilon_{\text{CO(3-2)},obs}$ we use the FWHM of the CO(1-0) line  as an approximation for the width of the CO(3-2) line ($W_{\rm CO}$).
  All the CO(3-2) spectra from \coldgass\ are shown in the appendix (Fig.\,\ref{fig:CO32_spectra}) and the measured line properties in Table\,\ref{tab:COLD_CO}.

\subsection{BASS}
\label{sec:BASS_CO}
 Both the CO(2-1) and CO(3-2) lines have been observed at the JCMT: the CO(3-2) with HARP and the CO(2-1) with the RxA instrument (beam size: 20"). 
The HARP observations took place in weather bands 3-4 (corresponding to an opacity $\tau_{225 \text{GHz}} = 0.07-0.21$), % median tau=0.12
 while the RxA observations took place in weather band 5 ($\tau_{225 \text{GHz}} = 0.20-0.32$). The observations and data reduction of the CO(2-1) line emission is explained in detail in Koss et al. (in prep). 
 
 The CO(3-2) observations were taken between and February 2011 and November 2012 in programs M11AH42C (P.I: E. Treister) and M12BH03E (PI: M. Koss). Additionally, we also include 13 spectra from archival observations. 
 Each galaxy was initially observed for 30 minutes. For weak detections, additional observations were obtained up to no more than two hours. The individual scans for a single galaxy were first-order baseline-subtracted and then co-added. We extract the CO(3-2) spectrum only from the pixel centred on the galaxy.
 We measure the CO(3-2) and CO(2-1) line fluxes using the same method as for the \coldgass\ sample, for consistency. We measure the $S_{\rm CO}$ line flux in [Jy \kms] by adding the signal within a spectral range that covers the entire width of the line. In the appendix (see online material),we show the CO(3-2) spectra from BASS, in which we highlight the spectral regions where we integrate the fluxes.
 All BASS objects have good detections (i.e. S/N$ >\ $3) of the CO(2-1) lines, while we have non-detections (i.e. S/N$ <\ $3) in the CO(3-2) line for 3/\Nbass\ galaxies.
  For these galaxies we use upper limits equal to five times the flux error: $S_{\rm CO, limit}= 5\cdot \epsilon_{\rm CO, obs}$.
  
Our set of observations is not homogeneous since for the \coldgass\ and SLUGS samples we compare the CO(3-2) to the CO(1-0) line, but for the BASS sample we have to estimate CO(1-0) from the CO(2-1) line. Therefore we need to assume a value for the ratio $r_{21}=L'_\text{CO(2-1)}/L'_\text{CO(1-0)}$.
The typical value observed for normal spiral galaxies is $r_{21}= 0.8$ \citep{Leroy2009, Saintonge2017}.
 \cite{Leroy2009} studied a sample of ten nearby spiral galaxies and found \ratiotwo\ values between 0.48 and 1.06, with most values in the range $0.6-1.0$. They found an average of 0.81. For the \coldgass\ survey, \cite{Saintonge2017} found a mean value of $r_{21} = 0.79 \pm 0.03$ using a sample of 28 galaxies.

Some of the AGN in our sample (12/\Nbass) have recently been observed with the IRAM 30m telescope as part of a programme to measure CO(1-0) line luminosity for 133 BAT AGN (P.I: T. Shimizu). % (check this number: ask Taro Shimizu. Based on the proposal, they should be 155)
We compute the $r_{21}$ line ratios for these 12 objects using the values from Shimizu et al. (in prep.). Since the difference in beam size is very small (IRAM: 22", JCMT RxA: 20"), we did not apply any beam corrections. 
The \ratiotwo\ line ratios for these 12 objects are in the range 0.4-2.1, with a median $r_{21} = 0.72$. We obtain a robust standard deviation by computing  the median absolute deviation $MAD= 0.17$. The robust standard deviation, under the assumption of a normal distribution, is given by $\sigma= 1.4826\cdot MAD = 0.26$ \citep{Hoaglin1983}.
 For the 12 objects with CO(1-0) observations, we use the CO(1-0) luminosities from  Shimizu et al. (in prep.) to compute the \ratio\ line ratio.
For the remaining AGN, we use a constant \ratiotwo = 0.72, and we assume an uncertainty of 0.26 on this value. The CO line fluxes for this sample are shown in Table\,\ref{tab:BASS_CO}.

\newpage
\subsection{SLUGS}
 The CO(1-0) observations were taken with the Nobeyama Radio Observatory (NRO) 45~m telescope (beam size: 14.6") and the CO(3-2) observations with the HARP instrument on the JCMT. 
We take the CO line luminosities and line ratios from \cite{Yao2003} and we refer to that paper for information about the observations and data reduction.

\subsection{Beam corrections}
We calculate beam corrections for two purposes:  1) to correct for the different beam sizes of the CO(3-2), CO(2-1), and CO(1-0) observations and 2) to extrapolate the CO luminosity measured within the beam to the total CO luminosity of the galaxy.

\textbf{1) Corrections for the different beam sizes:}\\
For the SLUGs sample the beam sizes are similar (14.6" for the CO(1-0) line and 14" for the CO(3-2) line), therefore the line luminosities can be directly compared without applying any corrections for the beam size. For the \coldgass\ and BASS samples instead, the beam sizes of the telescopes used for the CO(3-2), CO(2-1) and CO(1-0) observations vary between 14" and 22", thus we need to apply beam corrections. 
 In order to compare the CO emission from different lines, we need first to ensure that we are comparing fluxes coming from the same part of the galaxy. 
To estimate the amount of flux that is missing in the observation done with the smaller beam, we use the following approach, which is based on the assumption that the dust emission in the infrared is a good tracer of the cold molecular gas distribution \citep[e.g.][]{Leroy2008}. Under this assumption, we can estimate the flux that would be observed from beams of different sizes by measuring the flux within different apertures in the infrared images. After that, we  apply an additional correction to take into account the fact that the infrared images have a point-spread function (PSF) that causes the observed flux  to appear more extended than the intrinsic emission.

To calculate the beam corrections from the infrared images, we apply the following procedure. We multiply the infrared image by a 2D Gaussian centred on the galaxy centre and with FWHM equal to the beam size, to mimic the effect of the beam sensitivity of the telescope that took the CO observations. Then we measure the total flux from the image multiplied by the 2D Gaussian.
 We repeat this measurement for the two beams, and we take the ratio of the fluxes:
\begin{multline}
C_{IR} = \frac{F\text{(inside\ the\ larger\ beam) }}{F \text{(inside\ the\ smaller\ beam)}} =\\
 = \frac{F\text{(inside\ the\ CO(1-0) or (CO(2-1) beam)}}{F \text{(inside\ the\ CO(3-2)\ beam)}}.
\end{multline}

For the \coldgass\  sample, we use the 22$\mu$m images from the WISE survey. Specifically, we use the co-added images from `unWISE' \footnote{\url{http://unwise.me/}} which have been systematically produced without blurring, retaining the intrinsic resolution of the data \citep{Lang2014, Meisner2016}.
 For \Nbass\ galaxies in our BASS sample there are \Herschel /PACS observations at 70\micron\ and 160\micron\  available \citep{Melendez2014,Shimizu2017}. We decide to use the PACS 160\micron\ images because the longer wavelength is less likely to be contaminated by AGN emission, which can still contribute for a significant fraction of the 70\micron\ emission \citep{Shimizu2017}.

 The point-spread-functions (PSF) of the WISE 22\micron\ and PACS 160\micron\ images are rather large (12") when compared with the size of the CO beams (14"-22"), and can therefore affect the measurement of the beam corrections. The images that we are using to trace the distribution of the FIR emission  are not maps of the `true' distribution, instead they are maps of the `true' distribution convolved with the PSF of the FIR telescope.
 To correct for the effect of the PSF, we use a simulated galaxy gas profile, following the procedure described in \cite{Saintonge2012}. 
For each galaxy, we create a model galaxy simulating a molecular gas disk following an exponential profile, with a scale length equivalent to its half-light radius.
 Then the profile is tilted according to the inclination of the galaxy and we measure the amount of flux that would be observed from this model galaxy, using an aperture corresponding to the size of the beam ($F_{sim}$).
  Then we convolve the galaxy profile with a 2D Gaussian with the FWHM equal to the size of the PSF of the image and we measure again the flux within the beam radius ($F_{sim, PSF}$). By taking the ratio of these two measurements, we estimate how much the flux changes due to the effect of the PSF:
\begin{equation}
C_{PSF} = \frac{F_{sim, PSF}}{F_{sim}}.
\end{equation}
This correction is in the range $1.04-1.27$. %($4-27\%$).
We apply this PSF correction to the beam correction obtained from the infrared images:
\begin{equation}
C_{IR, PSF} =  \frac{C_{IR}}{C_{PSF}}.
\end{equation}
We finally apply this factor to the $r_{31}$ ratios:
\begin{equation}
r_{31, corr} = r_{31}\cdot C_{IR, PSF}.
\end{equation}

The final beam corrections ($C_{IR, PSF}$) for the BASS sample are in the range  $1.05-1.70$, with a mean value of 1.27.
For the \coldgass\ sample they span a similar range between 1.08 and 1.80, with a mean of 1.31. 
 The corrections for the \coldgass\ samples are larger because of the larger difference between the two beams (22" for the CO(1-0) beam vs. 14" for the CO(3-2) beam), compared to the BASS sample (20" for the CO(2-1) beam vs. 14" for the CO(3-2) beam).
 In order to check that the beam corrections do not have an effect on our analysis, we look at the relation between \ratio\ and galaxy angular size or the beam corrections value. We do not find any dependence of the \ratio\ on the beam corrections or on the angular size of the galaxies (see Appendix\,\ref{sec:r31_vs_beam_corr}).
 
We note that the line ratios presented in this paper are measured in the central region of the galaxies, and may not be representative of the line ratio of the entire galaxy.
Resolved studies of the CO line ratios in nearby galaxies find that the excitation tend to be higher in the central part than at larger radii \citep{Leroy2009, Wilson2009}. With the beam corrections, we want to correct for the fact that the beams of the two transitions have different sizes, but they still represent only the central part of the galaxy. \\

\textbf{2) Beam-to-total luminosity corrections:}\\
To calculate the total CO(1-0) emission and molecular gas mass, we  need to apply a correction to extrapolate the CO(1-0) emission within the beam to the total CO(1-0) luminosity.
For the \coldgass\ sample, we retrieve these values from the \coldgass\ catalogue \citep{Saintonge2017}.
 They are in the range $1.02-1.95$.
For the BASS sample, we use the method describe above to estimate the total amount of CO emission. 
We measure the total infrared 160\micron\ emission of the galaxy within a radius big enough to include the entire galaxy, paying attention not to include any emission not related to the galaxy. 
We determine the radius until which we integrate the flux based on the curve of growth of the galaxy profile. 
For compact sources the radius extends until $\sim$60", while for the more extended and nearby galaxies, we measure the flux within a radius up to 140".

Then we take the ratio between the flux from the map multiplied by the CO(2-1) beam sensitivity,
 measured as explained above, and the total infrared flux and we use this value to extrapolate the total CO(2-1) flux. The beam corrections for BASS are in the range $1.46-15.66$. 
For the analysis in Section\,\ref{sec:r31_results}, we use only galaxies with angular diameter $D < 100"$, for which the beam corrections are $< 2.4$, to avoid galaxies for which the CO emission within the beam is not representative of the total CO emission. 
 For the angular size $D$ of  \coldgass\ and SLUGS we use $D=D_{25}$, i.e. the optical diameter derived from SDSS g-band. For BASS we use $D=2\times R_{k20}$, where $R_{k20}$ is the isophotal radius at 20mag arcsec$^{-2}$ in the K-band. We expect the sizes measured in the g-band and in the K-band to be similar \citep{Casasola2017}. 
 The beam correction values can be found in Tables \ref{tab:COLD_CO} and \ref{tab:BASS_CO}.

\subsection{Total molecular gas mass}
We use two different CO-to-H$_2$ conversion factors: for normal star-forming galaxies we adopt a Galactic conversion factor $\alpha_{CO}= 4.3$ \Msun /(K km s$^{-1}$ pc$^2$) \citep{Strong1996, Abdo2010, Bolatto2013} and for ``ULIRGs-type" galaxies we use $\alpha_{CO} = 1$ \Msun /(K km s$^{-1}$ pc$^2$) \citep{Bolatto2013}.
 To distinguish between normal SFGs and ``ULIRGs-type" galaxies, we apply the selection criterion described in \cite{Saintonge2012}, which is based on the FIR luminosity and on the dust temperature. According to this criterion, we apply the ``ULIRGs-type" conversion factor to galaxies with  $\log L_\text{FIR}/L_\odot > 11.0$ and $S_{60\mu m}/S_{100\mu m} > 0.5$.
For the other galaxies, we use the Galactic conversion factor.
For the BASS sample, we also need to apply a conversion from CO(2-1) to CO(1-0) line luminosity, which is explained in Section\,\ref{sec:BASS_CO}.

\section{CO line ratios}
\label{sec:r31_results}
\subsection{$r_{31}$ and star-formation}
In this section we look at the \ratio\ distribution for AGN and SFGs and investigate the relation between \ratio\ and galaxy global properties. 
 For this part of the analysis, we exclude from the sample the galaxies with large angular size (diameter $D> 100"$), in order to avoid galaxies for which the luminosity measured within the beam is not representative of its total emission.
 The sample used in this section consists of 25 galaxies from \coldgass, 20 from BASS, and 8 from SLUGS. 

 The \ratio\ values in the \coldgass\ sample are in the range $0.25-1.15$ and the mean value is $ 0.55 \pm 0.05$, with a standard deviation of 0.22. This value is consistent with observations of low redshift galaxies. 
\citet{Mao2010} found a mean value $r_{31}=0.61\pm0.16$ in their sample of normal SFGs.
  \citet{Papadopoulos2012} found a higher mean value $r_{31}=0.67$ in a sample of nearby LIRGs, which are expected to have higher \ratio\ given their higher SSFR and SFE. Also, \citet{Yao2003} found a higher mean value $r_{31}=0.66$ in their sample of infrared luminous galaxies. 
 The \ratio\ values in the BASS AGN sample span a very similar range to the \coldgass\ sample $0.22-1.23$, with a mean value $0.53 \pm 0.06$ (standard deviation 0.25). 
 For the SLUGS sample, the \ratio\ values are in the range $0.32-0.89$ with a mean value $0.58 \pm 0.07$ (standard deviation 0.20). The mean value of the total sample is  $<r_{31}>=0.55\pm 0.03$ (standard deviation 0.23).

\begin{figure*}
\centering
\includegraphics[width=0.98\textwidth]
{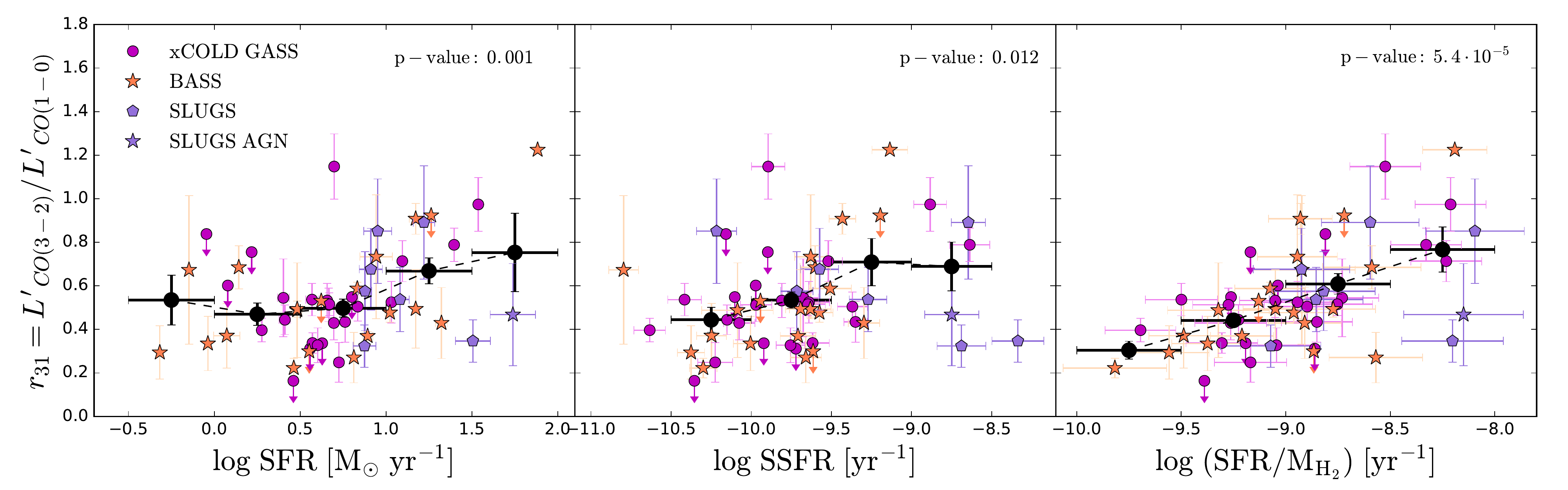}
\caption{ Ratio
\ratio = \LCOthree / \LCOone\ as a function of star-formation rate (SFR), specific star formation rate (SSFR = SFR/M$_*$) and star-formation efficiency (SFE = SFR/\MHtwo ) for the \coldgass, BASS and SLUGS samples.  The black points show the mean values of the total sample in bins of 0.5 dex, with the error bars showing the standard errors on the mean values. The dashed line connects the mean values to help to visualize the trends. In each plot, we show the $p$-value of the null hypothesis that there is no correlation, calculated using the Kendall's rank correlation test for censored data. }
\label{fig:r31_vs_SFE}
\end{figure*}

We investigate how the ratio \ratio\ evolves as a function of SFR, SSFR and SFE (Fig.\,\ref{fig:r31_vs_SFE}).
 We find a general trend for \ratio\ to increase as these quantities increase (Pearson correlation coefficients $R=0.26-0.60$). To illustrate the evolution of \ratio, we divide the total sample in bins of 0.5 dex according to the quantity on the x-axis (SFR, SSFR, or SFE), and calculate the mean values of \ratio\ in these bins. The mean values are shown as black points in the plots, with the error bars showing the standard errors on the mean values. For bins that contain less than three objects we do not show the mean values.

In order to properly take into account the upper limits on the \ratio\ values, we apply the principles of survival analysis \citep{Feigelson1985}. 
 We perform the Kendall's rank correlation test for censored data (i.e. data with upper limits) as given in \cite{Brown1974}. The  test gives $p\text{-value}= 9.1\cdot 10^{-4}, 1.2\cdot 10^{-2}, 5.4\cdot 10^{-5}$ for the relation of \ratio\ with SFR, SSFR and SFE, respectively. The $p$-values of the correlation with SFR and SFE are $<$ 0.05, meaning that we can reject the null hypothesis that there is no association between the two quantities.
 The strongest relation is the one with the SFE (largest Pearson correlation coeff. $R=0.6$). The correlation of \ratio\ with SFE is significantly different from the correlation of \ratio\ with SFR and SSFR, according to the Fisher Z-test ($p$-value=0.03 and $p$-value = $9.6\cdot10^{-5}$, respectively).  This trend has already been reported  by \cite{Yao2003} and \cite{Leech2010} for samples of infrared luminous galaxies and LIRGs.
If we consider the $r_{31}$ ratio to be a proxy for the ratio of relatively dense to very diffuse molecular gas,  the correlation between \ratio\ and SFE suggests that galaxies with a higher fraction of dense molecular gas tend to have higher SFE. The connection between \ratio\ and gas density is investigated further in Section\,\ref{sec:uclpdr}. 
 We find that the \ratio\ ratio tends to increase with SFE, but there is a large scatter in the relation. It is then likely that other factors contribute to regulate the \ratio\ ratio.

\subsection{Comparison of star-forming galaxies and AGN}
\begin{figure}
\centering
\includegraphics[width=0.37\textwidth]
{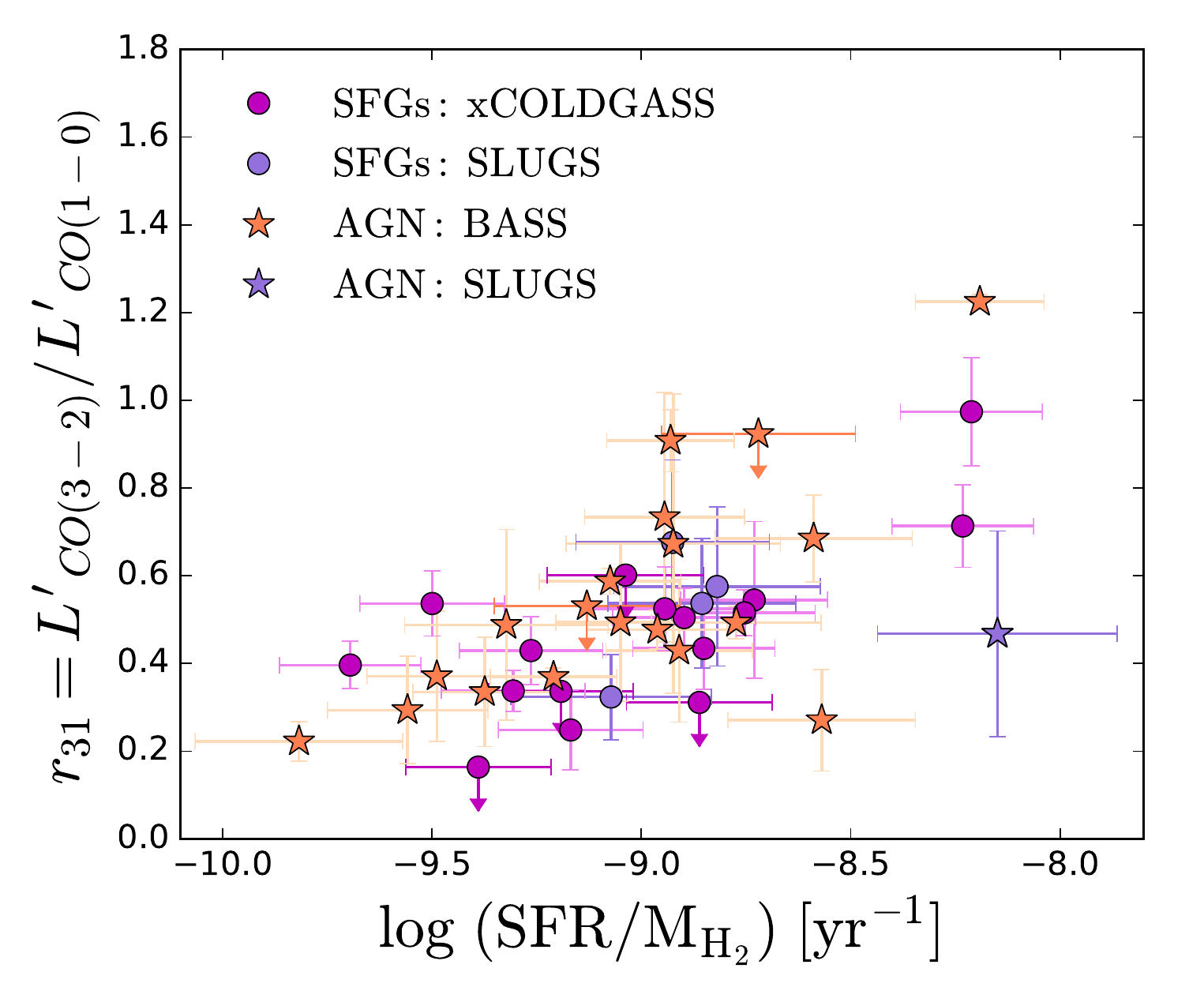}
\caption{ Ratio \ratio = \LCOthree / \LCOone as a function of star-formation efficiency (SFE = SFR/M(H$_2$))  for SFGs (circles) and AGN (stars). The SFG and AGN samples are matched in SSFR: at every SFG corresponds the AGN host galaxy with the most similar value of SSFR.}
\label{fig:r31_vs_SFE_matched_in_SSFR}
\end{figure}

We divide the sample into AGN (20 BASS objects and one AGN from SLUGS) and star-forming galaxies (25 \coldgass\ galaxies and the remaining 7 SLUGS galaxies), to investigate whether we see any difference in the \ratio\ values between these two classes of objects.
The two samples have different distributions of specific-star formation rate (SSFR= SFR/M$_{*}$): the AGN host galaxies have  lower values of SSFR $(-10.8 < \log \text{SSFR/[yr}^{-1}] < -8.8) $  than the star-forming galaxies $(-10.6 < \log \text{SSFR/[yr}^{-1}] < -8.3) $.
  To remove the effect of the different SSFR in the two samples, we match the samples in SSFR, and we look again at the distribution of \ratio\ in SFGs and AGN. This is important because of the correlation between SSFR and SFE \citep{Saintonge2011b, Saintonge2016}. We pair every SFG  with the AGN host galaxy which has the most similar value of SSFR. The results are shown in  Fig.\,\ref{fig:r31_vs_SFE_matched_in_SSFR}.
The mean \ratio\ for the matched samples are consistent with each other: \ratio = $0.52 \pm 0.04$   for SFGs and $0.53 \pm 0.06$ for AGN.
To test whether the two samples have different \ratio\ distributions at the same SSFR, we do a Two Sample test using the survival analysis package ASURV \citep{Feigelson1985}, which allows to take into account upper limits. 
  We find that the two samples are not significantly different according to the Gehan's, Logrank and Peto-Prentice's Two Sample Tests ($p$-value=0.57-0.79).
 So our results suggest that  there is no clear difference in the \ratio\ values due to the AGN contribution. 
 
\citet{Mao2010} find a higher $r_{31}=0.78\pm0.08$ in AGN  than in normal star-forming galaxies ($r_{31}=0.61\pm0.16$). They however do not control for the SSFR, so it is possible that the difference in \ratio\ is partly due to differences in SSFR between the two samples and not to the effect of the AGN. They also find higher \ratio\ values in  starbursts ($r_{31}=0.89\pm0.11$) and in ULIRGs ($r_{31}=0.96\pm0.14$) than in AGN.
Additionally, most of the galaxies in their sample have rather large angular size (optical diameter $D_{25} >$ 100") and thus the CO beam is sampling a smaller region around the nucleus. Therefore it is reasonable to expect that the AGN could have a large impact on the observed \ratio\ line ratio.

We look at the relation between \ratio\ and hard X-ray luminosity (14-195 keV) for the BASS sample, but we do not find a clear trend between the two quantities (see Fig.~\ref{fig:r31_vs_xray} in the appendix), which suggests that the X-ray flux is not the main parameter affecting this line ratio. Even though the X-ray radiation may contribute to enhance the \ratio\ ratio in the nuclear region, as is shown later in Section\,\ref{sec:xray}, it is probably not enough to regulate the CO excitation in the entire galaxy.

We conclude that there is no significant difference between the values of \ratio\ of AGN and SFGs.

\newpage
\section{Modelling: \uclpdr}
\label{sec:uclpdr}
In order to better understand which physical parameters influence the line ratios $r_{21}$ and $r_{31}$, we model the CO emission lines using a photon-dissociation region (PDR) code. 
 Our goal is to test which are the physical quantities that have the largest effect on the CO line ratios, and which values of these quantities can reproduce our observations.

We employ the 1D {\tt UCL-PDR} code, developed by \cite{Bell2005, Bell2006} and upgraded by \cite{Bayet2011}. The latest version of the code is presented in \cite{Priestley2017}.
The code models the gas cloud as a semi-infinite slab with a constant density, illuminated from one side by a far-ultraviolet (FUV) radiation field.  At each depth point in the slab, the code calculates the chemistry and thermal balance of the gas self-consistently and returns, for every element, the gas chemical abundances, emission line strengths and gas temperature. Surface reactions on dust grains are not included.

The gas is cooled  by the emission from collisionally excited atoms and molecules and by the interactions with the cooler dust grains \citep{Bell2006}. We include in our model the cooling from the following lines: 
Lyman $\alpha$, $^{12}$C$+$, $^{12}$C, $^{16}$O, $^{12}$CO, and the para and ortho H$_2$ and H$_2$O states.
Table\,\ref{tab:chem_network} shows the elements included in the chemical network and their initial abundances relative to hydrogen, where depletion in the dust by some elements is already taken into consideration.  For the values of the initial elemental abundances we follow \cite{Bell2006}.  We set $n(\text{H}_2)/n_\text{H}=0.4$ (where \nH\ is the volume density of hydrogen nuclei \nH$=n(\text{H}) + 2\cdot n(\text{H}_2)$) following \cite{Bell2005}.

\begin{table}
\centering
\caption{Initial elemental abundances used in the \uclpdr\ code relative to the hydrogen nuclei.}
\begin{tabular}{lc} 
\hline
Element & Abundance \\ \hline 

He & $7.50\cdot 10^{-2}$ \\
O & $3.19\cdot 10^{-4}$  \\
C+ & $1.42\cdot 10^{-4}$  \\
N & $6.50\cdot10^{-5}$  \\
Mg(+) & $5.12\cdot 10^{-6}$  \\
S(+) & $1.43\cdot 10^{-6}$  \\
\hline 
\end{tabular}
\label{tab:chem_network}
\end{table}

We calculate the integrated line intensity of the CO emission lines as described in \cite{Bell2006}.
 The opacity is included in the calculation of each coolant transition along each path \citep{Bell2006, Banerji2009}.   The intensity $I$ in units of erg s$^{-1}$ cm$^{-2}$ sr$^{-1}$  is calculated by integrating the line emissivity $\Lambda$ over the depth into the cloud $L$:
\begin{equation}
 I = \frac{1}{2 \pi} \int \Lambda(L) dL, 
\end{equation}
where $\Lambda$ has units of erg s$^{-1}$ cm$^{-3}$, and the factor of $2\pi$ takes into account the fact that the photons only emerge from the edge of the cloud/slab.

The velocity-integrated antenna temperature in units of K km s$^{-1}$ is calculated from the intensity as:
\begin{equation}
T_{int} = \int T dv  = \frac{c^3}{2 k_B \nu^3} I , 
\end{equation}
where $c$ is the speed of light, $\nu$ the frequency of the line, and $k_B$ is the Boltzmann constant.
 
 The model is computed from $A_V=0$ to $A_V=10$. We choose the maximum $A_V$ value to be representative of the average visual extinction measured in dark molecular clouds.  At these $A_V$, the temperature is already $\leq$ 10~K and the gas  enters the dense molecular cloud regime, where the freeze out starts to be efficient and it can not be considered a PDR anymore \citep{Bergin2007}.

We define the \ratio\ line ratio as the ratio between the integrated antenna temperatures:
\begin{equation}
r_{31} = \frac{T_{int,\text{CO(3-2)}}}{T_{int,\text{CO(1-0)}}} = \left(\frac{\nu_\text{CO(3-2)}}{\nu_\text{CO(1-0)}} \right)^{-3}\frac{I_\text{CO(3-2)}}{I_\text{CO(1-0)}},
\end{equation}
where $\nu$ is the frequency of the line.
In an analogous way we calculated \ratiotwo. This ratio is equivalent to the observed $L'_{CO}$ luminosity ratio that we studied in the previous section.

\subsection{CO line ratios from modelling}
We define a grid of models, varying three parameters: the volume density of hydrogen nuclei (\nH$=n(\text{H}) + 2\cdot n(\text{H}_2)$),  the FUV radiation field, and the cosmic ray ionization rate ($c.r.$). The values assumed in our models are summarized in Table\,\ref{tab:para_model}. The standard Galactic value of the cosmic ray ionization rate is  $2.5\cdot  10^{-17}$ s$^{-1}$ \citep{Shaw2008}. We select a range up to two orders of magnitude higher, to take into account the fact that in AGN the cosmic ray density is higher \citep[][ and references therein]{George2008}.
 Recent studies found that cosmic ray ionization rates can be up to 100 times the Galactic value in particular regions of the interstellar medium \citep{Indriolo2012, Indriolo2015,  Bisbas2015, Bisbas2017}. Even though these extreme conditions may happen close to the source of cosmic rays, i.e. the AGN, the cosmic ray ionization rate will decrease quickly with increasing H$_2$ column density  \citep{Padovani2009, Schlickeiser2016}.
 Since we are studying integrated CO fluxes within a beam that has a minimum size of $\sim$ 2 kpc,  we do not expect to have an average cosmic ray ionization rates higher than 10 times the Galactic value in the region covered by the CO beam.

We note that a limitation of our approach is the degeneracy of the low-J CO line ratios to the average state of the ISM \citep{Aalto1995}. Using only two low-J CO line ratios to derive physical properties of the gas can lead to large uncertainties.
 Additionally, it is possible that models with line ratios that match the observations have  individual intensities that are unrealistic. We compare the individual line intensities from the \uclpdr\ models which match the observed \ratio\ line ratios, with the observed line intensities (both for CO(3-2) and for CO(1-0)). For all galaxies, we find that the line intensities from the models are higher than the observed line intensities (by a factor that varies between 1.7 and 124). This can be explained by  beam dilution effects. The \uclpdr\ models assume a 100\% filling factor. The observed PDR regions typically do not fill the entire beam and thus the emission from the PDR regions is diluted when averaging over the beam. As a result, the observed intensities are lower than the ones predicted from the models.
Even given these limitations, qualitatively the \uclpdr\ models can provide an indication of which physical parameters have the highest impact in regulating the CO line ratios.

\begin{figure*}
\centering
\includegraphics[width=0.48\textwidth]{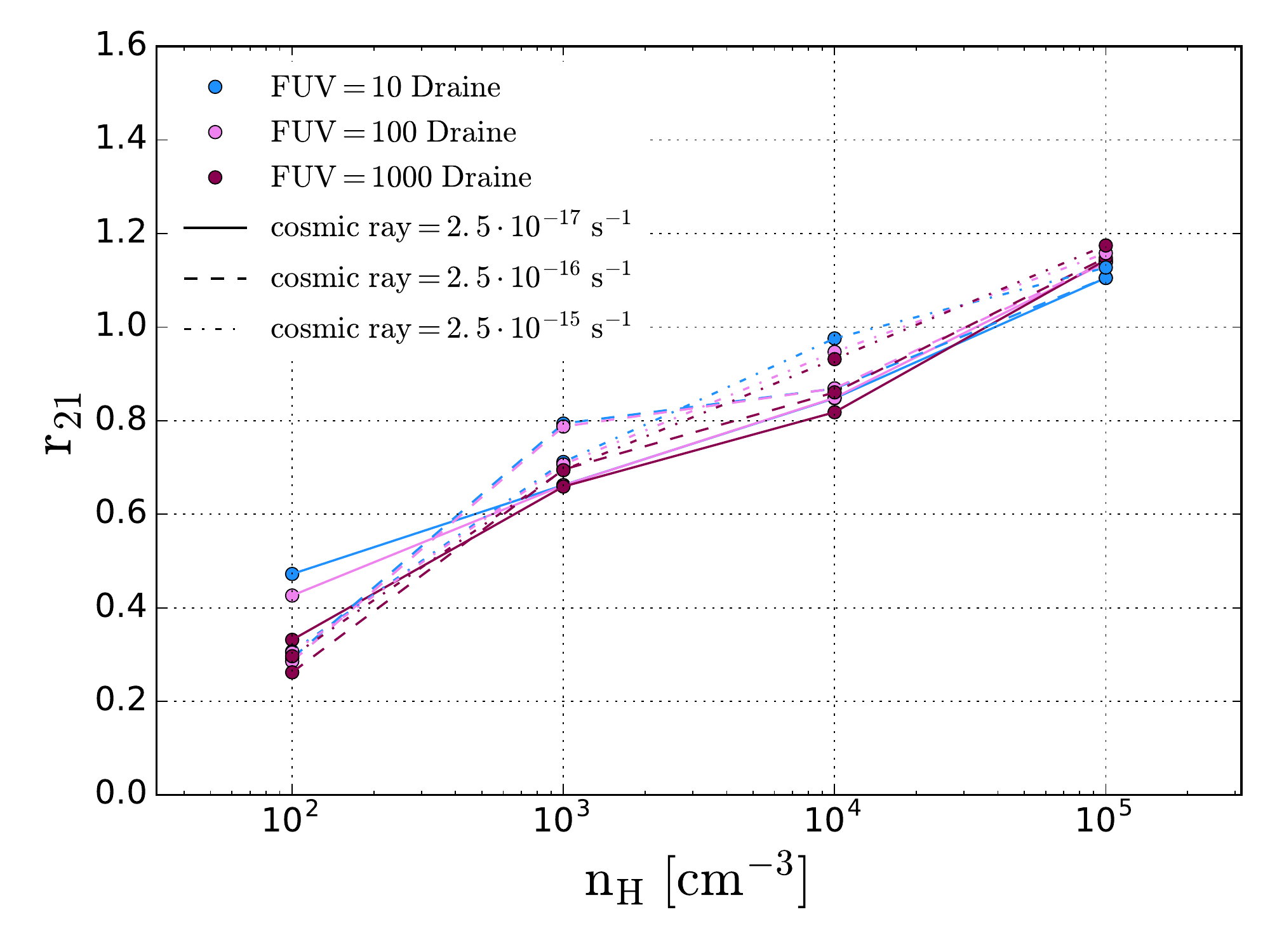}
\includegraphics[width=0.48\textwidth]{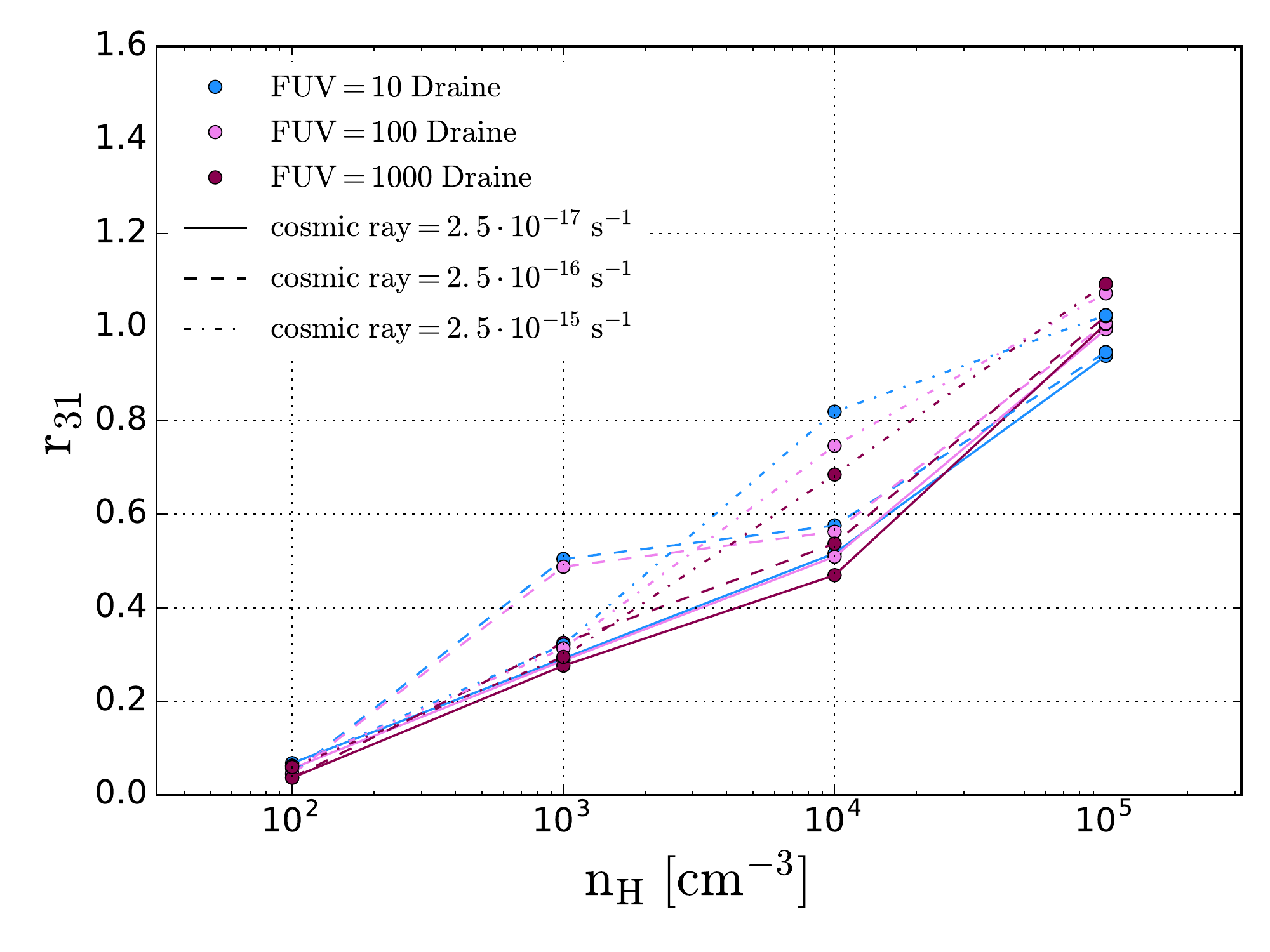}

\caption{CO line ratios $r_{21}= L'_\text{CO(2-1)}/L'_\text{CO(1-0)}$ (left) and $r_{31}= L'_\text{CO(3-2)}/L'_\text{CO(1-0)}$ (right) predicted from \uclpdr\ as a function of gas density \nH. 
 The colors indicate models with different FUV values: $10^1$ Draine (blue),  $10^2$ Draine (orange), $10^3$ Draine (magenta). The line styles indicate models with different cosmic ray ionization rate:  $2.5\cdot 10^{-17}$ s$^{-1}$ (full line), $2.5\cdot 10^{-16}$ s$^{-1}$(dashed line), $2.5\cdot 10^{-15}$ s$^{-1}$ (dotted-dashed line).} 
\label{fig:model_grid}
\end{figure*}

Figure\,\ref{fig:model_grid} shows the modelled line ratios \ratiotwo\ (left) and \ratio\ (right) as a function of \nH. The colors indicate different values of the FUV radiation field and different line types correspond to different cosmic ray ionization rates. The parameter that has the largest effect on the line ratios is the density \nH. As expected, there is a clear increase in both line ratios with \nH. The \ratiotwo\ values are in the range 0.3-1.1.
 The \ratio\ value goes from 0.01 at \nH = 10$^2$ cm$^{-3}$ to 1 at \nH = 10$^5$ cm$^{-3}$.

The FUV radiation field has  very little effect on the line ratio. The only visible difference is for the \ratiotwo\ ratio: at \nH$ = 10^2$ cm$^{-3}$
 it decreases from $\sim 0.45$ for FUV = 10 Draine\footnote{1 Draine $= 9.41\cdot  10^{-4}$ erg s$^{-1}$ cm$^{-2}$. The FUV radiation is defined by the standard Draine field \citep{Draine1978, Bell2006}.} to $\sim 0.3$ for FUV = 1000 Draine.  At low density, the high FUV field suppresses the CO emission in all J-levels. 
 This is due to the fact that a stronger FUV field will increase the photo-dissociation of CO and consequently the CO abundance will decrease. The $J = 2-1 $ level is slightly more suppressed that the $J=1-0$ level, causing a decrease in the \ratiotwo\ line ratio.

 We note also that the cosmic ray ionization rate does not have a big impact on the CO line ratios. We see an effect only at \nH = 10$^4$ cm$^{-3}$, where there is an enhancement of $\sim 0.2$ in both line ratios when the cosmic ray ionization rate is two order of magnitude above the Galactic value ($2.5 \cdot 10^{-15}$ s$^{-1}$).
 
So we conclude that both CO line ratios are mainly tracing the gas density. The range of variation of \ratiotwo\ is smaller than the range of \ratio, but it is still significant.

The mean  \ratio\ line ratio in the combined \coldgass, BASS and SLUGS samples is 0.55, which corresponds to a density of \nH $\sim 10^4$ cm$^{-3}$. We note that this value should be interpreted as the average gas density of the gas traced by CO, and not as the average gas density of the ISM in giant molecular clouds.
 The \ratiotwo\  value at that density from \uclpdr\ model is 0.8, which is consistent with the mean values reported by \citet{Saintonge2017} and \citet{Leroy2009}.

One possible caveat of our analysis is that the FUV radiation field is modelled as the standard Draine field in the range 912-2000 \AA, but the shape and intensity of the SED in the UV is different in AGN and in SFGs. 
 This effect is not considered in our current model. However, we consider a wide range for the strength of the FUV field, in order to take into account the stronger UV field due to the accretion disk of AGN.

\begin{table}
\centering
\caption{Parameters used in the grid of {\tt UCL-PDR} models.}
\begin{tabular}{ccc} % four columns, alignment for each
\hline
Gas density  & FUV radiation field & cosmic ray\\
 (\nH) & (FUV)  &ionization rate ($c.r.$) \\
$[$cm$^{-3}]$ &  [Draine (a)] & [$ 10^{-17}$ s$^{-1}$]\\ \hline
$10^2$ & $10$ &  $2.5$ (b)\\ 
$10^3$ & $10^2$& $25$\\
$10^4$ & $10^3$ & $250$\\
$10^5$ & &\\
\hline %\hline
\end{tabular}
\tablecomments{(a) 1 Draine $= 9.41\cdot  10^{-4}$ erg s$^{-1}$ cm$^{-2}$ . The FUV radiation is defined by the standard Draine field \citep{Draine1978, Bell2006}. (b) Standard Galactic value \citep{Shaw2008}.}
\label{tab:para_model}
\end{table}

\subsection{Effect of the X-rays}
\label{sec:xray}

We consider also the effect of the X-rays on the observed CO line ratios. AGN can be a strong source of X-rays and this could potentially affect the excitation of the CO molecules. The BASS sample is selected in the hard X-rays, and therefore we know that our sources are strong X-ray emitters. 

The X-ray chemistry and physics are implemented in the latest version of the \uclpdr\ code following \citet{Meijerink2005} and \citet{Stauber2005}. 
 The shape and intensity of the X-ray spectrum can be defined to describe the spectrum of an AGN or of a young stellar object \citep{Priestley2017}. In the case of an AGN, the X-ray spectrum is modelled in the range 1-10 keV as a black-body with a temperature of $1.16\cdot 10^{7}$ K, corresponding to an energy $kT = 1$~keV.
The intensity of the X-ray can be specified. 

We estimate the X-rays flux that would be observed at a distance of 1 kpc from the AGN, based on the observed fluxes measured in the 2-10 keV energy band from \citet{Ricci2017}.
 For our sample, this flux ranges from $10^{-4}$ to $3\cdot 10^{-1}$ erg s$^{-1}$ cm$^{-2}$, with a median of $10^{-2}$  erg s$^{-1}$ cm$^{-2}$.

\begin{figure*}
\centering
\includegraphics[width=0.48\textwidth]{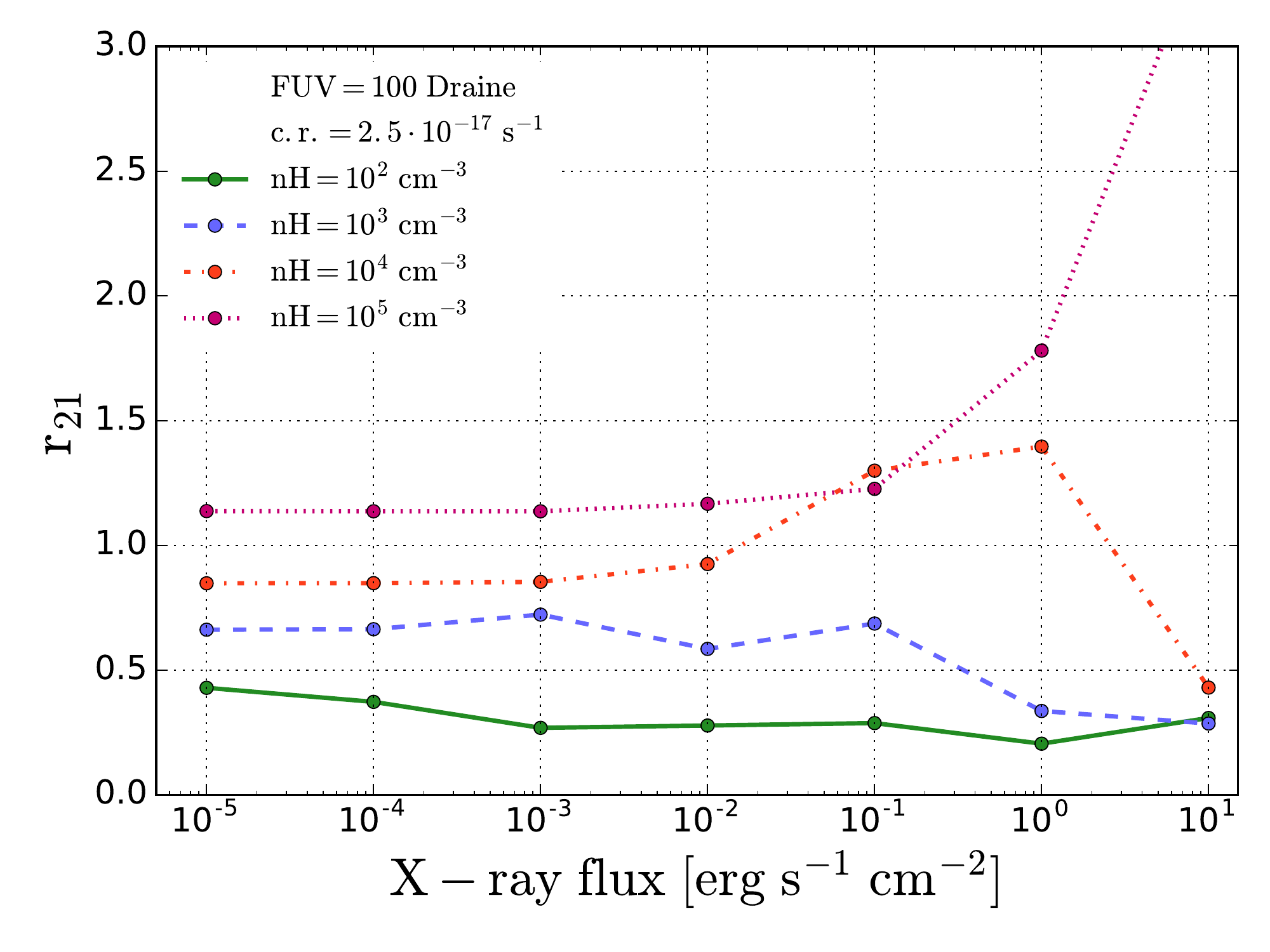}
\includegraphics[width=0.48\textwidth]{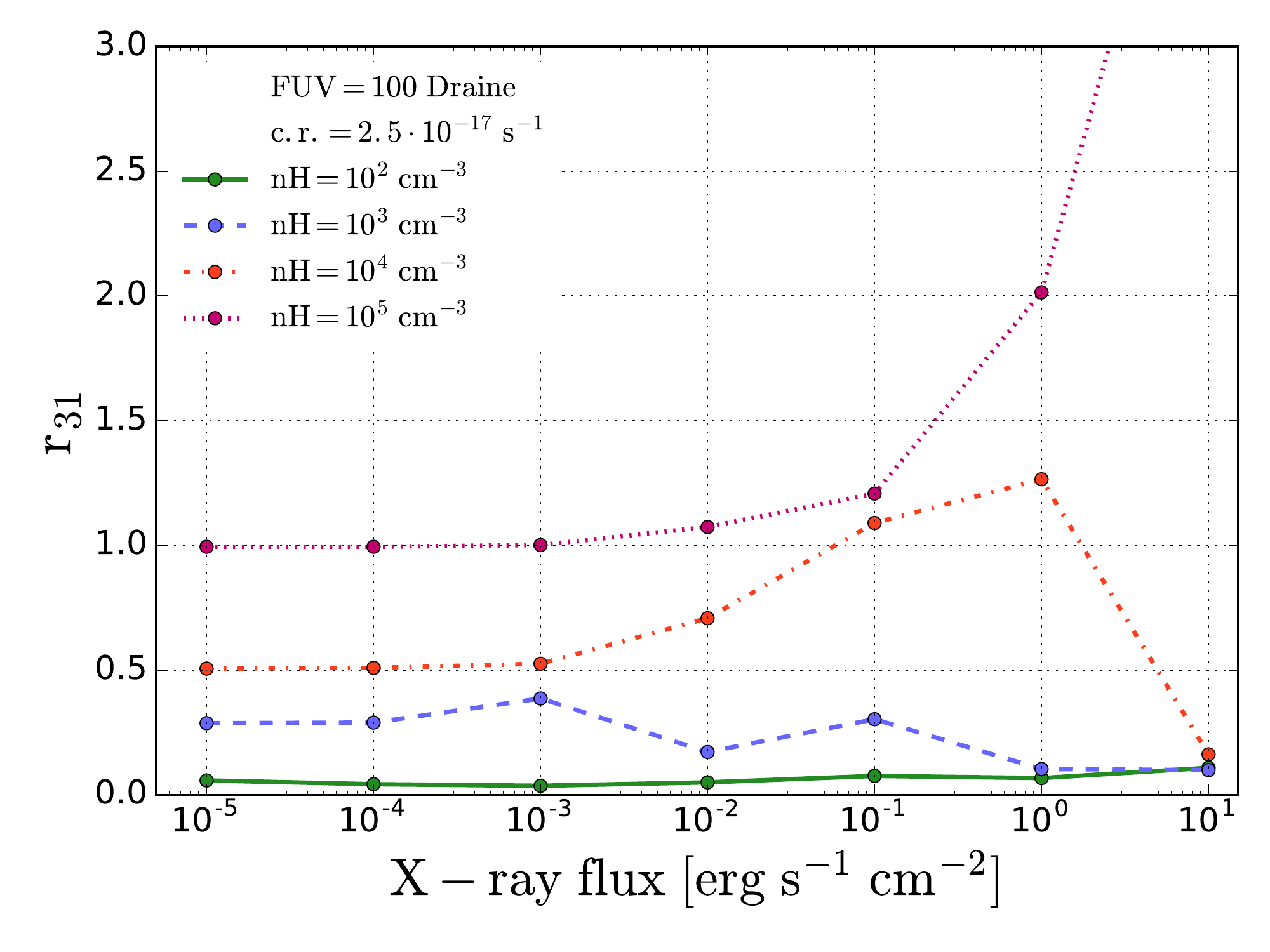}

\caption{Predicted CO line ratios from \uclpdr\ as a function of X-ray flux $r_{21}= L'_\text{CO(2-1)}/L'_\text{CO(1-0)}$ of the left and
$r_{31}= L'_\text{CO(3-2)}/L'_\text{CO(1-0)}$ on the right. The colors indicate models with different gas density  values:  \nH =$10^2$ cm$^{-3}$ (green solid line), $10^3$ cm$^{-3}$ (blue dashed line), $10^4$ cm$^{-3}$  (red dotted-dashed line), and $10^5$ cm$^{-3}$ (magenta dotted line). For all models the FUV radiation field is set at 100 Draine and the cosmic ray ionization rate is set at the standard Galactic value $2.5\cdot 10^{-17}$ s$^{-1}$. }
\label{fig:model_xray}
\end{figure*}

Figure\,\ref{fig:model_xray} shows the modelled CO ratios \ratiotwo\ and \ratio\ as a function of X-ray flux.
For X-ray flux $< 10^{-2}$ erg s$^{-1}$ cm$^{-2}$ 
 the effect on the CO ratios is negligible. This flux corresponds to an X-ray luminosity of $\sim10^{42}$ erg s$^{-1}$ in the 2-10 keV band, assuming that the flux is observed at 1 kpc from the nucleus.
  For higher X-ray fluxes in the range from 10$^{-2}$ to 1 erg s$^{-1}$ cm$^{-2}$, both \ratiotwo\ and \ratio\ are enhanced if they are combined with high densities ($n_\text{H}= 10^4-10^5$ cm$^{-3}$). If instead they are combined with lower densities ($n_\text{H} = 10^2-10^3$ cm$^{-3}$), the ratios stay constant or decrease.

If we consider an even higher X-ray flux of 10 erg s$^{-1}$ cm$^{-2}$ (corresponding to an X-ray luminosity of $\sim10^{45}$ erg s$^{-1}$), then the behaviour is clearly different for high and low densities. For \nH$<\ 10^5$ cm$^{-3}$, both line ratios decrease to $r_{21} <\ 0.5$ and $r_{31} <\ 0.3$. For the highest density considered \nH$= 10^5$ cm$^{-3}$, both line ratios increase to very high values ($>\ 3$). This can be explained by the fact that for low density gas the high X-ray flux reduces the CO abundance, due to photo-dissociation of CO. Thus the overall CO emission is weak and the CO ladder peaks at J=1. Only when the density is high enough can the X-rays start to excite the higher CO levels, causing the $r_{21}$ and $r_{31}$ levels to increase.

We conclude that the X-rays can affect the CO line ratios only for very high density and high X-ray flux. This is likely to occur only in a region very close to the active nucleus, but not in the rest of the galaxy. Thus if we consider the total CO emission of a galaxy, we do not expect to see a difference due to the presence of an AGN.

\section{Molecular Kennicutt-Schmidt relation}
\label{sec:KS_relation}
In this section we investigate how the relation between the SFR  and the molecular gas mass changes when the latter is derived from the CO(3-2) luminosity instead of from  the CO(1-0) luminosity.
Since CO(3-2) is tracing only the denser gas, we expect that the Kennicutt-Schmidt (KS) relation \citep{Kennicutt1998} measured from CO(3-2) will be tighter. 
Past studies found that the CO(3-2) emission correlates more strongly than the CO(1-0) emission with SFR \citep{Muraoka2007, Komugi2007, Wilson2009}.

For this analysis we consider all properties measured within the beam, applying inverse beam corrections to scale the total SFR to the SFR measured within the beam. In this way we can include also the galaxies with large angular size ($D > 100"$).

\begin{figure*}
\centering
\includegraphics[width=0.48\textwidth]{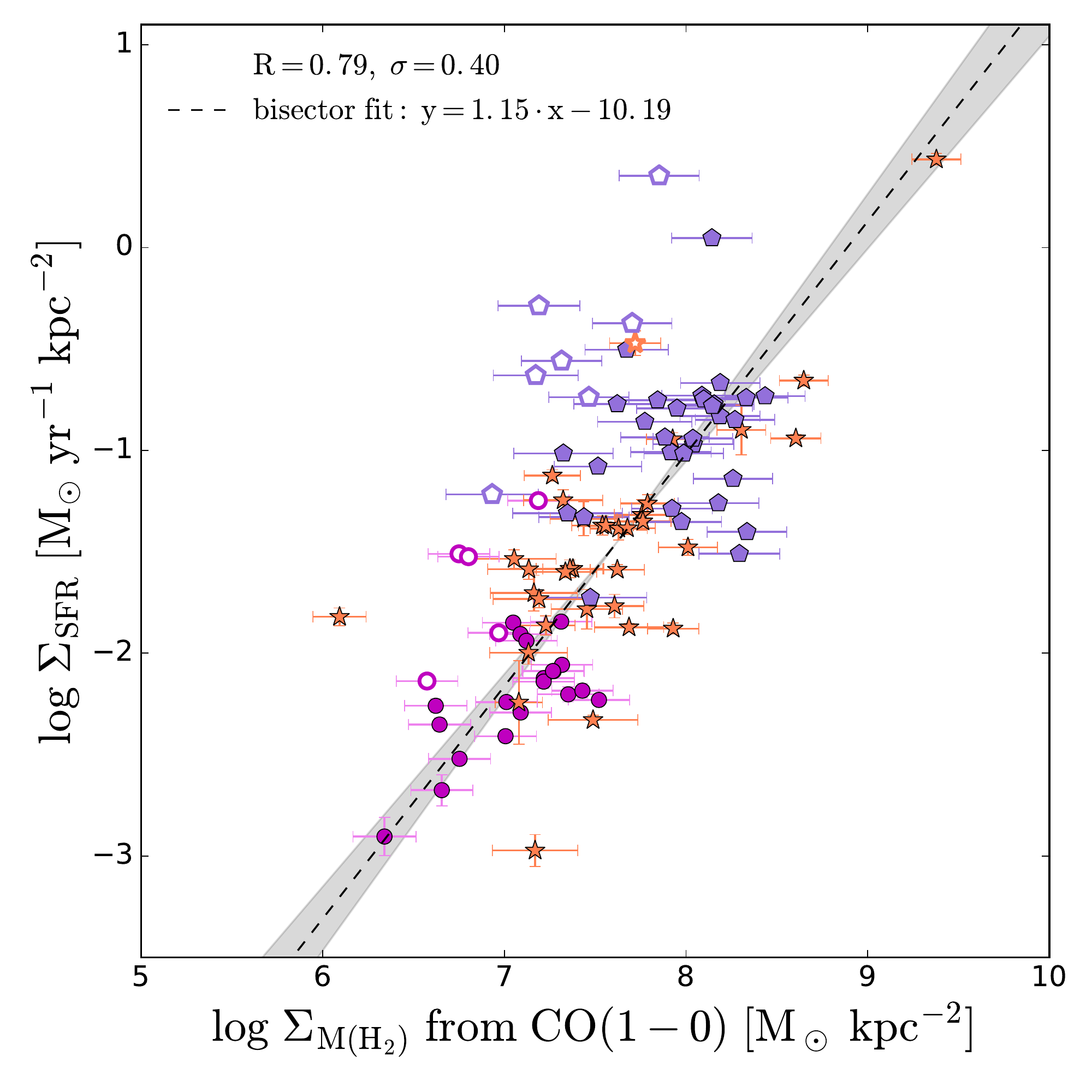}
\includegraphics[width=0.48\textwidth]{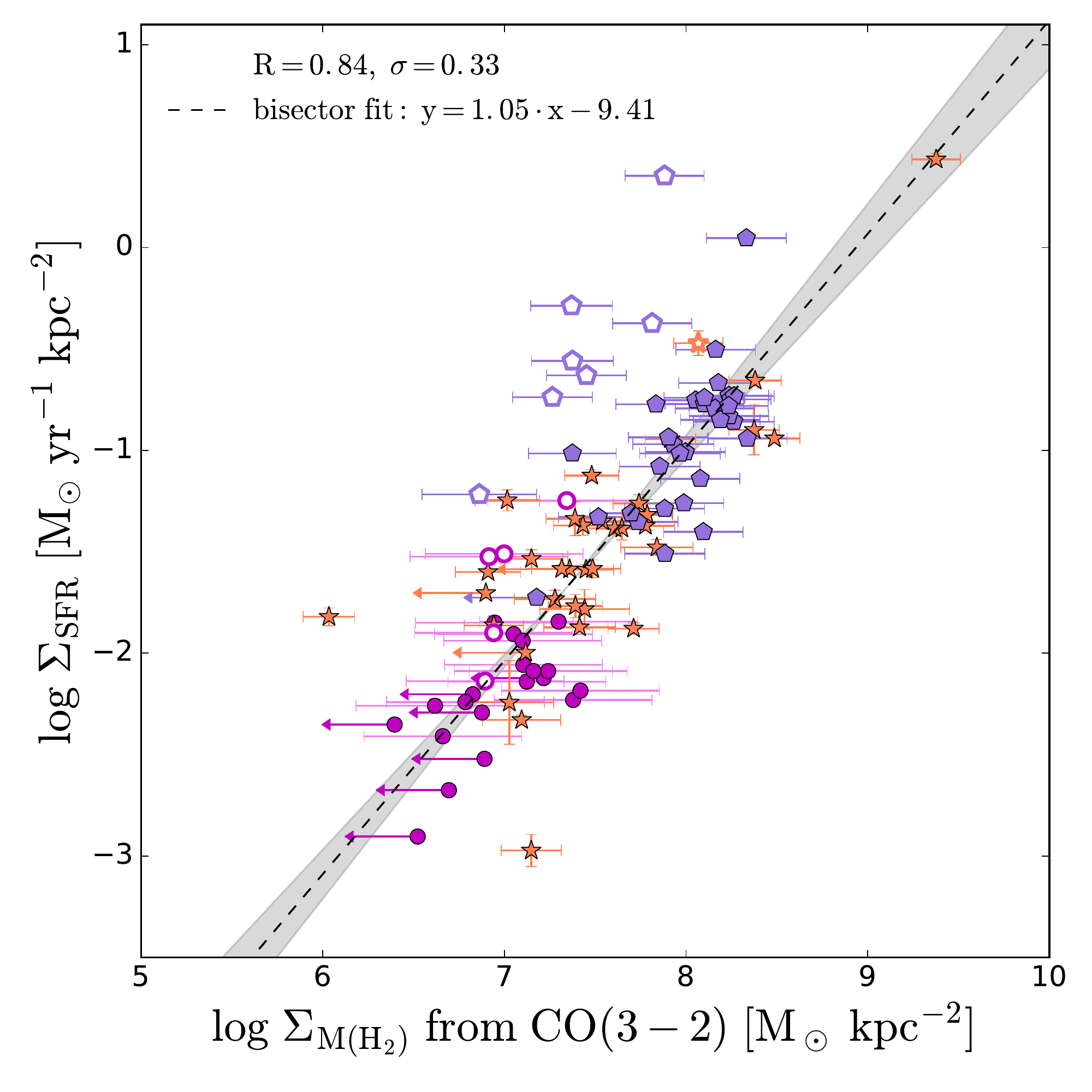}
\caption{Kennicutt-Schmidt relation  for the molecular gas measured from the CO(1-0) luminosity (left), and the molecular gas measured from the CO(3-2) luminosity (right). The CO(3-2) luminosities have been converted to CO(1-0) luminosities using a constant \ratio =0.55, before applying the same CO-to-H$_2$ conversion factor ($\alpha_{CO}$) as used for the CO(1-0) lines. The surface densities are calculated for the quantities within the telescope beams. The dashed line shows the ordinary least-squares bisector fit taking into account the upper limits. Circles are xCOLDGASS galaxies, the star symbol is used for AGN from BASS, and pentagons are galaxies from SLUGS. The empty symbols are the sources for which the ``ULIRG-type" CO-to-H$_2$ conversion factor ($\alpha_{CO}$)  was used and which are not included in the fit.}
\label{fig:KS_relation}
\end{figure*}

Figure\,\ref{fig:KS_relation} shows the KS relation with the total molecular gas mass measured from the CO(1-0) luminosity and the mass of the `dense' molecular mass measured from the CO(3-2) luminosity. 
 The CO(3-2) luminosities have been converted to CO(1-0) luminosities using a constant \ratio =0.55, before applying the same CO-to-H$_2$ conversion factor ($\alpha_{CO}$) used for the CO(1-0) luminosities.
 We did not include in the fit the galaxies for which we used the ``ULIRGs-type" $\alpha_{CO}$ conversion factor  (empty symbols in Fig.\,\ref{fig:KS_relation}). By assigning to them a different conversion factor, we implicitly assume that they have a different star-formation mechanism and do not follow the same relation between the amount of molecular gas and the SFR.
 The correlation of SFR surface density with the molecular gas mass derived from CO(3-2) (measured by the Pearson correlation coefficient $R=0.84$) is only slightly higher than  the correlation with the molecular gas mass measured from CO(1-0) ($R=0.79$). The two correlation coefficients are not significantly different, according to the Fisher Z-test ($p$-value = 0.06).

We fit the KS relation $\log \Sigma_{SFR} = a\cdot \log \Sigma_{M(H_2)} +b$  using the ordinary least-squares bisector fit \citep{Isobe1990} taking into account the upper limits and including an intrinsic scatter. The fit to the molecular gas derived by CO(1-0) has a slope $a = 1.15\pm 0.10$ with an intrinsic scatter of 0.48, while the fit to the molecular gas derived by CO(3-2) gives a slightly lower value $a = 1.05 \pm 0.09$ with intrinsic scatter 0.42.  
The two slopes are consistent with each other, within the uncertainties.
We find that the KS relation becomes tighter when we consider only the dense molecular gas traced by the CO(3-2) transition. The intrinsic scatter decreases from 0.40 to 0.33, but it is still quite large also for the dense molecular gas.
 Thus the fact that CO(1-0) is also tracing the diffuse molecular gas is probably not the only cause of the scatter in the KS relation.
 The CO(3-2) emission line is commonly used to measure the molecular gas content of galaxies at redshift $z> 1$, for which observations of the CO(1-0) line  are more time consuming. 
 Despite the fact that CO(3-2) is tracing denser gas than CO(1-0), the KS relations obtained from CO(3-2) and from CO(1-0) are similar, with slopes that are consistent with each other and similar scatters. It is important to note that we have excluded from this analysis the `ULIRGs'-type of galaxies.
 The similar KS slope of CO(3-2) and CO(1-0) suggests that there is no systematic trend in SFE along the KS relation for `normal' star-forming galaxies in the parameter space studied in this paper. This result may not hold for objects above the main-sequence (ULIRGs, starbursts), that have higher SFE with respect to MS galaxies.
  For `normal' star-forming galaxies (with $\log \text{SFR}< 1$), we do not observe a systematic variation of the mean \ratio\ line ratio as a function of SFR (Fig.~\ref{fig:r31_vs_SFE}). Thus we do not expect systematic variations in the relation between the emission of `dense' and `total' molecular gas in these galaxies. Therefore the KS relation derived from CO(3-2) can be directly compared to the KS relation derived from CO(1-0), once a constant offset due to the \ratio\ line ratio is taken into account. 
 
 For galaxies with higher SFR ($\log \text{SFR}> 1$), the \ratio\ ratio increases as a function of SFR. Thus for galaxies above the main sequence, the systematic increase of the \ratio\ values with SFR will cause the KS relation for CO(3-2) to be different from the CO(1-0) KS relation. Since we have excluded the ULIRG-type of galaxies from our analysis, this effect is not present in our result.
  We also note the that SFE measured in our samples is similar to the SFE of main-sequence galaxies at higher redshift (z$\sim$1-3). For example \cite{Aravena2019} find a typical depletion time of 1 Gyr (log SFE = -9)  in galaxies with $\log SFR = 1-1.5$, and \citet{Tacconi2013} find a mean depletion time of 0.7 Gyr, in a sample of galaxies at z$\sim$1-2.
 
 Our result suggests that the CO(3-2) line can be used to study the relation between SFR and molecular gas for high-redshift `main-sequence' galaxies, and to compare it with studies of low-redshift galaxies.

\section{Summary and conclusions}

In this paper we study the ratio between the CO(1-0) and CO(3-2) emission of star-forming galaxies and AGN using observations and modelling. 

Simulations from \uclpdr\  show that the main parameter regulating the \ratio\ ratio is the gas density. The FUV radiation field and X-rays play only a secondary role.

 We find a relation between the \ratio\ line ratio and the star-formation efficiency using  data from the \coldgass, BASS and SLUGS survey.
 This relation was already reported for the full SLUGS sample by \citet{Yao2003}, and in spatially resolved observations of M83, NGC 3627, and NGC 5055 \citep{Muraoka2007, Morokuma2017}.
 If the CO(1-0) emission traces the total molecular gas and the CO(3-2) emission traces the denser gas, then \ratio\ can be interpreted as a measure of the fraction of molecular gas which is in the dense star-forming molecular clouds.
If this fraction is higher, then the efficiency of a galaxy in forming stars  will be higher.
The same effect is reflected in the tightening of the Kennicutt-Schmidt relation when we consider only the dense molecular gas, traced by CO(3-2), instead of the total molecular gas, traced by CO(1-0).

We have shown that the SFE is related to the amount of molecular gas which is in the dense phase, but we 
do not know which factors cause the variation of the dense molecular gas fraction. 
 The presence of spiral arms and bars  may be connected to higher fraction of dense molecular gas \citep[e.g.][]{Sakamoto1999, Sheth2005}. 
The presence of a bulge may also have an impact, with SFE that may be different in bulge or disk-dominated galaxies \citep{Martig2009, Saintonge2012}. However, we do not see a relation between \ratio\ and the concentration index of the galaxies.

We also compare the \ratio\ values in star-forming galaxies and active galaxies, to test whether the presence of an AGN has an impact on the \ratio\ ratio. 
We do not see a difference in the distribution of  the \ratio\ values of AGN and SFGs. This is not surprising, as the effect of the AGN is expected to become relevant at higher J-levels \citep[J$>$10;][]{Lu2017}. The \uclpdr\ models show that the X-rays emitted from an AGN can have an impact on the \ratio\ values at higher gas density. However, the X-ray flux needs to be high ($>10^{-1}$ erg s$^{-1}$ cm$^{-2} $) and thus the X-rays can affect the condition of the ISM only close to the nucleus. This explains why we do not see this effect if we consider the total CO emission of the host galaxy.
This can be different at high redshift, where we can find both more luminous quasars (with bolometric luminosities $L_{bol}>10^{45}$ erg s$^{-1}$) and higher fraction of dense gas. In these conditions, the presence of an active nucleus could significantly impact the \ratio\ line ratio.

We do not find large variations in the \ratio\ line ratio in our sample of `normal' star-forming galaxies. However, based on our modelling, we expect to observe higher \ratio\ values in galaxies  with a larger fraction of dense gas, as for example in starburst galaxies, ULIRGs, or in sub-millimeter galaxies at high redshift \citep[e.g.][]{Gao2004a, Carilli2013, Riechers2013, Casey2014}.
%Therefore our results are not valid for these types of objects/
 If we were to study the \ratio\ ratio in a sample of starbursts or ULIRGs, we would probably find different results.
 Indeed \cite{Mao2010} found higher \ratio\ in starburst and ULIRGs ($0.89\pm0.11$ and $0.96\pm 0.14$ respectively) than in normal star-forming galaxies ($0.61\pm0.16$). 
 
In summary, the main conclusions of this paper are:
\begin{itemize}
\item The mean value of the $r_{31}$  ratio in our sample is $r_{31}= 0.55 \pm 0.03 $. There is no significant difference in the $r_{31}$ values of star-forming galaxies and AGN. \\

\item We model the $r_{31}$ using the \uclpdr\ code and find that the main parameter regulating the $r_{31}$ ratio is the gas density. The mean value $r_{31}= 0.55$ corresponds to a volume density of hydrogen nuclei   $n_{\text{H}}\sim 10^4$ cm$^{-3}$.\\

\item There is a trend for the $r_{31}$ ratio to increase with SFE ($p$-value=$5.4\cdot 10^{-5}$). % $P$-value=0.006 for the null hypothesis of no association.
 We find that the correlation with SFE is stronger than with SFR and SSFR.\\

\item  The correlation of the Kennicutt-Schmidt relation increases  when we consider molecular gas mass traced by CO(3-2) ($R=0.84$), instead of the molecular gas mass traced by CO(1-0) ($R=0.79$). However, the difference is not statistically significant ($p=0.06)$. This suggests that the CO(3-2) emission line can be used to study the relation between SFR and molecular gas for `normal' star-forming galaxies at high redshift, and to compare it with studies of low-redshift galaxies.
\end{itemize}

\newpage
\section*{acknowledgments}

We thank Felix Priestley for the help in installing and running the \uclpdr\ code.
A.S. acknowledges support from the Royal Society through the award of a University Research Fellowship.
M.K. acknowledges support from NASA through ADAP award NNH16CT03C.
E.T. acknowledges support from FONDECYT Regular 1160999, CONICYT PIA ACT172033 and Basal-CATA PFB-06/2007 and AFB170002 grants.
This work is based on observations carried out with the IRAM 30m telescope. IRAM is supported by INSU/CNRS (France), MPG (Germany) and IGN (Spain).
The James Clerk Maxwell Telescope is operated by the East Asian Observatory on behalf of The National Astronomical Observatory of Japan; Academia Sinica Institute of Astronomy and Astrophysics; the Korea Astronomy and Space Science Institute; the Operation, Maintenance and Upgrading Fund for Astronomical Telescopes and Facility Instruments, budgeted from the Ministry of Finance (MOF) of China and administrated by the Chinese Academy of Sciences (CAS), as well as the National Key R\& D Program of China (No. 2017YFA0402700). Additional funding support is provided by the Science and Technology Facilities Council of the United Kingdom and participating universities in the United Kingdom and Canada.
The {\tt Starlink} software \citep{Currie2014} is currently supported by the East Asian Observatory.
The Pan-STARRS1 Surveys (PS1) and the PS1 public science archive have been made possible through contributions by the Institute for Astronomy, the University of Hawaii, the Pan-STARRS Project Office, the Max-Planck Society and its participating institutes, the Max Planck Institute for Astronomy, Heidelberg and the Max Planck Institute for Extraterrestrial Physics, Garching, The Johns Hopkins University, Durham University, the University of Edinburgh, the Queen's University Belfast, the Harvard-Smithsonian Center for Astrophysics, the Las Cumbres Observatory Global Telescope Network Incorporated, the National Central University of Taiwan, the Space Telescope Science Institute, the National Aeronautics and Space Administration under Grant No. NNX08AR22G issued through the Planetary Science Division of the NASA Science Mission Directorate, the National Science Foundation Grant No. AST-1238877, the University of Maryland, Eotvos Lorand University (ELTE), the Los Alamos National Laboratory, and the Gordon and Betty Moore Foundation.
Funding for the SDSS and SDSS-II has been provided by the Alfred P. Sloan Foundation, the Participating Institutions, the National Science Foundation, the U.S. Department of Energy, the National Aeronautics and Space Administration, the Japanese Monbukagakusho, the Max Planck Society, and the Higher Education Funding Council for England. The SDSS Web Site is http://www.sdss.org/.
The SDSS is managed by the Astrophysical Research Consortium for the Participating Institutions. The Participating Institutions are the American Museum of Natural History, Astrophysical Institute Potsdam, University of Basel, University of Cambridge, Case Western Reserve University, University of Chicago, Drexel University, Fermilab, the Institute for Advanced Study, the Japan Participation Group, Johns Hopkins University, the Joint Institute for Nuclear Astrophysics, the Kavli Institute for Particle Astrophysics and Cosmology, the Korean Scientist Group, the Chinese Academy of Sciences (LAMOST), Los Alamos National Laboratory, the Max-Planck-Institute for Astronomy (MPIA), the Max-Planck-Institute for Astrophysics (MPA), New Mexico State University, Ohio State University, University of Pittsburgh, University of Portsmouth, Princeton University, the United States Naval Observatory, and the University of Washington.
This research has made use of the NASA/IPAC Extragalactic Database (NED) which is operated by the Jet Propulsion Laboratory, California Institute of Technology, under contract with the National Aeronautics and Space Administration.
This research made use of Astropy, a community-developed core Python package for Astronomy \citep{astropy}, {\tt Matplotlib} \citep{Hunter2007} and {\tt NumPy} \citep{VanDerWalt2011}.  This research made use of  {\tt APLpy}, an open-source plotting package for Python \citep{Robitaille2012}. This research used the {\tt TOPCAT} tool for catalogue cross-matching \citep{Taylor2005}. This research used the {\tt Stan} interface for Python {\tt PyStan} \citep{pystan}. 


\vspace{5mm}
\facilities{JCMT, IRAM:30m}


\software{Astropy \citep{astropy},
  		{\tt Matplotlib} \citep{Hunter2007},
  		{\tt NumPy} \citep{VanDerWalt2011},
  		{\tt TOPCAT} \citep{Taylor2005},
  		{\tt PyStan} \citep{pystan}.
        {\tt Starlink} \citep{Currie2014} 
         }


%\bibliography{sample63}{}
\bibliography{Biblio/CO32_paper_biblio}{}
\bibliographystyle{aasjournal}


%% command and will number appendix equations (A1), (A2), etc. The Figure and Table counter will not reset.

\appendix

\section{$r_{31}$ dependence on galaxy size and beam corrections}
\label{sec:r31_vs_beam_corr}

\begin{figure*}
\centering
\includegraphics[width=0.44\textwidth]
{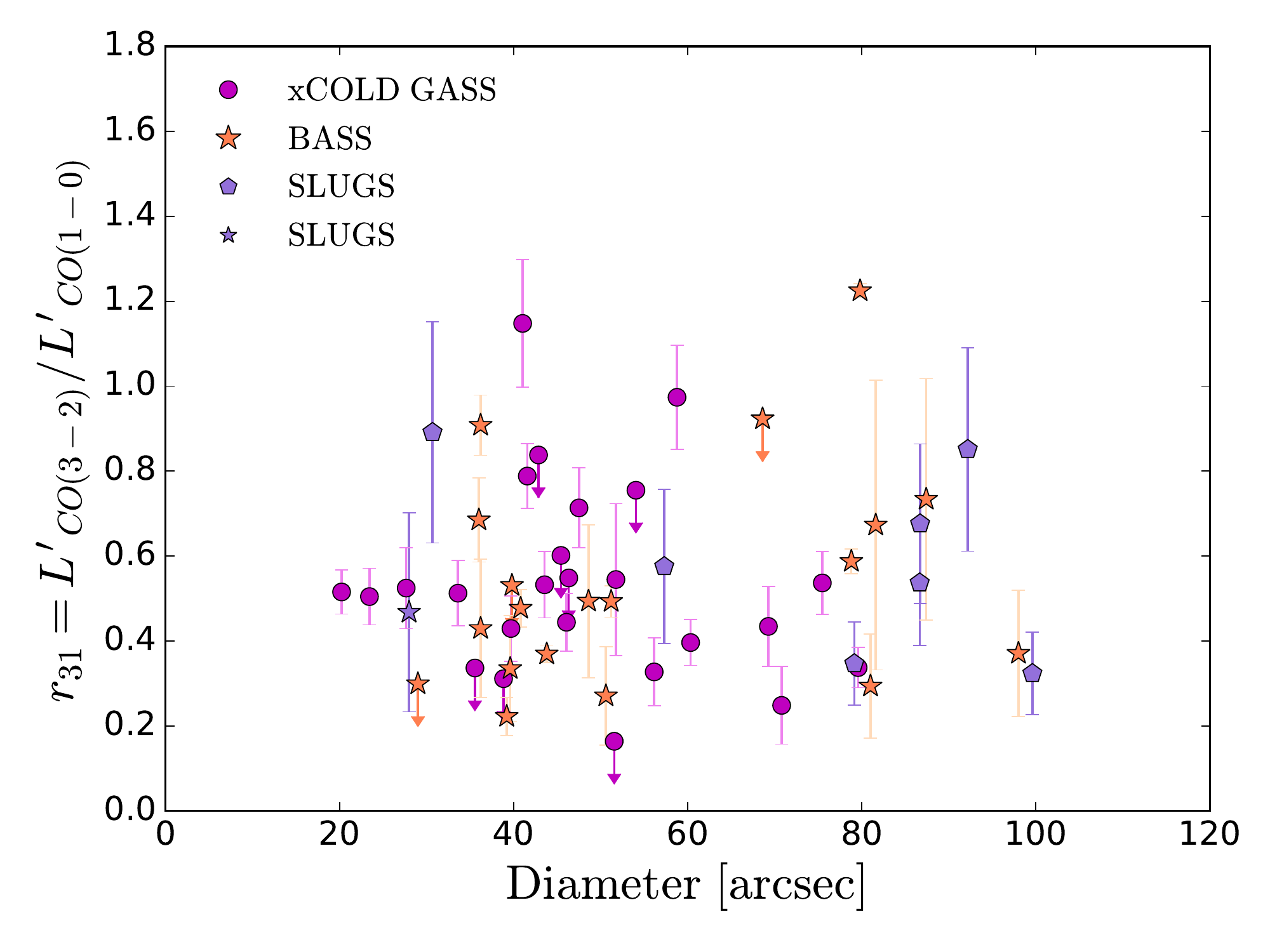}
\includegraphics[width=0.44\textwidth]
{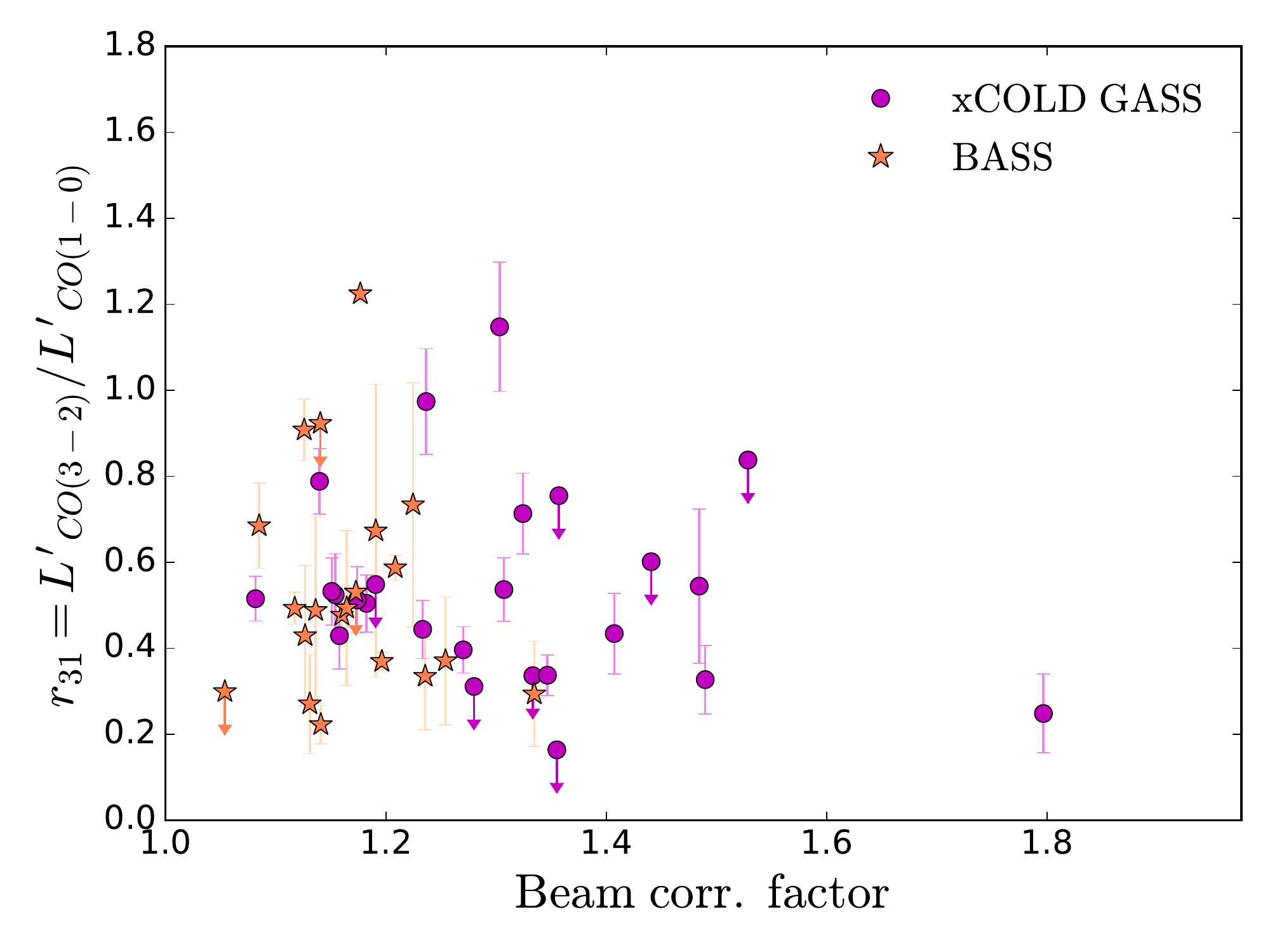}
\caption{\textit{Left:} Ratio
$r_{31}= L'_\text{CO(3-2)}/L'_\text{CO(1-0)}$ as a function of galaxy angular diameter. For the angular diameter of  \coldgass\ and SLUGS we use $D=D_{25}$, i.e. the optical diameter from SDSS g-band. For the angular diameter of BASS we use $D=2\times R_{k20}$, where $R_{k20}$ is the isophotal radius at 20mag arcsec$^{-2}$ in the K-band. \textit{Right:}  Ratio $r_{31}$ as a function of the beam correction applied to account for the different beam sizes of the CO(3-2) and CO(1-0) beam. The beam corrections extrapolate the CO(3-2) flux to the area of the CO(1-0) beam.
}
\label{fig:r31_vs_size}
\end{figure*}

In order to check that the beam corrections and beam sizes  do not have an effect on our analysis, we investigate if there is any relation between \ratio\ and galaxy angular size. For the angular size of  \coldgass\ and SLUGS we use $D=D_{25}$, i.e. the optical diameter from SDSS g-band. For BASS we use $D=2\times R_{k20}$, where $R_{k20}$ is the isophotal radius at 20mag arcsec$^{-2}$ in the K-band. %20mag/sq." isophotal K fiducial ell. ap. semi-major axis.
 If there is a correlation between \ratio\ and the galaxy angular size, that could mean that the part of the galaxy that we are sampling is affecting the \ratio\ measurements (i.e. the difference in the CO(1-0) and CO(3-2) beam sizes are affecting the \ratio\ measurements.)
For galaxies with large angular size, the telescope beam is only sampling a small part of the galaxy. If the gas is denser in the central part of the galaxy, \ratio\ will be higher, and thus we expect to observe a higher \ratio\ for galaxies with large angular sizes. On the other hand, if the region of the galaxy included in the beam is large enough, we would not find this trend.
The left panel of Fig.\,\ref{fig:r31_vs_size} shows \ratio\ as a function of galaxy angular size. We do not find any trend of \ratio\ increasing or decreasing with angular size ($R$ = -0.03). Thus we can rule out the possibility that the angular size plays a significant role in the \ratio\ variations.
 We note that BASS objects have in general larger angular size that the \coldgass\ galaxies, due to their lower redshift (most objects in the BASS sample have $z < 0.025$ with respect to $z = 0.026-0.05$ for \coldgass).
We also look at the distribution of \ratio\ with respect to the beam correction factor $C_{IR, PSF}$ (right panel of Fig.\,\ref{fig:r31_vs_size}). We do not see any evidence of \ratio\ increasing with the beam correction factor.

\begin{figure}
\centering
\includegraphics[width=0.4\textwidth]
{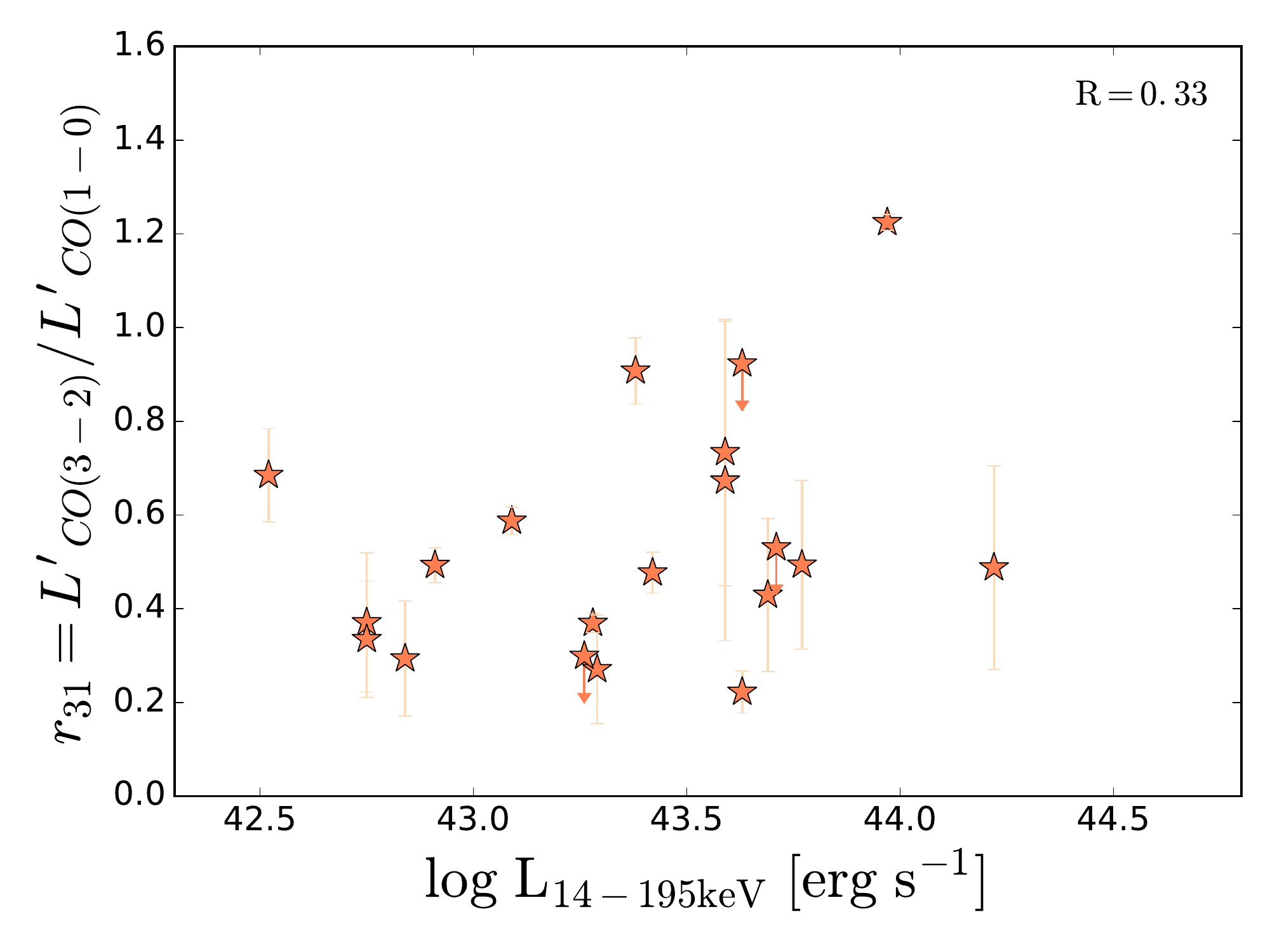}
\caption{Ratio \ratio = \LCOthree / \LCOone as a function of hard X-ray luminosity measured in the 14-195 keV band for the BASS sample.}
\label{fig:r31_vs_xray}
\end{figure}


%%%%%%%%%%%%%%%%%%%%%%%%%%%%%%%%%%%%%%%%%%%%%%%%%%
\section{Tables}

\begin{table*}

\centering
\caption{Properties of the \coldgass\ sample.}
%\begin{tabular}{rlrlrrr}
\begin{tabular}{rccccccccc}
\hline
Index & RA  & Dec & z &  D$_{25}$ & log M$_*$ & log SFR & \ $\alpha_{CO}$ & log M(H$_2$)  \\ 
   & [deg] & [deg] & & [arcsec] &[log \Msun] & [log \Msun yr$^{-1}$]& $\left[\frac{\text{M}_\odot}{\text{K km s}^{-1}\text{pc}^2}\right]$ &  [log \Msun] \\  
  \hline \hline
  
 522 & 171.07767 & 0.64373 & 0.02637 & 51.7 & 10.08 & 0.40 & 4.35 & 9.13 \\ 
1115 & 214.56213 & 0.89111 & 0.02595 & 69.3 & 10.11 & 0.76 & 4.35 & 9.61 \\ 
1137 & 215.81121 & 0.97835 & 0.04007 & 38.8 & 10.28 & 0.56 & 4.35 & 9.42 \\ 
1221 & 218.85558 & 0.33433 & 0.03455 & 23.4 & 10.20 & 0.84 & 4.35 & 9.73 \\ 
3819 & 25.42996 & 13.67579 & 0.04531 & 27.7 & 10.67 & 1.03 & 4.35 & 9.98 \\ 
3962 & 30.99646 & 14.31038 & 0.04274 & 46.3 & 10.90 & 0.80 & 4.35 & 10.06 \\ 
4045 & 32.88983 & 13.91716 & 0.02651 & 43.5 & 10.47 & 0.66 & 4.35 & 9.71 \\ 
7493 & 216.83387 & 2.83838 & 0.02644 & 46.1 & 10.56 & 0.41 & 4.35 & 9.64 \\ 
9551 & 216.88492 & 4.82163 & 0.02688 & 60.3 & 10.91 & 0.28 & 4.35 & 9.97 \\ 
11112 & 345.66796 & 13.32907 & 0.02765 & 51.6 & 10.81 & 0.46 & 4.35 & 9.85 \\ 
11223 & 346.56850 & 13.98231 & 0.03554 & 33.6 & 10.64 & 0.67 & 4.35 & 9.94 \\ 
11408 & 350.61417 & 13.81586 & 0.026 & 45.4 & 10.05 & 0.08 & 4.35 & 9.12 \\ 
14712 & 139.74192 & 5.88840 & 0.03827 & 35.5 & 10.55 & 0.63 & 4.35 & 9.82 \\ 
15155 & 160.22996 & 5.99141 & 0.02773 & 79.6 & 10.19 & 0.57 & 4.35 & 9.88 \\ 
22436 & 139.97725 & 32.93328 & 0.04916 & 58.8 & 10.42 & 1.54 & 1.00 & 9.75 \\ 
23194 & 158.38929 & 11.87138 & 0.03404 & 41.0 & 10.59 & 0.70 & 1.00 & 9.22 \\ 
23245 & 160.91283 & 12.06066 & 0.02623 & 42.9 & 10.11 & -0.05 & 4.35 & 8.76 \\ 
24973 & 218.82654 & 35.11868 & 0.0285 & 47.5 & 10.61 & 1.09 & 1.00 & 9.33 \\ 
25327 & 203.10100 & 11.10636 & 0.03144 & 41.6 & 10.03 & 1.40 & 1.00 & 9.72 \\ 
25763 & 135.79688 & 10.15197 & 0.02962 & 54.0 & 10.11 & 0.22 & 4.35 & 9.39 \\ 
26221 & 154.15996 & 12.57738 & 0.03166 & 75.5 & 10.98 & 0.57 & 4.35 & 10.07 \\ 
28365 & 235.34412 & 28.22975 & 0.03209 & 56.1 & 10.36 & 0.60 & 4.35 & 9.65 \\ 
40439 & 196.06267 & 9.22346 & 0.03501 & 70.8 & 10.95 & 0.73 & 4.35 & 9.89 \\ 
42013 & 229.01862 & 6.84763 & 0.03681 & 39.7 & 10.77 & 0.70 & 4.35 & 9.96 \\ 
48369 & 167.80421 & 28.71190 & 0.02931 & 20.2 & 10.32 & 0.67 & 1.00 & 9.42 \\ 

\hline
\end{tabular}
\label{tab:COLD_prop}
\end{table*}

\begin{table*}
\centering
\caption{CO(3-2) measurements for the \coldgass\ sample.}
\begin{tabular}{rcccccccc}
\hline
Index & $\sigma_\text{CO32}$ rms & S/N$_\text{CO32}$ & flag$_\text{CO32}$
 & $S_\text{CO32}$ & $\log L'_\text{CO(32)}$
  & $\log L'_\text{CO(10)}$ &  \ratio\ & beam corr. \\    
  & $[$mK] &   &   & [Jy km s$^{-1}$] & [log K km s$^{-1}$pc$^2$] &[log K km s$^{-1}$pc$^2$] &   &  14" to 22" \\ 
 (1)  & (2) &  (3) & (4)  & (5) & (6) &(7) & (8)  & (9)   \\ 
         
 \hline \hline
522 & 3.00 & 3.47 & 1 & 22.63 $\pm$3.67 & 7.90 $\pm$0.13 & 8.34 $\pm$0.07 & 0.54 $\pm$ 0.18 & 1.48\\ 
1115 & 2.73 & 6.62 & 1 & 50.81 $\pm$8.24 & 8.24 $\pm$0.07 & 8.75 $\pm$0.07 & 0.43 $\pm$ 0.09 & 1.41\\ 
1137 & 1.80 & -1.83 & 2 & -7.37 $\pm$3.58 & $<$ 8.10 & 8.72 $\pm$0.05 & $<$ 0.31 & 1.28\\ 
1221 & 2.22 & 12.03 & 1 & 76.36 $\pm$8.69 & 8.67 $\pm$0.04 & 9.04 $\pm$0.04 & 0.50 $\pm$ 0.07 & 1.18\\ 
3819 & 4.56 & 6.53 & 1 & 86.85 $\pm$9.42 & 8.96 $\pm$0.07 & 9.30 $\pm$0.04 & 0.52 $\pm$ 0.10 & 1.15\\ 
3962 & 6.52 & 1.33 & 2 & 29.70 $\pm$11.12 & $<$ 9.01 & 9.35 $\pm$0.04 & $<$ 0.55 & 1.19\\ 
4045 & 7.50 & 8.56 & 1 & 139.84 $\pm$13.38 & 8.70 $\pm$0.05 & 9.03 $\pm$0.04 & 0.53 $\pm$ 0.08 & 1.15\\ 
7493 & 2.24 & 9.32 & 1 & 87.24 $\pm$12.32 & 8.49 $\pm$0.05 & 8.93 $\pm$0.05 & 0.44 $\pm$ 0.07 & 1.23\\ 
9551 & 3.06 & 10.97 & 1 & 152.23 $\pm$22.25 & 8.75 $\pm$0.04 & 9.25 $\pm$0.04 & 0.40 $\pm$ 0.05 & 1.27\\ 
11112 & 2.07 & 0.04 & 2 & 0.38 $\pm$16.14 & $<$ 8.19 & 9.11 $\pm$0.05 & $<$ 0.16 & 1.36\\ 
11223 & 3.24 & 8.65 & 1 & 118.99 $\pm$12.47 & 8.88 $\pm$0.05 & 9.24 $\pm$0.04 & 0.51 $\pm$ 0.08 & 1.17\\ 
11408 & 3.99 & -0.09 & 2 & -0.79 $\pm$3.63 & $<$ 7.98 & 8.36 $\pm$0.06 & $<$ 0.60 & 1.44\\ 
14712 & 3.87 & -0.38 & 2 & -4.21 $\pm$8.72 & $<$ 8.53 & 9.12 $\pm$0.04 & $<$ 0.34 & 1.33\\ 
15155 & 1.14 & 16.03 & 1 & 76.42 $\pm$14.23 & 8.47 $\pm$0.03 & 9.07 $\pm$0.05 & 0.34 $\pm$ 0.05 & 1.35\\ 
22436 & 2.88 & 16.83 & 1 & 277.51 $\pm$16.61 & 9.54 $\pm$0.03 & 9.64 $\pm$0.05 & 0.97 $\pm$ 0.12 & 1.24\\ 
23194 & 4.32 & 12.70 & 1 & 209.96 $\pm$11.39 & 9.09 $\pm$0.03 & 9.15 $\pm$0.05 & 1.15 $\pm$ 0.15 & 1.30\\ 
23245 & 3.68 & 2.18 & 2 & 17.48 $\pm$2.17 & $<$ 7.79 & 8.05 $\pm$0.06 & $<$ 0.84 & 1.53\\ 
24973 & 8.11 & 12.31 & 1 & 218.30 $\pm$18.02 & 8.95 $\pm$0.04 & 9.22 $\pm$0.05 & 0.71 $\pm$ 0.09 & 1.32\\ 
25327 & 7.48 & 23.79 & 1 & 677.88 $\pm$43.03 & 9.53 $\pm$0.02 & 9.69 $\pm$0.04 & 0.79 $\pm$ 0.08 & 1.14\\ 
25763 & 5.56 & -0.15 & 2 & -1.77 $\pm$4.66 & $<$  8.31 & 8.57 $\pm$0.07 & $<$ 0.75 & 1.36\\ 
26221 & 3.17 & 12.09 & 1 & 161.63 $\pm$17.63 & 8.91 $\pm$0.04 & 9.30 $\pm$0.05 & 0.54 $\pm$ 0.07 & 1.31\\ 
28365 & 3.22 & 4.65 & 1 & 33.01 $\pm$7.23 & 8.24 $\pm$0.09 & 8.89 $\pm$0.05 & 0.33 $\pm$ 0.08 & 1.49\\ 
40439 & 1.92 & 3.15 & 1 & 20.48 $\pm$7.69 & 8.11 $\pm$0.14 & 8.96 $\pm$0.08 & 0.25 $\pm$ 0.09 & 1.80\\ 
42013 & 3.81 & 6.81 & 1 & 100.05 $\pm$13.38 & 8.84 $\pm$0.06 & 9.27 $\pm$0.04 & 0.43 $\pm$ 0.08 & 1.16\\ 
48369 & 3.99 & 18.72 & 1 & 283.35 $\pm$26.96 & 9.09 $\pm$0.02 & 9.41 $\pm$0.04 & 0.52 $\pm$ 0.05 & 1.08\\

\hline
\label{tab:COLD_CO}
\end{tabular}

(1)~Index.
(2)~Standard deviation of the noise.
(3)~Integrated S/N of the CO(3-2) line.
(4)~Flag for the detection of the CO(3-2) line based on a peak S/N $>$ 3. 1: detected, 2: non-detected.
(5)~Velocity-integrated flux $S_\text{CO32}$ within the 14" JCMT HARP beam.
(6)~CO(3-2) luminosity within the 14" JCMT HARP beam.
(7)~CO(1-0) luminosity within the 22" IRAM beam.
(8)~Beam corrected luminosity ratio $r_{31} = L'_{CO(32)}/L'_{CO(10)} \cdot \text{beam correction}$.
(9)~Beam correction factor for extrapolating the CO(3-2) flux from the 14" JCMT HARP to the 22" IRAM beam.
\end{table*}

\begin{table*}
\centering
\caption{Properties of the BASS sample.}
\begin{tabular}{rlcccccccc}
\hline
BAT  & Name & RA  & Dec & z &  Rk20 & log M$_*$ & log SFR & \ $\alpha_{CO}$ & log M(H$_2$)  \\ 
index   &  & [deg] & [deg] & & [arcsec] &[log \Msun] & [log \Msun yr$^{-1}$]& $\left[\frac{\text{M}_\odot}{\text{K km s}^{-1}\text{pc}^2}\right]$ &  [log \Msun] \\  
  \hline \hline
144 & NGC1068 &40.66960 & -0.01330 & 0.00303 & 95.0 & 10.54 & ... &4.35 & 9.79 \\ 
173 & NGC1275 &49.95070 & 41.51170 & 0.01658 & 62.9 & 11.13 & ... &4.35 & ...\\ 
228 & Mrk618 &69.09300 & -10.37600 & 0.03464 & 18.1 & 10.62 & 1.19 & 4.35 & 10.19 \\ 
308 & NGC2110 &88.04740 & -7.45620 & 0.00739 & 54.9 & 10.56 & 0.15 & 4.35 & 8.62 \\ 
310 & MCG+08-11-011 &88.72340 & 46.43930 & 0.02019 & 54.8 & 10.72 & -0.43 & 4.35 & 9.84 \\ 
316 & IRAS05589+2828 &90.54365 & 28.47205 & 0.03309 & 0.0 & 10.57 & 0.35 & 4.35 & 9.75 \\ 
337 & VIIZW073 &98.19654 & 63.67367 & 0.04042 & 34.3 & 10.46 & 1.13 & 4.35 & 9.98 \\ 
382 & Mrk79 &115.63670 & 49.80970 & 0.02213 & 30.9 & 10.49 & 0.46 & 4.35 & ...\\ 
399 & 2MASXJ07595347+2323241 &119.97280 & 23.39010 & 0.02894 & 21.1 & 10.72 & ... &4.35 & 10.32 \\ 
400 & IC0486 &120.08740 & 26.61350 & 0.02656 & 19.9 & 10.57 & 0.49 & 4.35 & 9.75 \\ 
404 & Mrk1210 &121.02440 & 5.11380 & 0.01354 & 15.2 & 9.93 & -0.18 & 4.35 & ...\\ 
405 & MCG+02-21-013 &121.19330 & 10.77670 & 0.03486 & 19.6 & 10.76 & 0.33 & 4.35 & 10.28 \\ 
439 & Mrk18 &135.49300 & 60.15200 & 0.01101 & 18.0 & 9.74 & 0.01 & 4.35 & 8.73 \\ 
451 & IC2461 &139.99200 & 37.19100 & 0.00753 & 40.5 & 10.06 & -0.45 & 4.35 & 9.20 \\ 
471 & NGC2992 &146.42520 & -14.32640 & 0.00757 & 50.9 & 10.22 & 0.34 & 4.35 & 9.30 \\ 
480 & NGC3081 &149.87310 & -22.82630 & 0.00763 & 54.3 & 9.96 & -0.33 & 4.35 & 8.85 \\ 
497 & NGC3227 &155.87740 & 19.86510 & 0.00329 & 92.6 & 10.06 & 0.35 & 4.35 & 9.74 \\ 
517 & UGC05881 &161.67700 & 25.93130 & 0.02048 & 14.5 & 10.17 & 0.42 & 4.35 & 9.42 \\ 
530 & NGC3516 &166.69790 & 72.56860 & 0.00871 & 40.8 & 10.65 & -0.28 & 4.35 & 8.73 \\ 
532 & IC2637 &168.45700 & 9.58600 & 0.02915 & 18.1 & 10.60 & 1.04 & 4.35 & 10.10 \\ 
548 & NGC3718 &173.14520 & 53.06790 & 0.00279 & 75.5 & 10.03 & -0.83 & 4.35 & 8.49 \\ 
552 & Mrk739E &174.12200 & 21.59600 & 0.02945 & 20.4 & 10.60 & 0.89 & 4.35 & 9.98 \\ 
560 & NGC3786 &174.92700 & 31.90900 & 0.00897 & 49.0 & 10.32 & -0.06 & 4.35 & 9.52 \\ 
585 & NGC4051 &180.79010 & 44.53130 & 0.00203 & 102.6 & 9.82 & 0.17 & 4.35 & 9.80 \\ 
588 & UGC07064 &181.18060 & 31.17730 & 0.02508 & 21.9 & 10.60 & 0.76 & 4.35 & 10.10 \\ 
590 & NGC4102 &181.59630 & 52.71090 & 0.00185 & 68.6 & 10.18 & 0.54 & 4.35 & 9.57 \\ 
599 & NGC4180 &183.26200 & 7.03800 & 0.00700 & 40.0 & 9.96 & 0.17 & 4.35 & ...\\ 
608 & Mrk766 &184.61050 & 29.81290 & 0.01292 & 25.6 & 10.11 & 0.35 & 4.35 & 9.26 \\ 
609 & M106 &184.73960 & 47.30400 & 0.00168 & 263.9 & 10.22 & -0.10 & 4.35 & 9.29 \\ 
615 & NGC4388 &186.44480 & 12.66210 & 0.00834 & 92.9 & 10.08 & -0.09 & 4.35 & 7.91 \\ 
631 & NGC4593 &189.91430 & -5.34430 & 0.00835 & 80.6 & 10.46 & ... &4.35 & 9.24 \\ 
669 & NGC5100NED02 &200.24830 & 8.97830 & 0.03259 & 21.9 & 10.71 & 1.19 & 4.35 & ...\\ 
670 & MCG-03-34-064 &200.60190 & -16.72860 & 0.01682 & 25.3 & 10.47 & 0.68 & 4.35 & 9.34 \\ 
688 & NGC5290 &206.32990 & 41.71260 & 0.00854 & 89.0 & 10.39 & -0.03 & 4.35 & 9.86 \\ 
703 & Mrk463 &209.01200 & 18.37210 & 0.05015 & 14.0 & 10.59 & ... &4.35 & ...\\ 
712 & NGC5506 &213.31190 & -3.20750 & 0.00609 & 74.6 & 9.92 & -0.26 & 4.35 & 8.84 \\ 
723 & NGC5610 &216.09540 & 24.61440 & 0.01691 & 39.4 & 10.34 & 0.70 & 4.35 & 9.91 \\ 
739 & NGC5728 &220.59970 & -17.25320 & 0.00990 & 80.4 & 10.31 & 0.19 & 4.35 & 9.34 \\ 
766 & NGC5899 &228.76350 & 42.04990 & 0.00844 & 67.8 & 10.37 & 0.54 & 4.35 & 9.74 \\ 
772 & MCG-01-40-001 &233.33630 & -8.70050 & 0.02285 & 43.7 & 10.57 & 0.81 & 4.35 & 9.84 \\ 
783 & NGC5995 &237.10400 & -13.75780 & 0.02442 & 24.3 & 10.87 & 1.04 & 4.35 & 10.18 \\ 
841 & NGC6240 &253.24540 & 2.40090 & 0.02386 & 39.9 & 11.02 & 1.75 & 1.00 & 10.07 \\ 
1042 & 2MASXJ19373299-0613046 &294.38800 & -6.21800 & 0.01036 & 19.8 & 9.97 & -0.17 & 4.35 & 9.29 \\ 
1046 & NGC6814 &295.66940 & -10.32350 & 0.00576 & 71.7 & 10.32 & 0.14 & 4.35 & 9.16 \\ 
1133 & Mrk520 &330.17242 & 10.55221 & 0.02753 & 14.4 & 10.33 & ... &4.35 & ...\\ 
1184 & NGC7479 &346.23610 & 12.32290 & 0.00705 & 87.9 & 10.41 & 0.57 & 4.35 & 9.86 \\ 

\hline
\end{tabular}
\label{tab:BASS_prop}
\end{table*}

\begin{table*}

\centering
\caption{CO(3-2) and CO(2-1) measurements for the BASS sample.}
\begin{tabular}{rcccccccccc}
\hline

BAT &  S/N & flag & $S_\text{CO32}$ & $\log L'_\text{CO(32)}$ & $\log L'_\text{CO(21)}$ &  $r_{31}$  &beam corr. & beam corr. tot \\ 
index &   \scriptsize{CO(32)}   & \scriptsize{CO(32)}  & [Jy km s$^{-1}$] &[log K km s$^{-1}$pc$^2$] & [log K km s$^{-1}$pc$^2$] & & 14" to 20"  &  20" to total\\ 
 (1)  & (2) &  (3) & (4)  & (5) & (6) &(7) & (8) & (9)  \\ 
  \hline \hline
   
144 & 87.30 &1 & 2787.72 $\pm$251.40 & 8.08 $\pm$0.00 & 8.49 $\pm$0.00 & 0.61 $\pm$ 0.00 & 1.95$^*$ & 3.62$^*$ \\ 
173 & 24.65 &1 & 198.29 $\pm$21.54 & 8.44 $\pm$0.02 & ... & ... & 1.80$^*$ & 2.32$^*$ \\ 
228 & 13.39 &1 & 159.38 $\pm$40.79 & 8.99 $\pm$0.03 & 9.26 $\pm$0.04 & 0.48 $\pm$ 0.13 & 1.13 & 1.54\\ 
308 & 23.52 &1 & 132.00 $\pm$11.37 & 7.65 $\pm$0.02 & 7.65 $\pm$0.02 & 1.01 $\pm$ 0.26 & 1.25 & 1.73\\ 
310 & 12.13 &1 & 72.27 $\pm$10.06 & 8.17 $\pm$0.04 & 8.32 $\pm$0.11 & 0.52 $\pm$ 0.07 & 1.31 & 4.33\\ 
316 & 3.83 &1 & 62.35 $\pm$10.48 & 8.54 $\pm$0.11 & 8.76 $\pm$0.00 & 0.54 $\pm$ 0.20 & 1.14 & 1.80\\ 
337 & -0.97 &2 &  -25.80 $\pm$40.08 & $<$ 9.04 & 8.86 $\pm$0.10 & $<$ 0.92 & 1.14 & 1.63\\ 
382 & 7.05 &1 & 51.22 $\pm$9.23 & 8.10 $\pm$0.06 & ... & ... & 1.31 & 2.50\\ 
399 & 7.70 &1 & 153.57 $\pm$18.33 & 8.81 $\pm$0.06 & 9.20 $\pm$0.04 & 0.25 $\pm$ 0.03 & 1.16$^*$ & 1.61$^*$ \\ 
400 & 2.09 &2 &  31.28 $\pm$21.74 & $<$ 8.43 & 9.09 $\pm$0.08 & $<$ 0.53 & 1.17 & 2.21\\ 
404 & 8.81 &1 & 49.19 $\pm$12.64 & 7.66 $\pm$0.05 & ... & ... & 1.09 & 1.92\\ 
405 & 5.08 &1 & 72.40 $\pm$19.39 & 8.65 $\pm$0.09 & 8.81 $\pm$0.12 & 0.22 $\pm$ 0.04 & 1.14 & 1.91\\ 
439 & 10.14 &1 & 88.24 $\pm$13.24 & 7.73 $\pm$0.04 & 7.80 $\pm$0.10 & 0.69 $\pm$ 0.10 & 1.08 & 1.46\\ 
451 & 6.49 &1 & 47.50 $\pm$10.23 & 7.58 $\pm$0.07 & 8.09 $\pm$0.06 & 0.33 $\pm$ 0.11 & 1.33 & 2.34\\ 
471 & 48.97 &1 & 469.43 $\pm$34.53 & 8.10 $\pm$0.01 & 8.33 $\pm$0.01 & 0.59 $\pm$ 0.15 & 1.27 & 1.69\\ 
480 & 13.18 &1 & 71.51 $\pm$11.22 & 7.13 $\pm$0.03 & 7.64 $\pm$0.03 & 0.32 $\pm$ 0.09 & 1.28 & 2.94\\ 
497 & 32.63 &1 & 700.29 $\pm$52.40 & 7.82 $\pm$0.01 & 8.33 $\pm$0.00 & 0.33 $\pm$ 0.08 & 1.33 & 4.71\\ 
517 & 2.69 &2 &  27.21 $\pm$11.45 & $<$ 8.03 & 8.34 $\pm$0.11 & $<$ 0.30 & 1.05 & 1.60\\ 
530 & 4.60 &1 & 55.56 $\pm$12.08 & 7.62 $\pm$0.09 & 7.72 $\pm$0.12 & 0.75 $\pm$ 0.32 & 1.19 & 1.88\\ 
532 & 13.95 &1 & 354.36 $\pm$30.97 & 9.18 $\pm$0.03 & 9.34 $\pm$0.02 & 0.91 $\pm$ 0.07 & 1.13 & 1.54\\ 
548 & 3.73 &1 & 63.49 $\pm$25.16 & 6.70 $\pm$0.12 & 7.00 $\pm$0.07 & 0.59 $\pm$ 0.24 & 1.48 & 5.77\\ 
552 & 12.19 &1 & 118.93 $\pm$33.69 & 8.72 $\pm$0.04 & 9.07 $\pm$0.04 & 0.48 $\pm$ 0.04 & 1.16 & 1.75\\ 
560 & 6.33 &1 & 217.02 $\pm$36.42 & 8.18 $\pm$0.07 & 8.57 $\pm$0.03 & 0.41 $\pm$ 0.13 & 1.25 & 1.64\\ 
585 & 42.09 &1 & 539.72 $\pm$64.87 & 7.50 $\pm$0.01 & 7.87 $\pm$0.01 & 0.47 $\pm$ 0.12 & 1.36 & 15.66\\ 
588 & 19.59 &1 & 157.45 $\pm$25.70 & 8.70 $\pm$0.02 & 9.05 $\pm$0.02 & 0.37 $\pm$ 0.02 & 1.20 & 1.80\\ 
590 & 123.24 &1 & 1978.99 $\pm$126.31 & 8.35 $\pm$0.00 & 8.56 $\pm$0.00 & 0.61 $\pm$ 0.15 & 1.25 & 1.87\\ 
599 & 13.42 &1 & 273.82 $\pm$36.99 & 7.83 $\pm$0.03 & ... & ... & 1.31 & 1.99\\ 
608 & 14.83 &1 & 148.61 $\pm$24.68 & 8.10 $\pm$0.03 & 8.26 $\pm$0.05 & 0.49 $\pm$ 0.04 & 1.12 & 1.47\\ 
609 & 29.31 &1 & 1166.84 $\pm$109.02 & 7.25 $\pm$0.01 & 7.66 $\pm$0.01 & 0.52 $\pm$ 0.13 & 1.67 & 7.88\\ 
615 & 37.34 &1 & 14.89 $\pm$1.29 & 6.25 $\pm$0.01 & 6.58 $\pm$0.02 & 0.53 $\pm$ 0.14 & 1.44 & 3.89\\ 
631 & 21.62 &1 & 115.15 $\pm$14.23 & 7.53 $\pm$0.02 & 8.03 $\pm$0.03 & 0.47 $\pm$ 0.00 & 1.89$^*$ & 2.99$^*$ \\ 
669 & 4.00 &1 & 56.64 $\pm$19.80 & 8.48 $\pm$0.11 & ... & ... & 1.48 & 1.21\\ 
670 & 9.10 &1 & 61.93 $\pm$10.29 & 7.95 $\pm$0.05 & 8.42 $\pm$0.09 & 0.30 $\pm$ 0.10 & 1.13 & 1.50\\ 
688 & 41.19 &1 & 251.32 $\pm$46.05 & 7.92 $\pm$0.01 & 8.44 $\pm$0.01 & 0.37 $\pm$ 0.09 & 1.54 & 4.87\\ 
703 & -2.11 &2 &  -26.24 $\pm$14.73 & $<$ 8.90 & ... & ... & 1.41$^*$ & 1.09$^*$ \\ 
712 & 35.71 &1 & 295.07 $\pm$29.45 & 7.66 $\pm$0.01 & 7.85 $\pm$0.03 & 0.64 $\pm$ 0.17 & 1.24 & 1.81\\ 
723 & 41.60 &1 & 341.24 $\pm$30.06 & 8.69 $\pm$0.01 & 8.98 $\pm$0.02 & 0.59 $\pm$ 0.03 & 1.21 & 1.84\\ 
739 & 50.02 &1 & 420.76 $\pm$36.26 & 8.02 $\pm$0.01 & 8.32 $\pm$0.02 & 0.52 $\pm$ 0.13 & 1.28 & 1.93\\ 
766 & 17.23 &1 & 131.36 $\pm$16.03 & 7.71 $\pm$0.03 & 8.27 $\pm$0.02 & 0.37 $\pm$ 0.10 & 1.64 & 5.48\\ 
772 & 17.93 &1 & 167.87 $\pm$18.64 & 8.65 $\pm$0.02 & 8.73 $\pm$0.06 & 0.82 $\pm$ 0.23 & 1.22 & 2.40\\ 
783 & 32.29 &1 & 317.22 $\pm$20.31 & 8.98 $\pm$0.01 & 9.21 $\pm$0.02 & 0.55 $\pm$ 0.14 & 1.16 & 1.70\\ 
841 & 87.81 &1 & 2897.27 $\pm$170.89 & 9.92 $\pm$0.00 & 9.85 $\pm$0.01 & 1.22 $\pm$ 0.02 & 1.18 & 1.48\\ 
1042 & 23.03 &1 & 135.31 $\pm$22.74 & 7.86 $\pm$0.02 & 8.29 $\pm$0.03 & 0.37 $\pm$ 0.10 & 1.24 & 1.86\\ 
1046 & 8.78 &1 & 37.78 $\pm$8.56 & 6.72 $\pm$0.05 & 7.51 $\pm$0.04 & 0.23 $\pm$ 0.07 & 1.73 & 8.32\\ 
1133 & 68.37 &1 & 494.40 $\pm$81.07 & 9.28 $\pm$0.01 & ... & ... & 1.41$^*$ & 1.10$^*$ \\ 
1184 & 52.72 &1 & 993.78 $\pm$131.74 & 8.49 $\pm$0.01 & 8.65 $\pm$0.00 & 0.72 $\pm$ 0.18 & 1.31 & 2.97\\

\hline
\label{tab:BASS_CO}
\end{tabular}
\scriptsize{
(1)~BAT index.
(2)~Integrated S/N of the CO(3-2) line.
(3)~Flag for the detection of the CO(3-2) line based on a peak S/N $>$ 3. 1: detected, 2: non-detected.
(4)~Velocity-integrated flux $S_\text{CO32}$ within the 14" JCMT HARP beam.
(5)~CO(3-2) luminosity within the 14" JCMT HARP beam.
(6)~CO(2-1) luminosity within the 20" JCMT RxA beam. Galaxies which do not have CO(2-1) observations have empty entries (...).
(7)~Beam corrected luminosity ratio $r_{31} = L'_{CO(32)}/L'_{CO(10)} \cdot \text{beam correction}$.
(8)~Beam correction factor for extrapolating the CO(3-2) flux from the 14" to the 22" beam. 
(9)~Beam correction factor for extrapolating the CO(2-1) flux from the 20" beam to the total flux. The star $^*$ indicates that the corrections are derived from simulated galaxy profiles, because the FIR images were not available.}
\end{table*}

\newpage

\begin{figure*}
\centering
\includegraphics[width=0.2\textwidth]{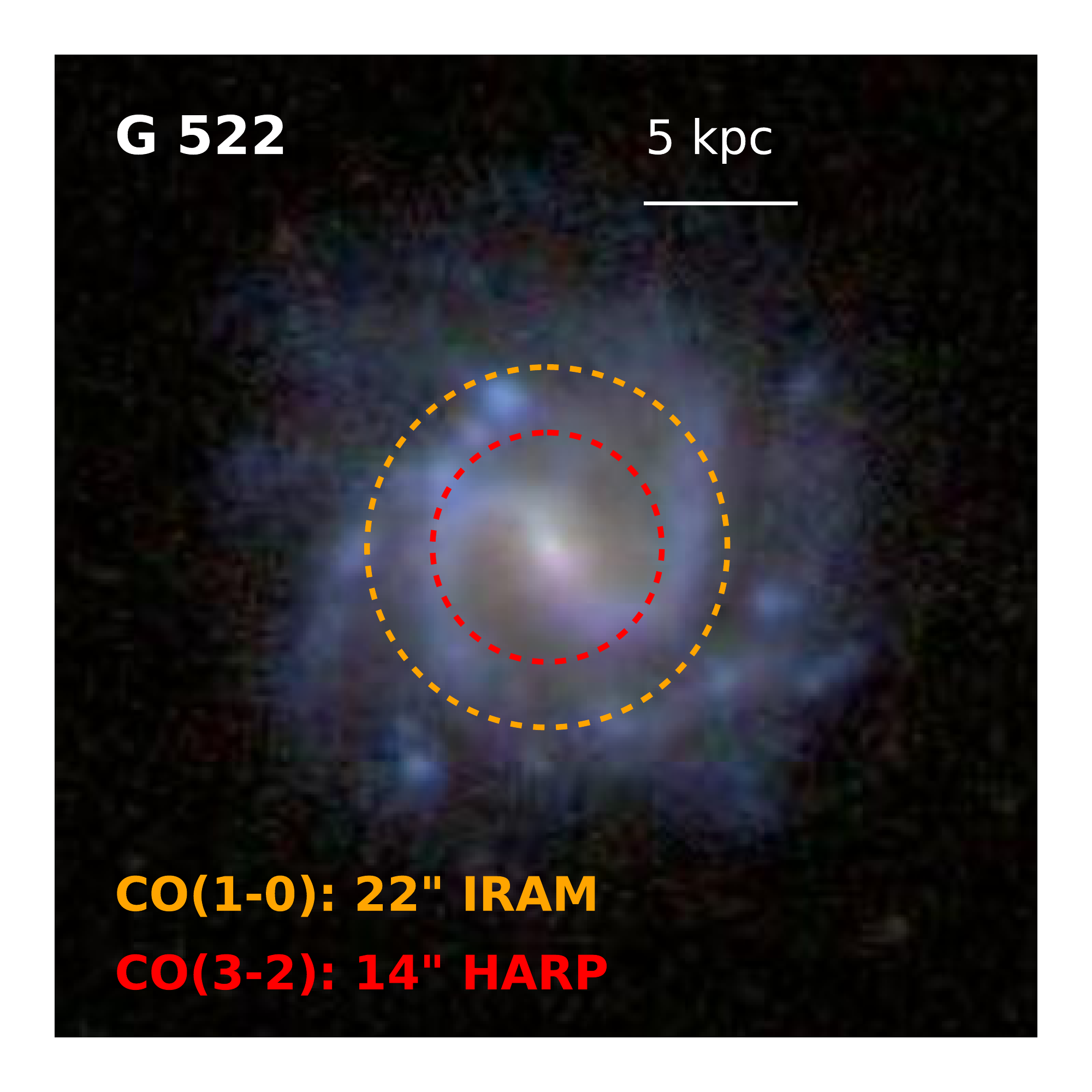}
\includegraphics[width=0.27\textwidth]{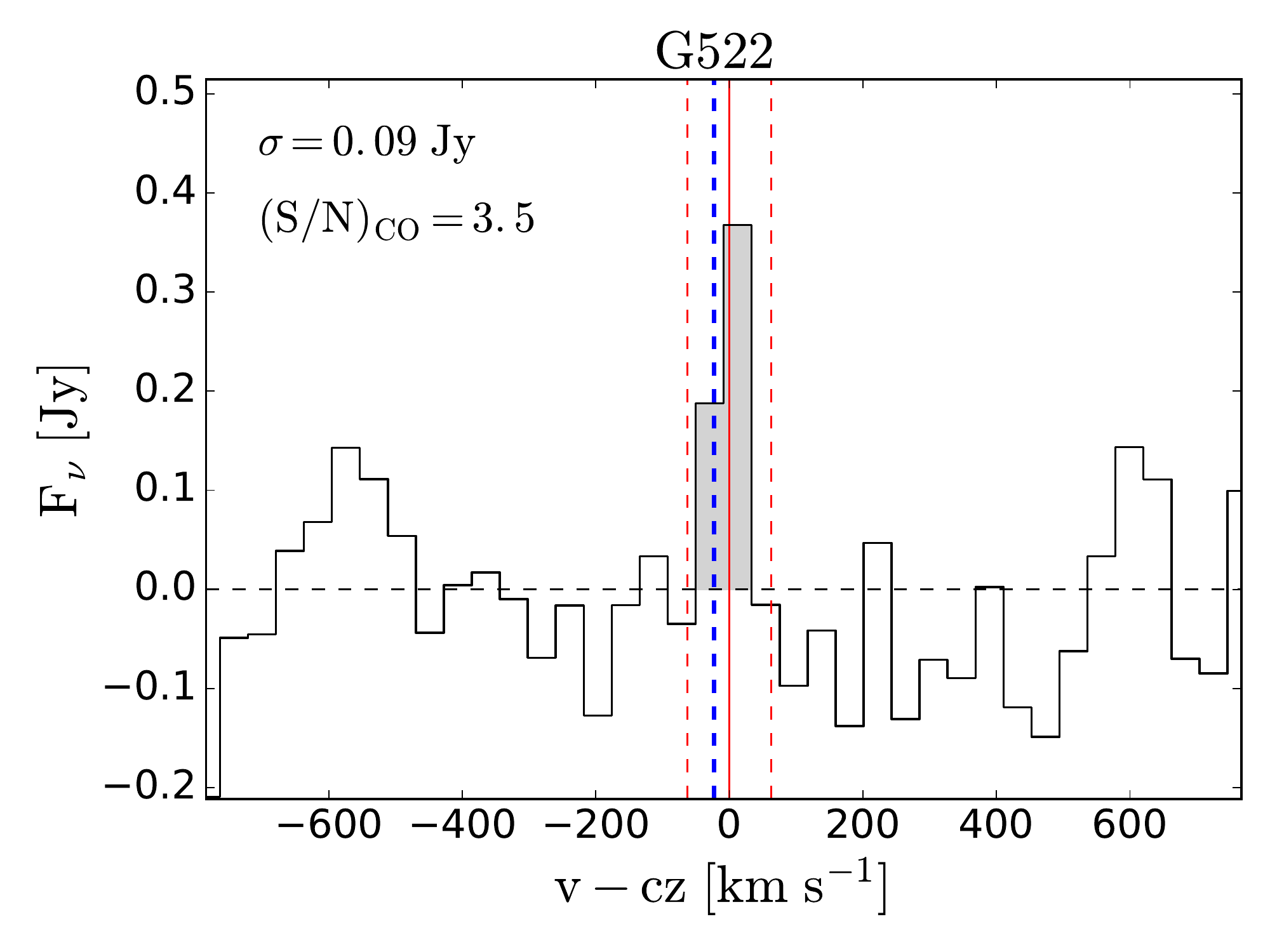}
\includegraphics[width=0.2\textwidth]{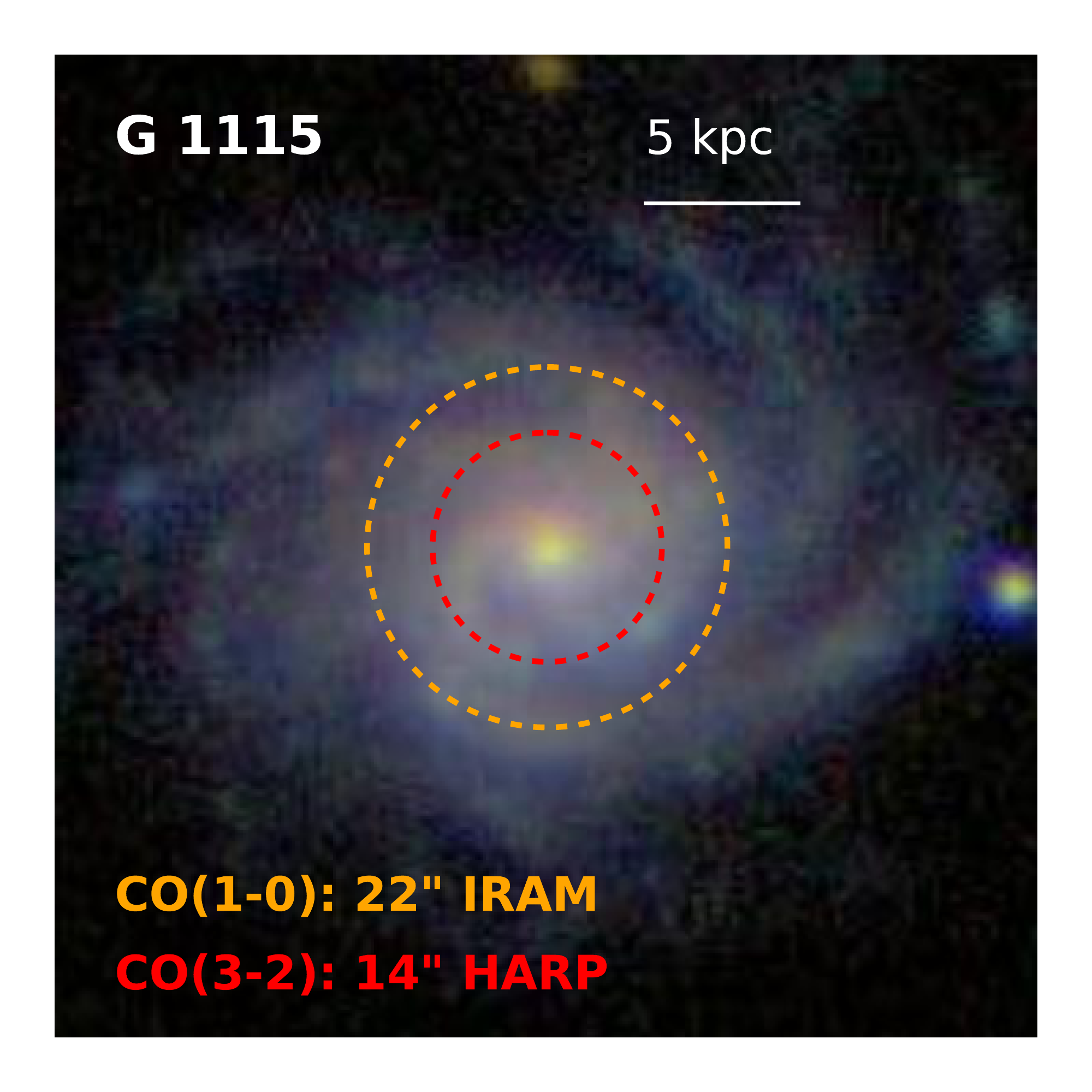}
\includegraphics[width=0.27\textwidth]{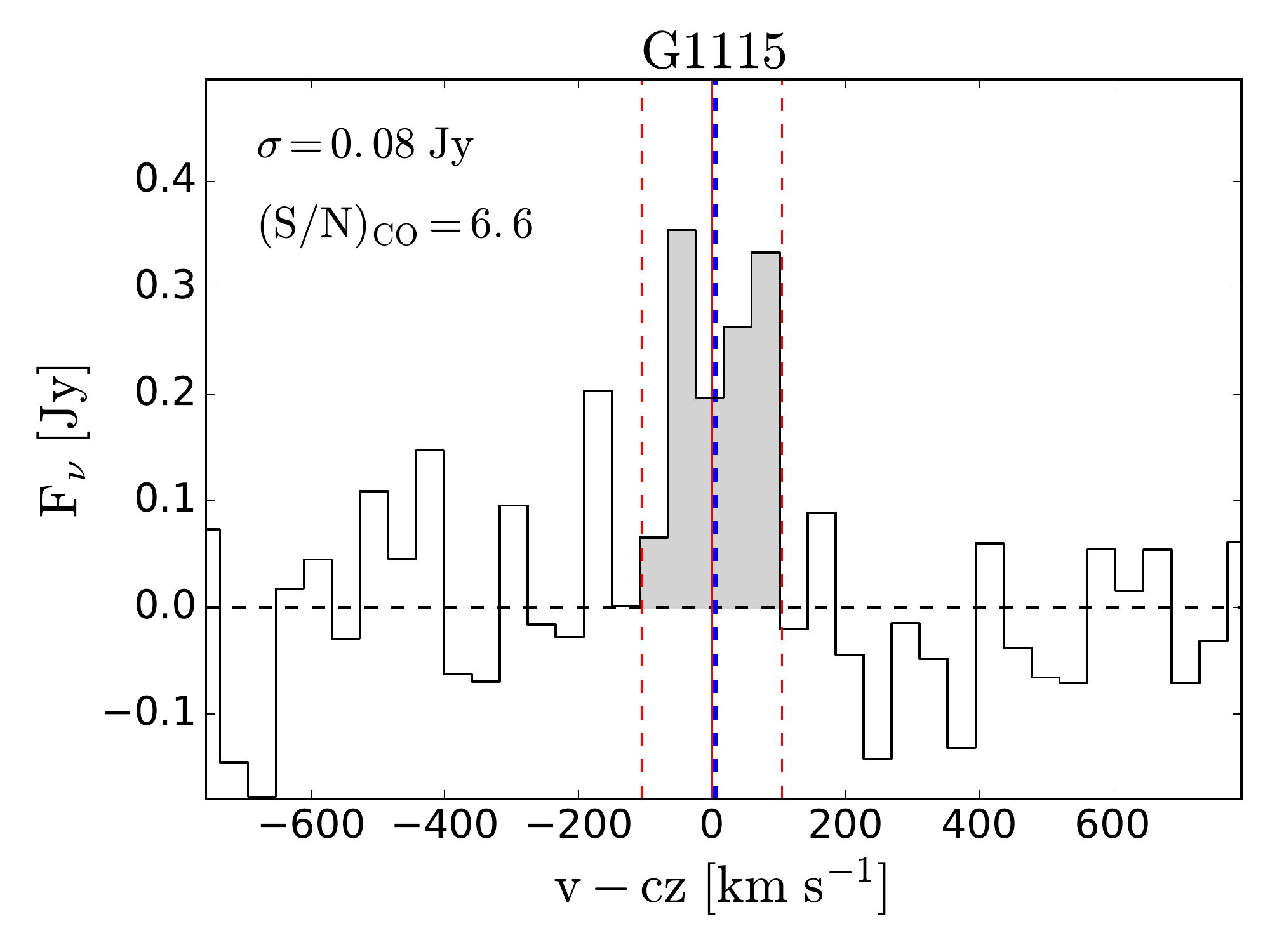}
\includegraphics[width=0.2\textwidth]{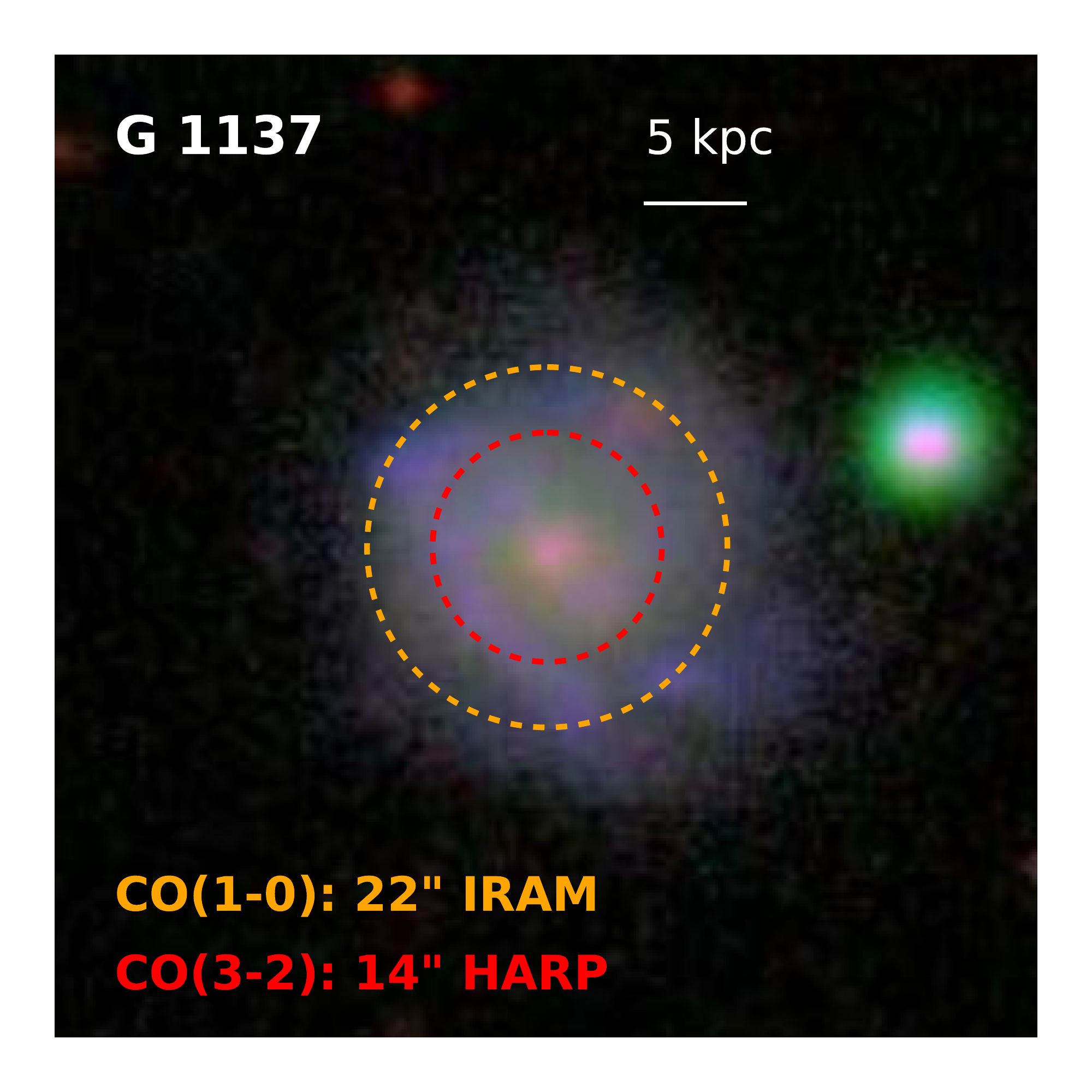}
\includegraphics[width=0.27\textwidth]{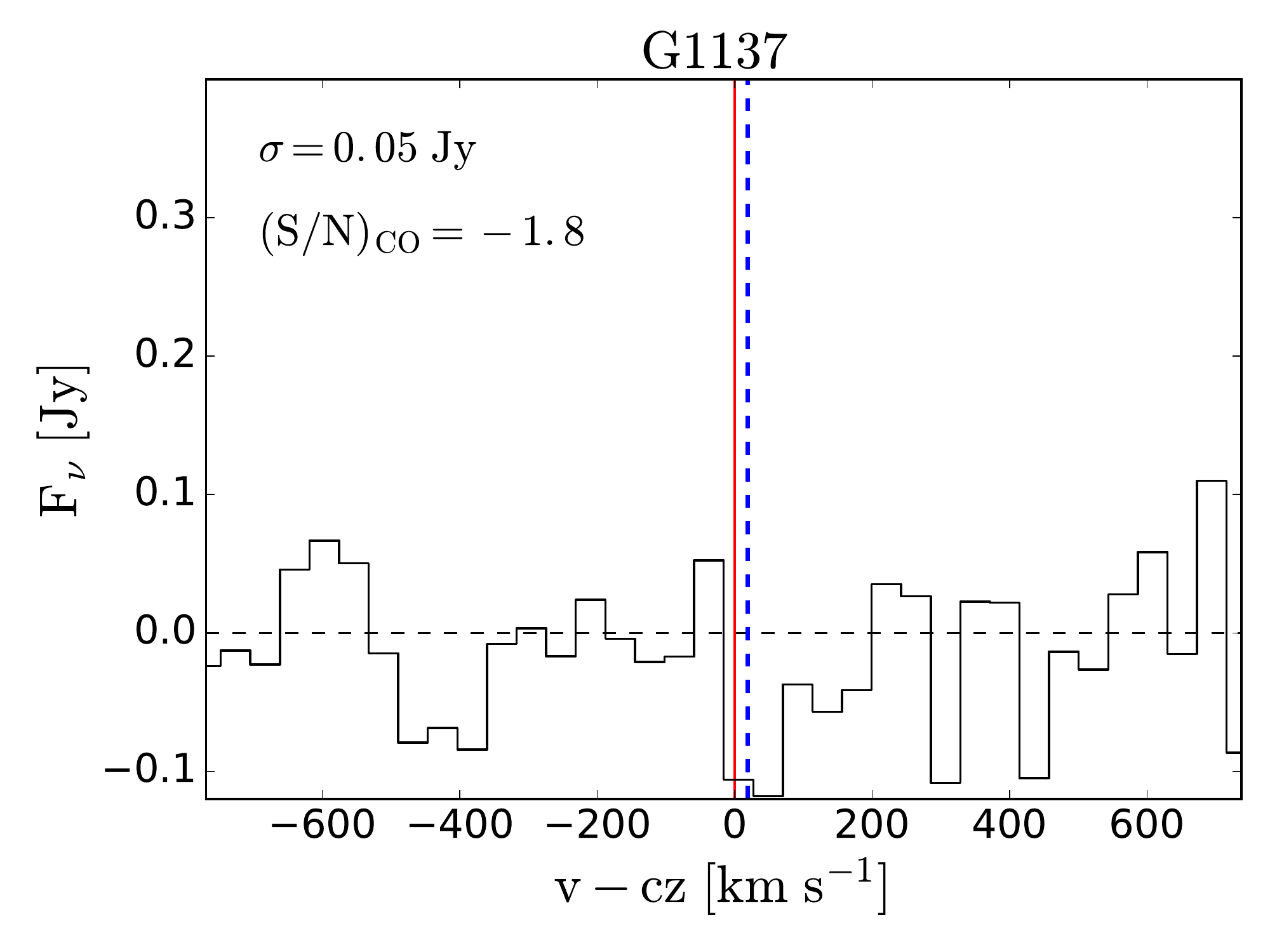}
\includegraphics[width=0.2\textwidth]{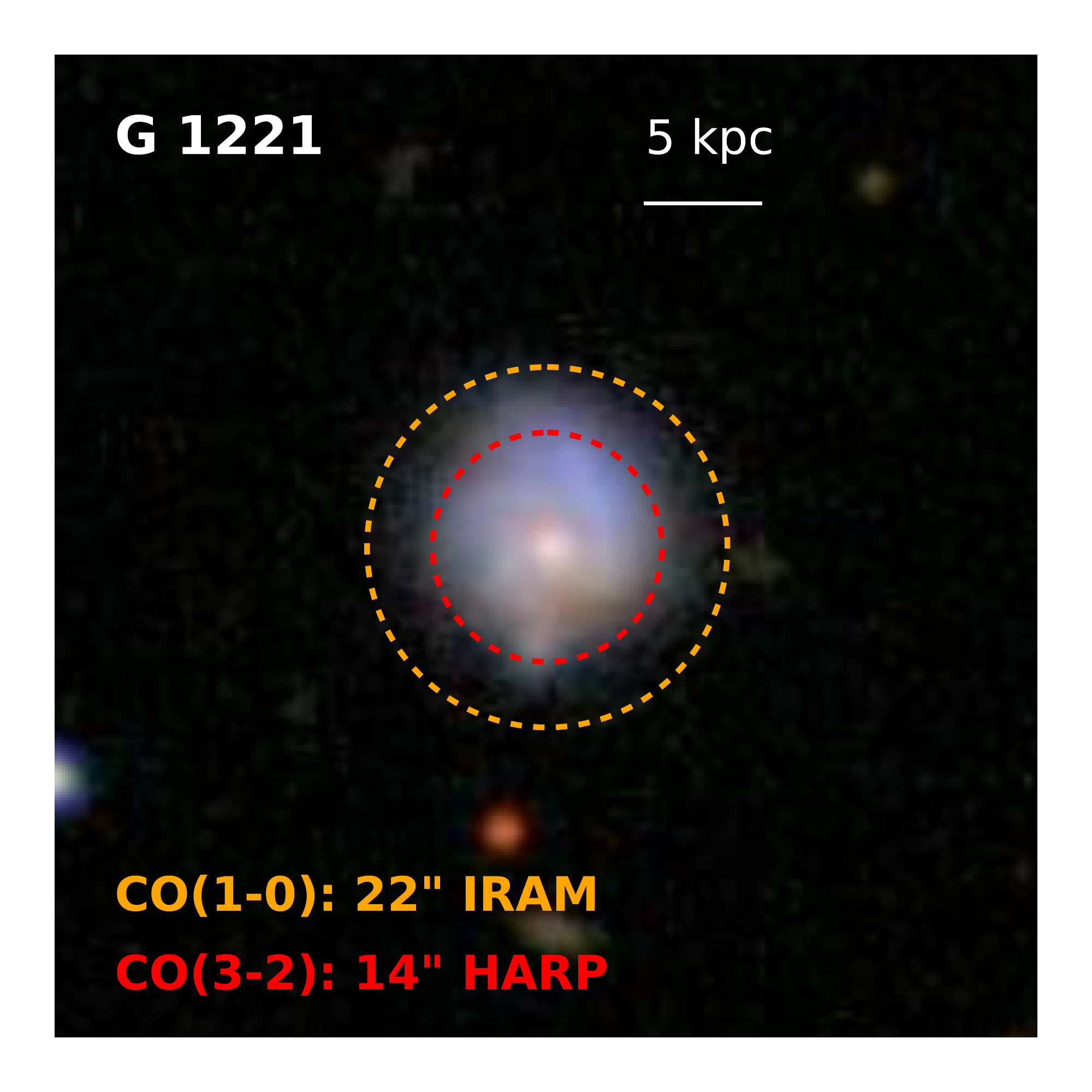}
\includegraphics[width=0.27\textwidth]{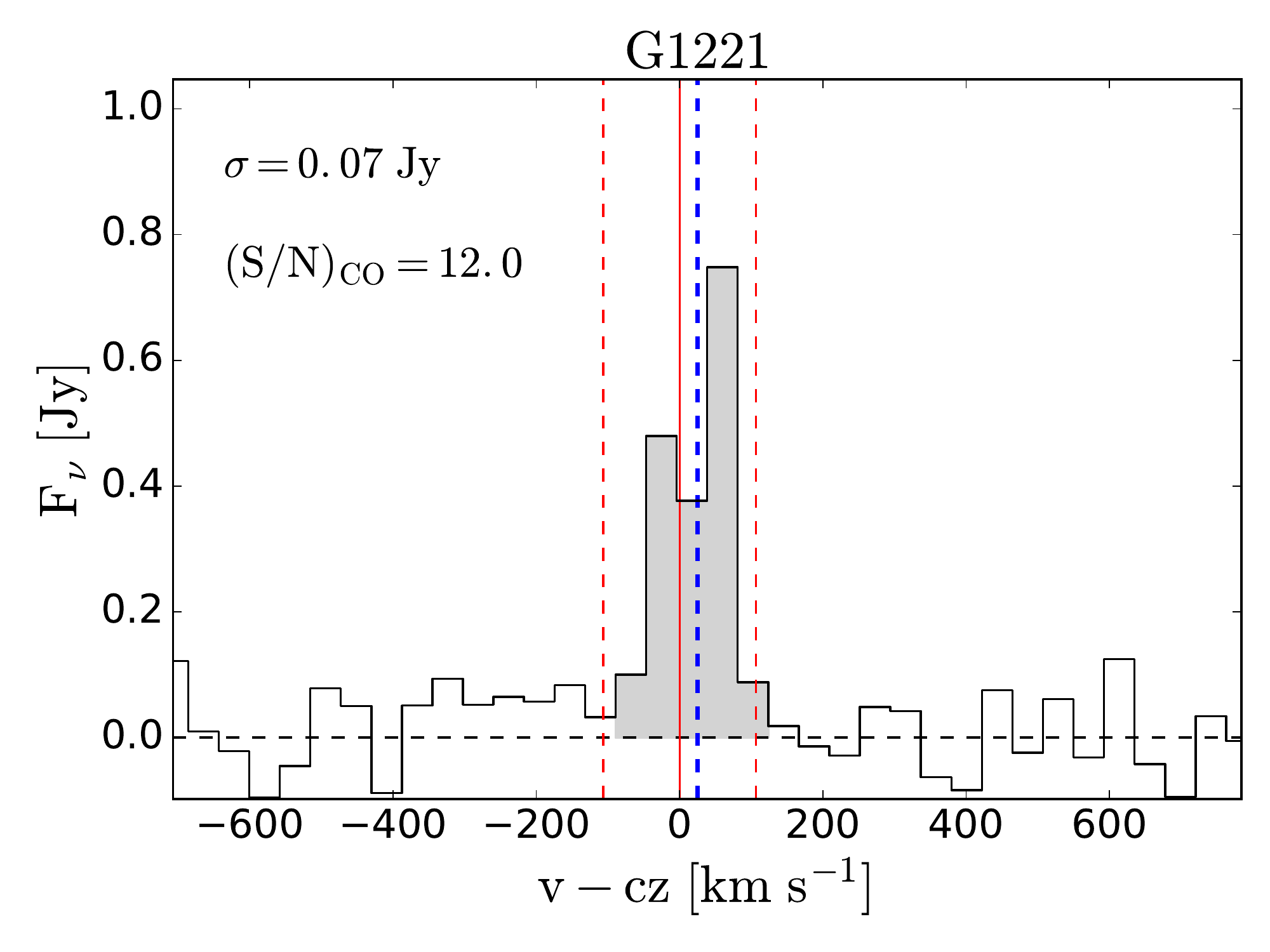}

\includegraphics[width=0.2\textwidth]{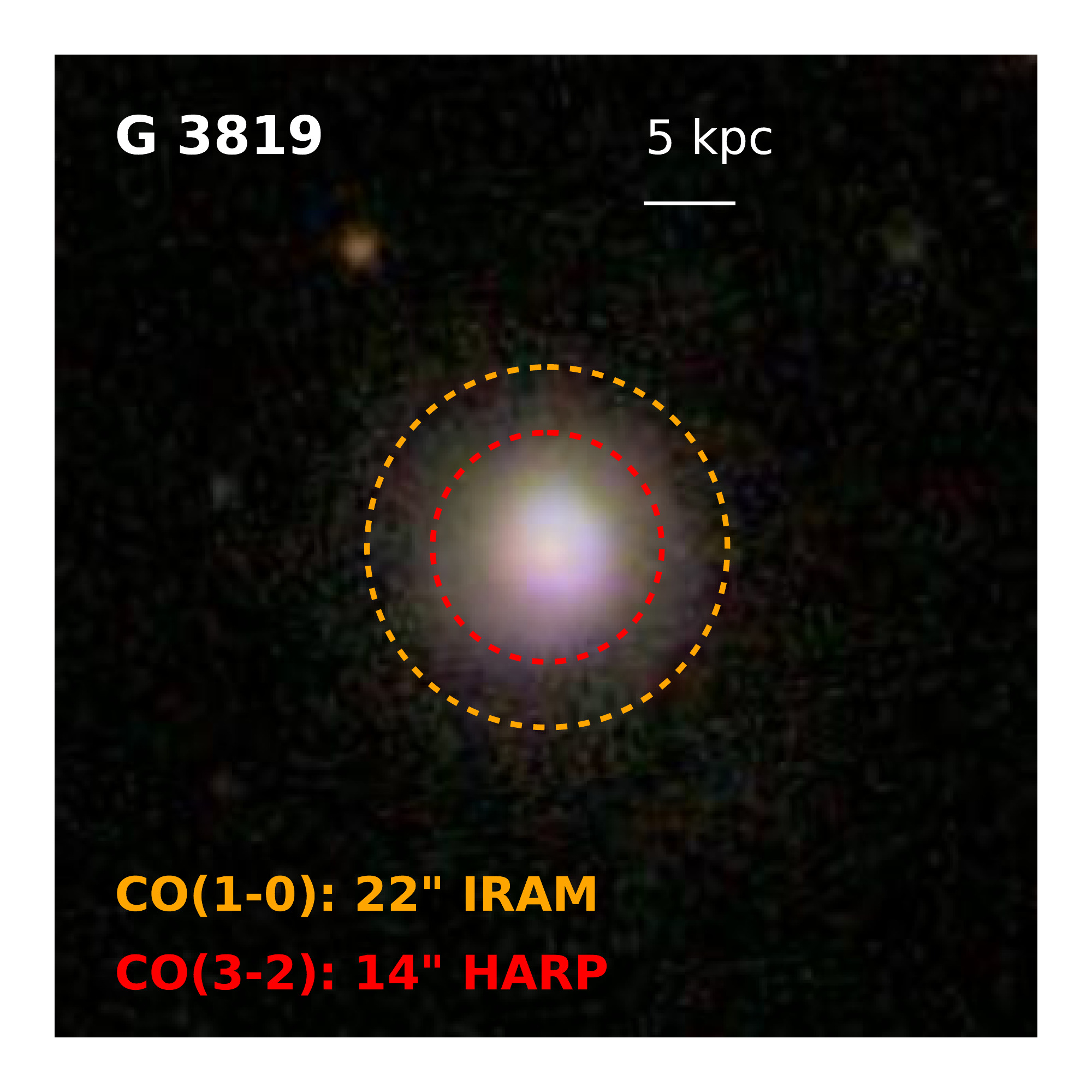}
\includegraphics[width=0.27\textwidth]{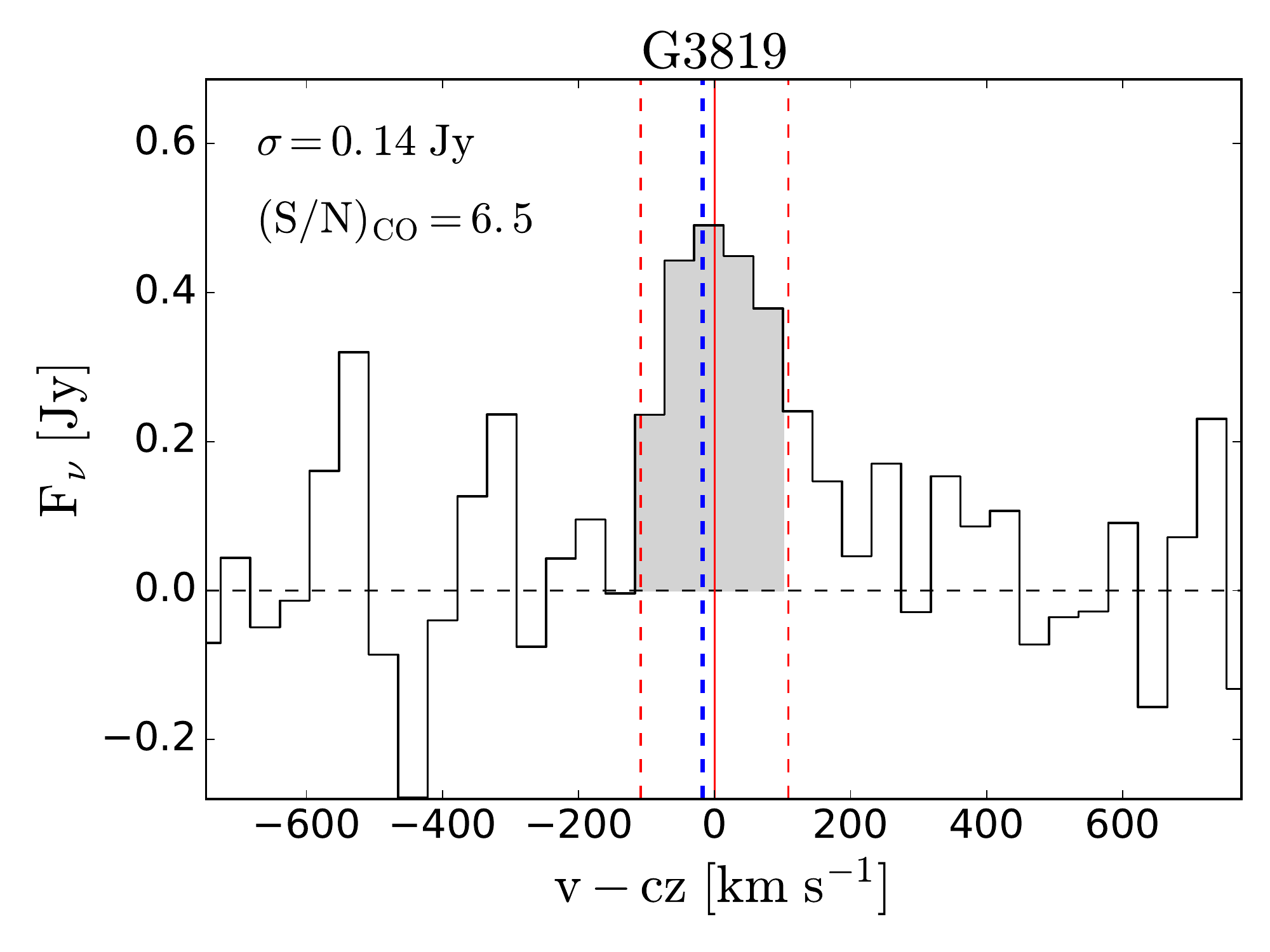}
\includegraphics[width=0.2\textwidth]{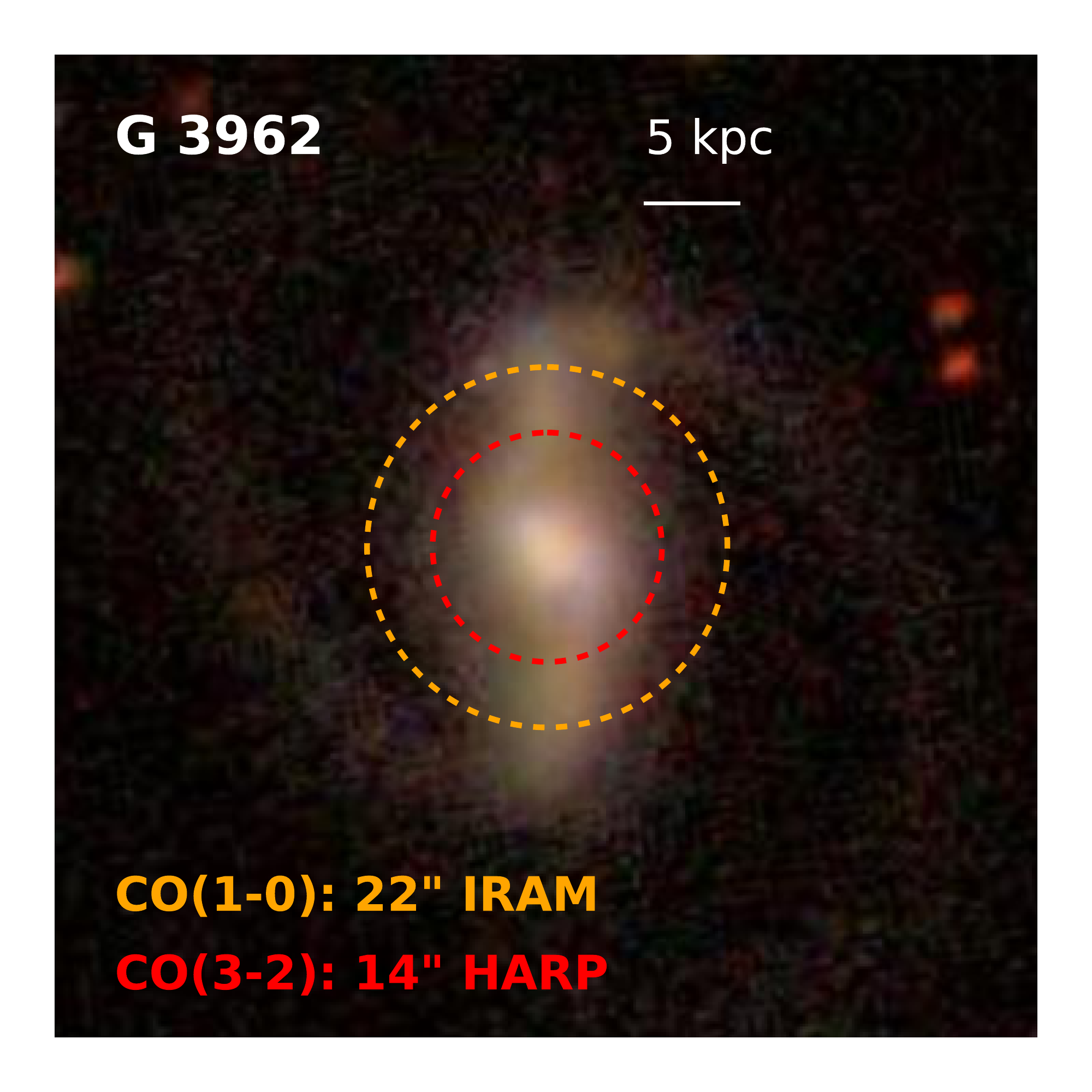}
\includegraphics[width=0.27\textwidth]{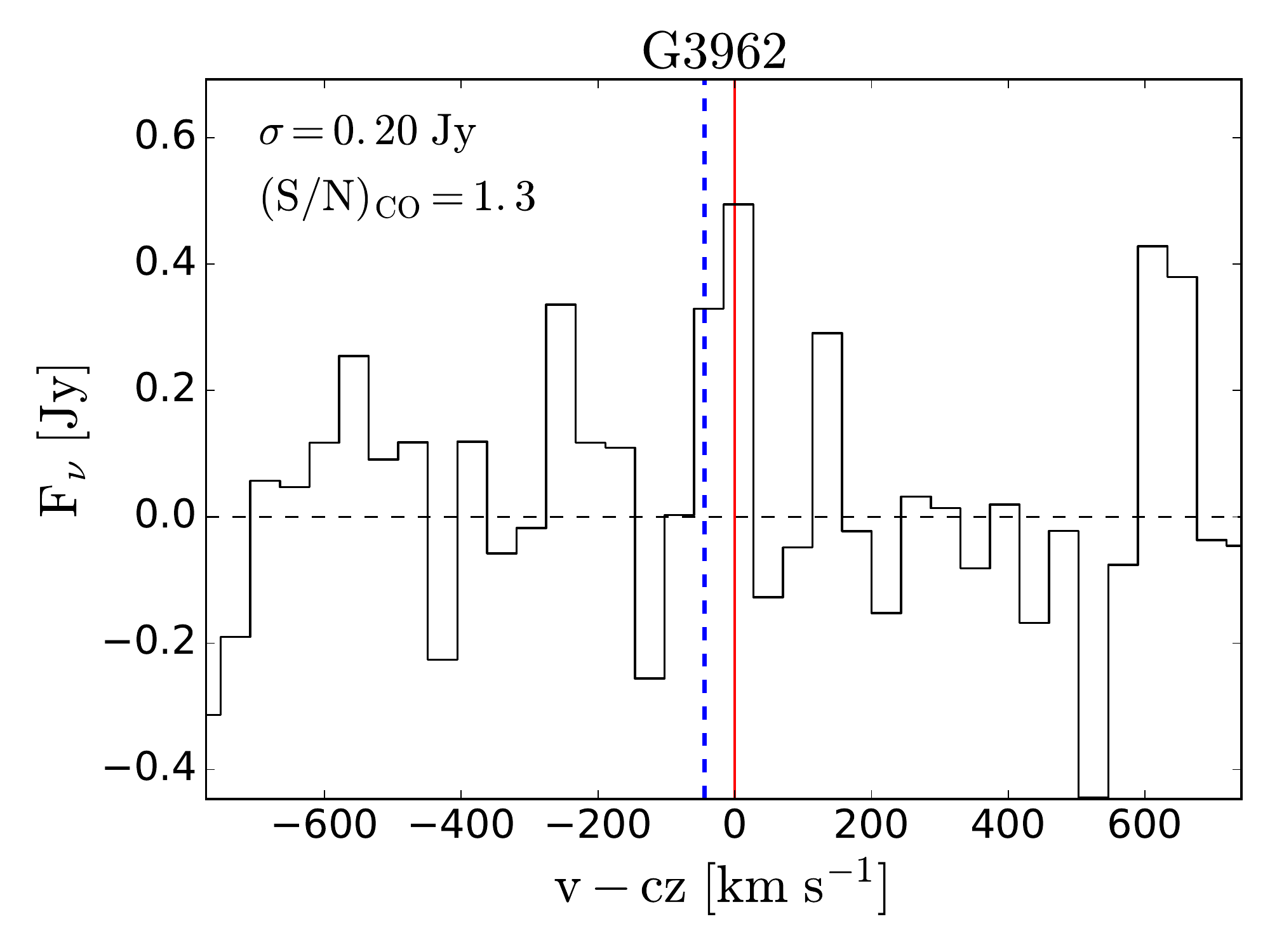}
\includegraphics[width=0.2\textwidth]{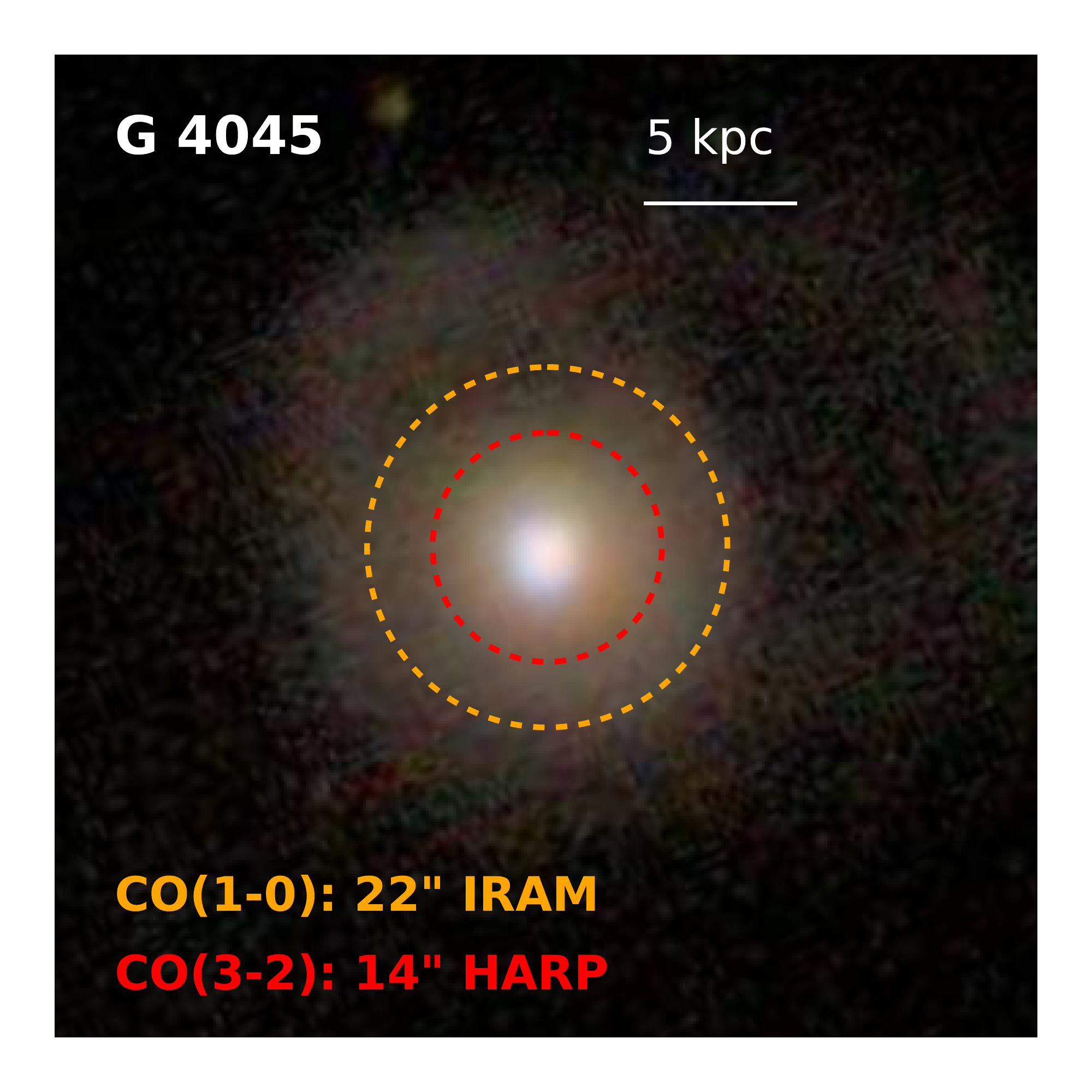}
\includegraphics[width=0.27\textwidth]{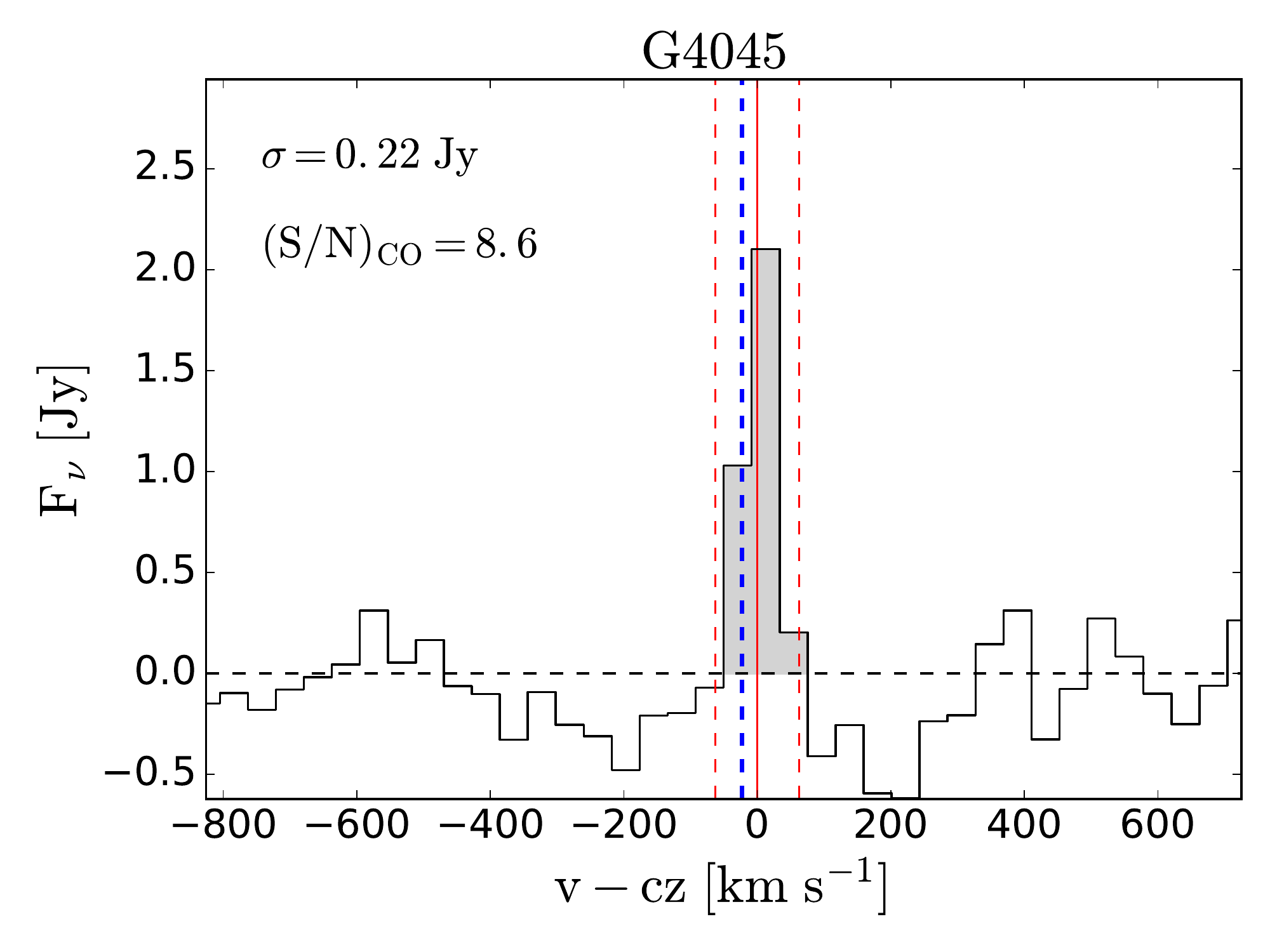}
\includegraphics[width=0.2\textwidth]{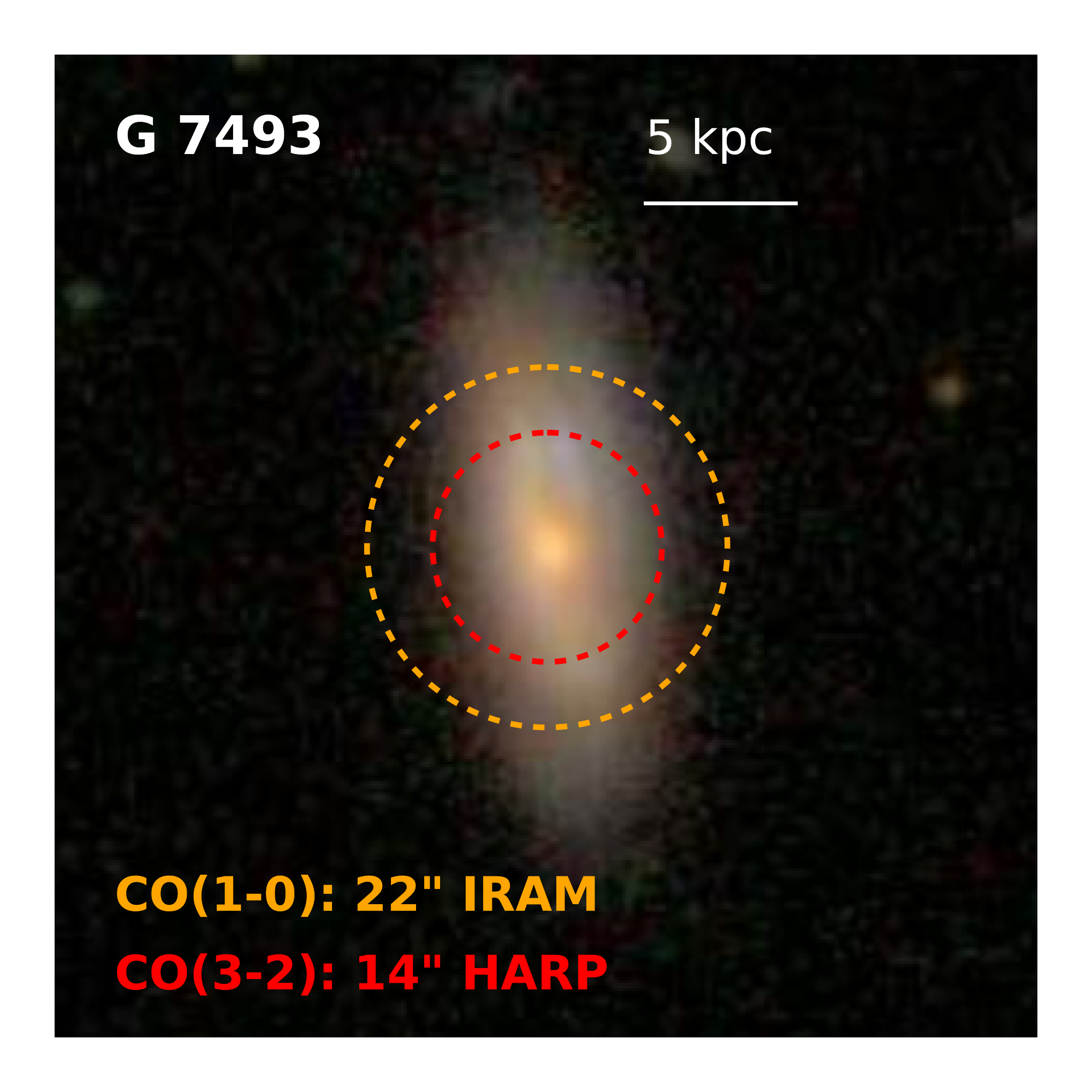}
\includegraphics[width=0.27\textwidth]{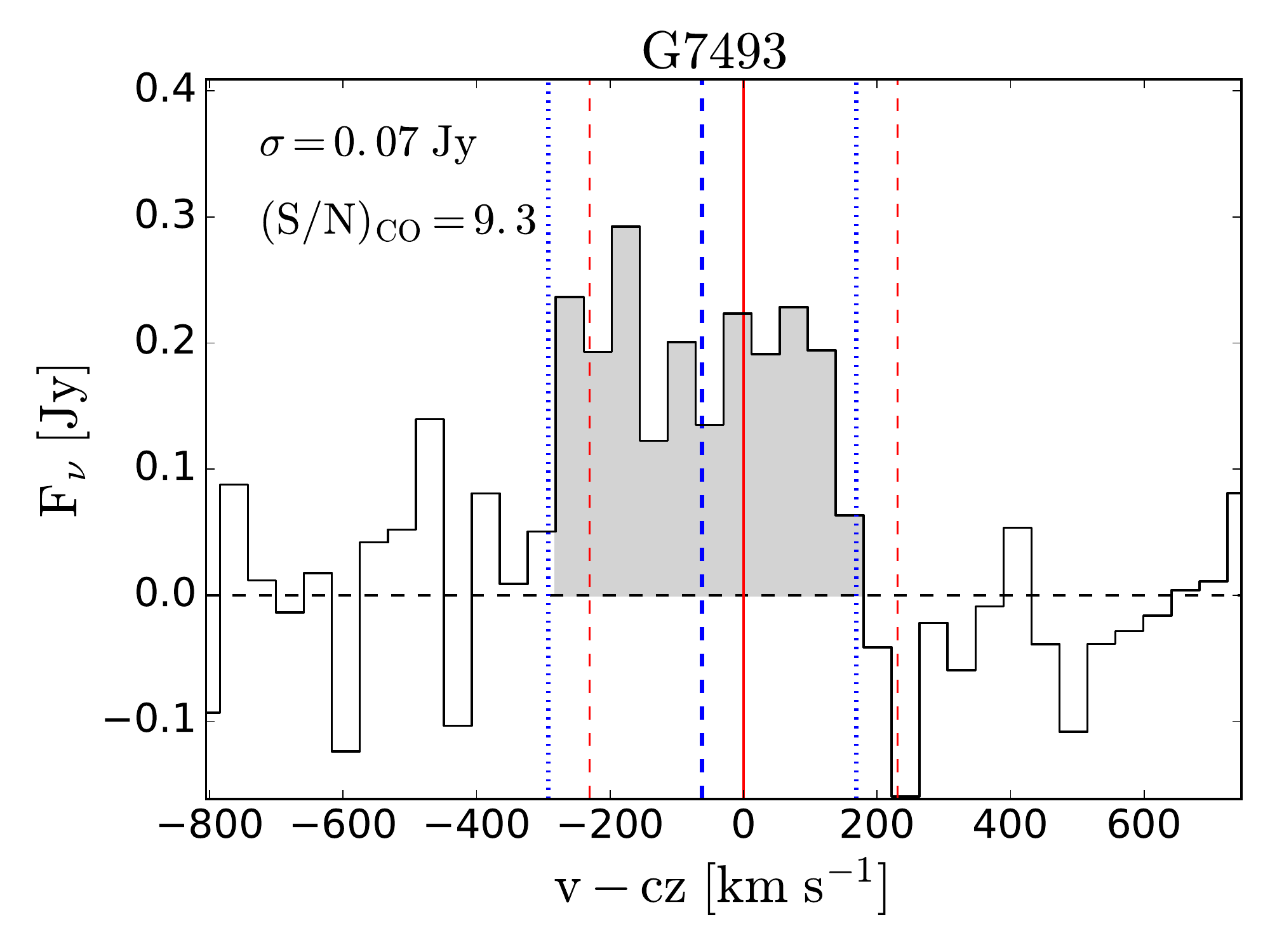}
\includegraphics[width=0.2\textwidth]{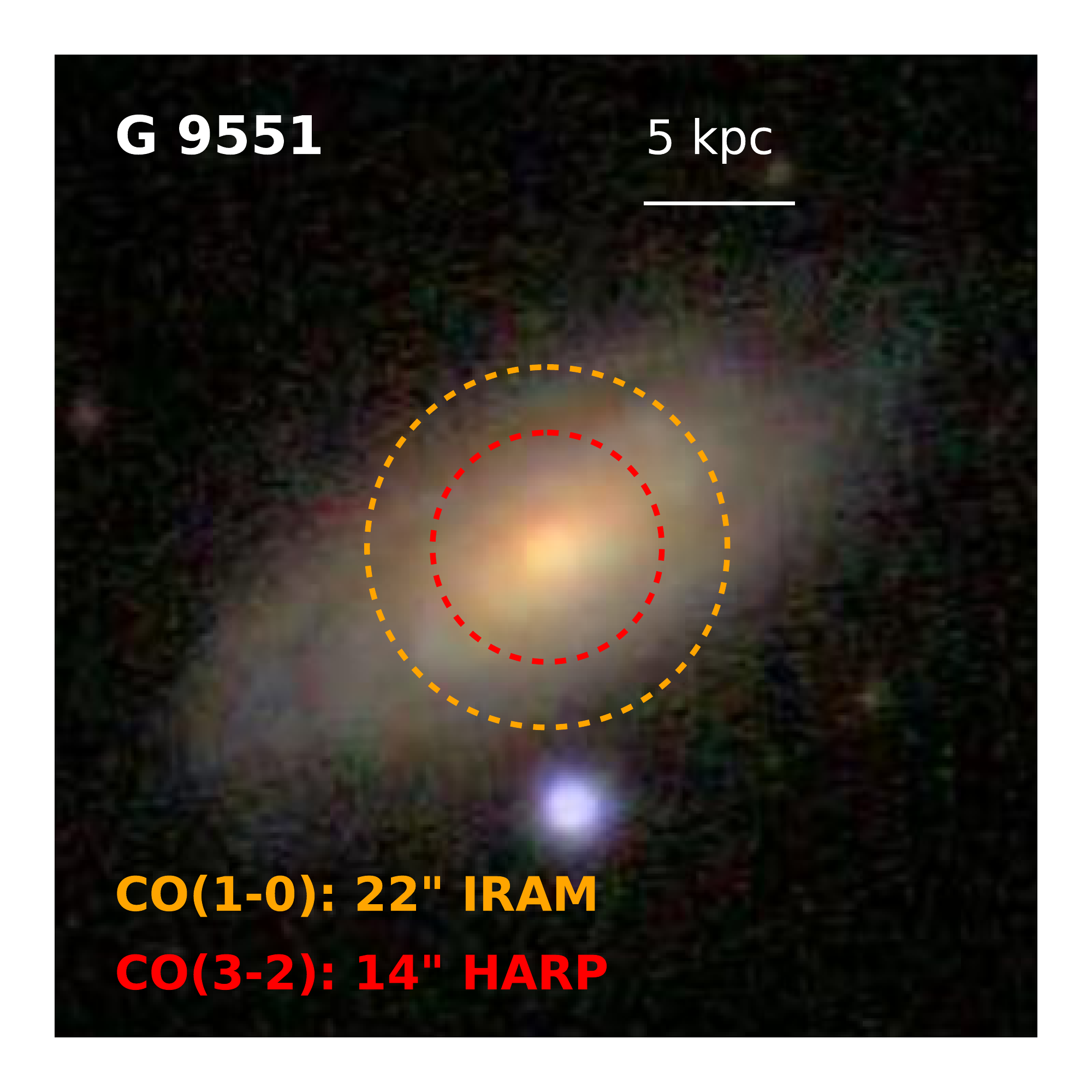}
\includegraphics[width=0.27\textwidth]{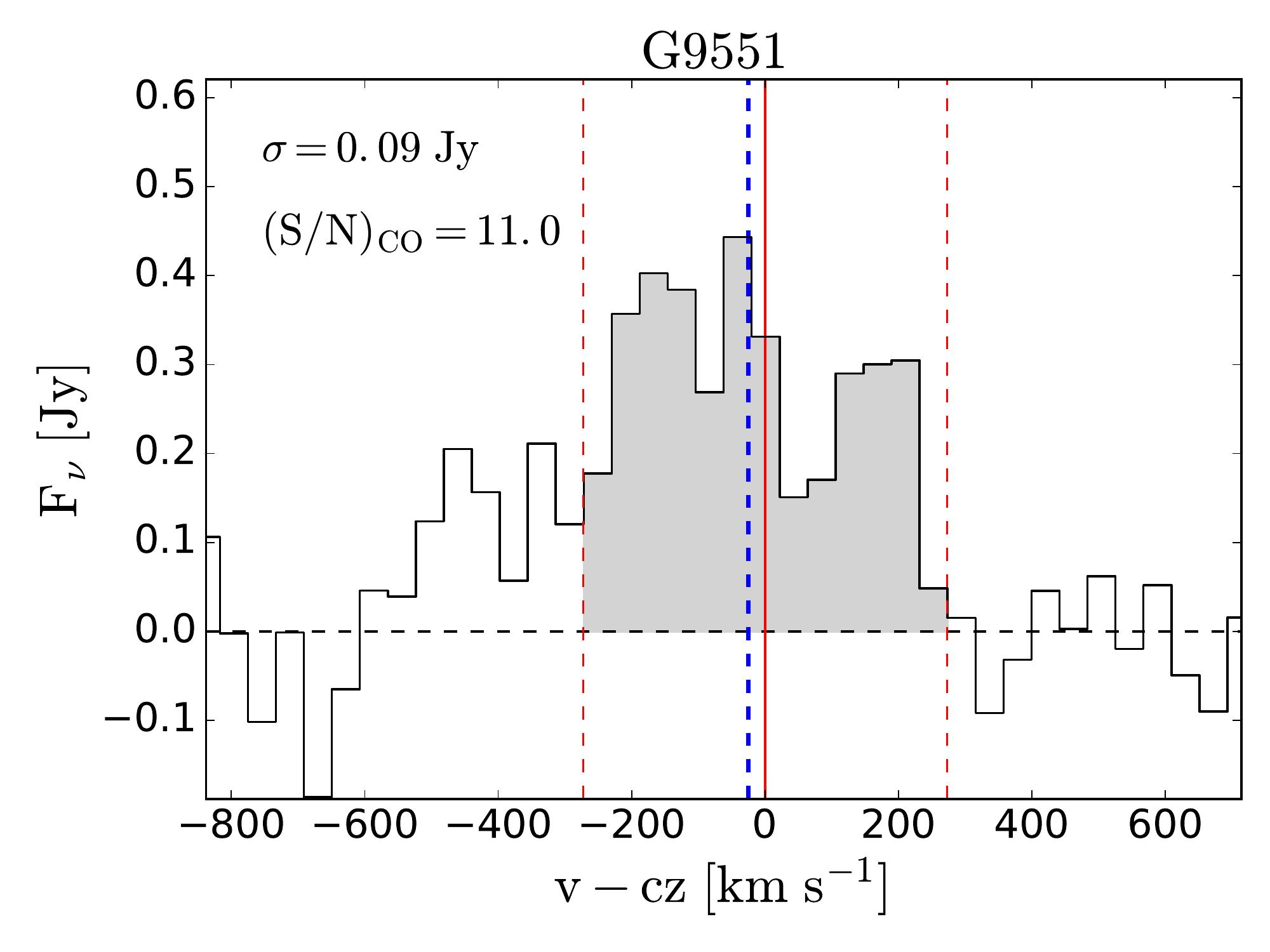}
\includegraphics[width=0.2\textwidth]{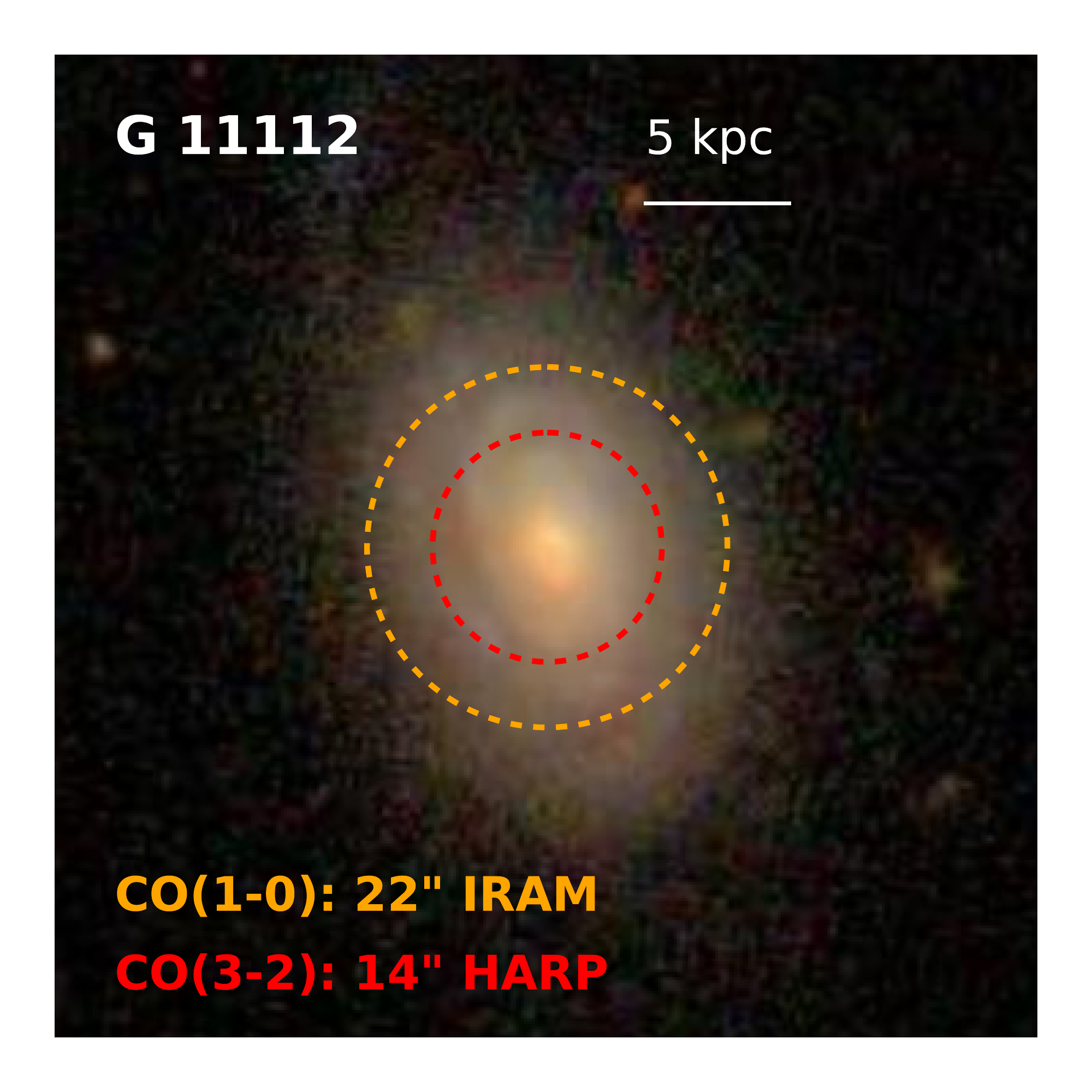}
\includegraphics[width=0.27\textwidth]{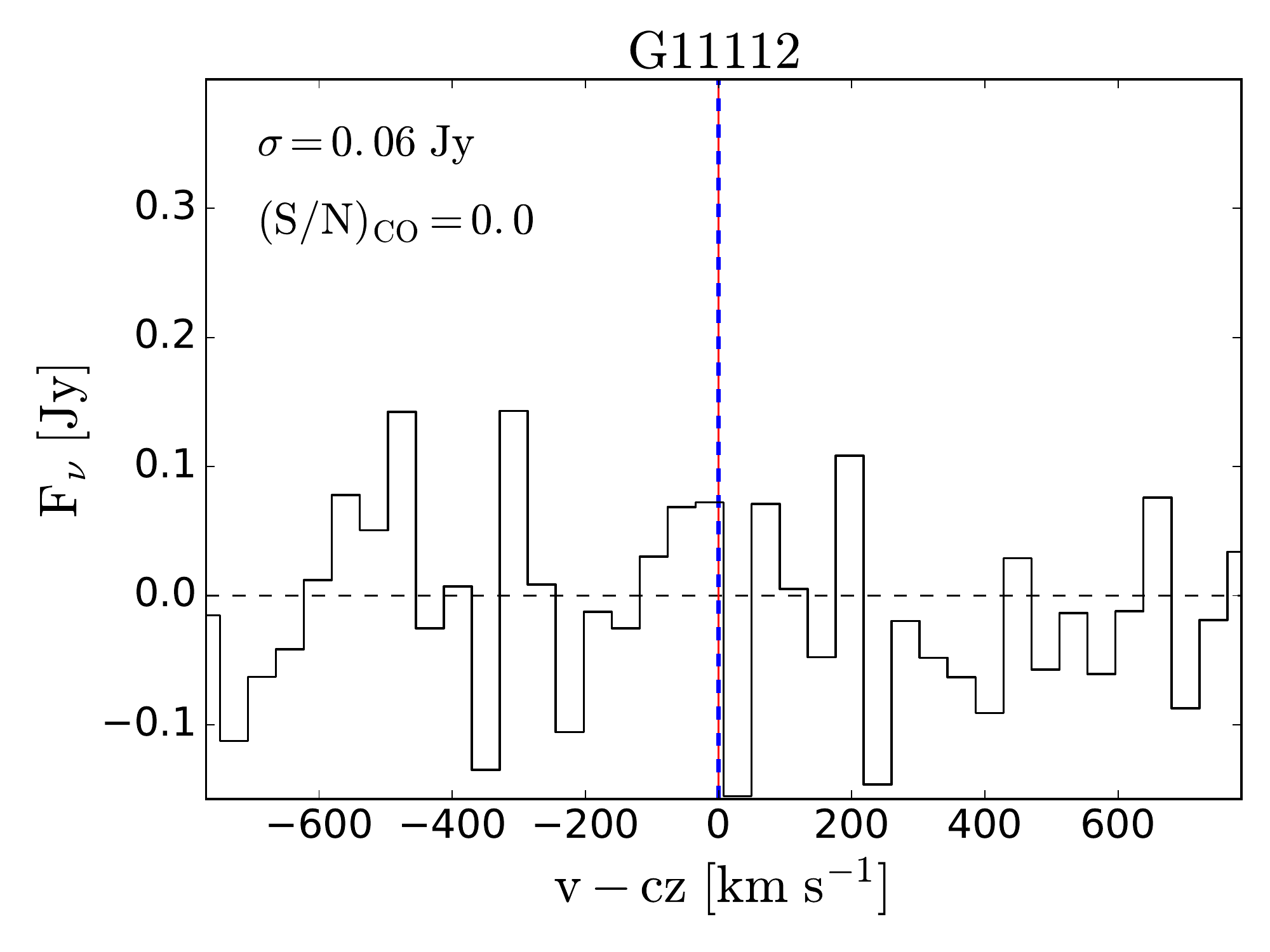}

\caption{\textit{Left}: SDSS \textit{gri} images of the \coldgass\ sample. Every image has dimension $60"\times60"$ ($1'\times1'$) and shows the the size of the IRAM-30m and JCMT HARP beams.  \textit{Right}: CO(3-2) spectra of the \coldgass\ sample taken with HARP on the JCMT. The spectra are centred at the position of the CO(3-2) line. The solid red line is the central velocity of the line based on the spectroscopic redshift from SDSS and the dashed red lines indicate the interval where the CO(3-2)  flux was integrated, based on the FWHM of the CO(1-0) line. The blue solid line indicates the central velocity of the CO(1-0) line. For the two galaxies (G7493 and G2527) where the CO(3-2) line flux was measured based on the position of the CO(1-0) line, the blue dotted line shows the interval where the CO(3-2) flux was integrated. Additional figures showing the remaining 15 galaxies of the xCOLDGASS sample and the full sample of 46 BASS objects are available online.} 
\label{fig:CO32_spectra}
\end{figure*}



\end{document}